\documentclass[aps,prd,twocolumn,showpacs,superscriptaddress,groupedaddress]{revtex4}
\usepackage{graphicx}
\usepackage{dcolumn}
\usepackage{bm}
\usepackage{amssymb}
\usepackage{amsmath}
\usepackage{xspace}
\usepackage{color}
\usepackage{multirow}
\usepackage{afterpage}

\suppressfloats[!]

\newcommand{\Ptg}{p_{T}^{\gamma}}

\newcommand{\ptg}{$p_T^{\gamma}$}

\newcommand{\lt}{\!<\!}
\newcommand{\gt}{\!>\!}

\newcommand{\tcs}{$\mathrm{d^3}\sigma / \mathrm{d}\Ptg\mathrm{d}y^\gamma\mathrm{d}y^\mathrm{jet}$\xspace}
\newcommand{\gpj}{$\gamma+{\rm jet}$\xspace}

\newcommand{\CHECK}[1]{\textbf{\color{red}[#1]}\xspace}
\newcommand{\chk}[1]{\CHECK{CHECK THIS!}}

\begin{document}

\hspace{5.2in} \mbox{FERMILAB-PUB-13-317-E}
\title{\boldmath Measurement of the differential cross section of photon plus jet production 
in $p\bar{p}$ collisions at $\sqrt{s}=1.96$~TeV}

%
\affiliation{LAFEX, Centro Brasileiro de Pesquisas F\'{i}sicas, Rio de Janeiro, Brazil}
\affiliation{Universidade do Estado do Rio de Janeiro, Rio de Janeiro, Brazil}
\affiliation{Universidade Federal do ABC, Santo Andr\'e, Brazil}
\affiliation{University of Science and Technology of China, Hefei, People's Republic of China}
\affiliation{Universidad de los Andes, Bogot\'a, Colombia}
\affiliation{Charles University, Faculty of Mathematics and Physics, Center for Particle Physics, Prague, Czech Republic}
\affiliation{Czech Technical University in Prague, Prague, Czech Republic}
\affiliation{Institute of Physics, Academy of Sciences of the Czech Republic, Prague, Czech Republic}
\affiliation{Universidad San Francisco de Quito, Quito, Ecuador}
\affiliation{LPC, Universit\'e Blaise Pascal, CNRS/IN2P3, Clermont, France}
\affiliation{LPSC, Universit\'e Joseph Fourier Grenoble 1, CNRS/IN2P3, Institut National Polytechnique de Grenoble, Grenoble, France}
\affiliation{CPPM, Aix-Marseille Universit\'e, CNRS/IN2P3, Marseille, France}
\affiliation{LAL, Universit\'e Paris-Sud, CNRS/IN2P3, Orsay, France}
\affiliation{LPNHE, Universit\'es Paris VI and VII, CNRS/IN2P3, Paris, France}
\affiliation{CEA, Irfu, SPP, Saclay, France}
\affiliation{IPHC, Universit\'e de Strasbourg, CNRS/IN2P3, Strasbourg, France}
\affiliation{IPNL, Universit\'e Lyon 1, CNRS/IN2P3, Villeurbanne, France and Universit\'e de Lyon, Lyon, France}
\affiliation{III. Physikalisches Institut A, RWTH Aachen University, Aachen, Germany}
\affiliation{Physikalisches Institut, Universit\"at Freiburg, Freiburg, Germany}
\affiliation{II. Physikalisches Institut, Georg-August-Universit\"at G\"ottingen, G\"ottingen, Germany}
\affiliation{Institut f\"ur Physik, Universit\"at Mainz, Mainz, Germany}
\affiliation{Ludwig-Maximilians-Universit\"at M\"unchen, M\"unchen, Germany}
\affiliation{Panjab University, Chandigarh, India}
\affiliation{Delhi University, Delhi, India}
\affiliation{Tata Institute of Fundamental Research, Mumbai, India}
\affiliation{University College Dublin, Dublin, Ireland}
\affiliation{Korea Detector Laboratory, Korea University, Seoul, Korea}
\affiliation{CINVESTAV, Mexico City, Mexico}
\affiliation{Nikhef, Science Park, Amsterdam, the Netherlands}
\affiliation{Radboud University Nijmegen, Nijmegen, the Netherlands}
\affiliation{Joint Institute for Nuclear Research, Dubna, Russia}
\affiliation{Institute for Theoretical and Experimental Physics, Moscow, Russia}
\affiliation{Moscow State University, Moscow, Russia}
\affiliation{Institute for High Energy Physics, Protvino, Russia}
\affiliation{Petersburg Nuclear Physics Institute, St. Petersburg, Russia}
\affiliation{Instituci\'{o} Catalana de Recerca i Estudis Avan\c{c}ats (ICREA) and Institut de F\'{i}sica d'Altes Energies (IFAE), Barcelona, Spain}
\affiliation{Uppsala University, Uppsala, Sweden}
\affiliation{Lancaster University, Lancaster LA1 4YB, United Kingdom}
\affiliation{Imperial College London, London SW7 2AZ, United Kingdom}
\affiliation{The University of Manchester, Manchester M13 9PL, United Kingdom}
\affiliation{University of Arizona, Tucson, Arizona 85721, USA}
\affiliation{University of California Riverside, Riverside, California 92521, USA}
\affiliation{Florida State University, Tallahassee, Florida 32306, USA}
\affiliation{Fermi National Accelerator Laboratory, Batavia, Illinois 60510, USA}
\affiliation{University of Illinois at Chicago, Chicago, Illinois 60607, USA}
\affiliation{Northern Illinois University, DeKalb, Illinois 60115, USA}
\affiliation{Northwestern University, Evanston, Illinois 60208, USA}
\affiliation{Indiana University, Bloomington, Indiana 47405, USA}
\affiliation{Purdue University Calumet, Hammond, Indiana 46323, USA}
\affiliation{University of Notre Dame, Notre Dame, Indiana 46556, USA}
\affiliation{Iowa State University, Ames, Iowa 50011, USA}
\affiliation{University of Kansas, Lawrence, Kansas 66045, USA}
\affiliation{Louisiana Tech University, Ruston, Louisiana 71272, USA}
\affiliation{Northeastern University, Boston, Massachusetts 02115, USA}
\affiliation{University of Michigan, Ann Arbor, Michigan 48109, USA}
\affiliation{Michigan State University, East Lansing, Michigan 48824, USA}
\affiliation{University of Mississippi, University, Mississippi 38677, USA}
\affiliation{University of Nebraska, Lincoln, Nebraska 68588, USA}
\affiliation{Rutgers University, Piscataway, New Jersey 08855, USA}
\affiliation{Princeton University, Princeton, New Jersey 08544, USA}
\affiliation{State University of New York, Buffalo, New York 14260, USA}
\affiliation{University of Rochester, Rochester, New York 14627, USA}
\affiliation{State University of New York, Stony Brook, New York 11794, USA}
\affiliation{Brookhaven National Laboratory, Upton, New York 11973, USA}
\affiliation{Langston University, Langston, Oklahoma 73050, USA}
\affiliation{University of Oklahoma, Norman, Oklahoma 73019, USA}
\affiliation{Oklahoma State University, Stillwater, Oklahoma 74078, USA}
\affiliation{Brown University, Providence, Rhode Island 02912, USA}
\affiliation{University of Texas, Arlington, Texas 76019, USA}
\affiliation{Southern Methodist University, Dallas, Texas 75275, USA}
\affiliation{Rice University, Houston, Texas 77005, USA}
\affiliation{University of Virginia, Charlottesville, Virginia 22904, USA}
\affiliation{University of Washington, Seattle, Washington 98195, USA}
\author{V.M.~Abazov} \affiliation{Joint Institute for Nuclear Research, Dubna, Russia}
\author{B.~Abbott} \affiliation{University of Oklahoma, Norman, Oklahoma 73019, USA}
\author{B.S.~Acharya} \affiliation{Tata Institute of Fundamental Research, Mumbai, India}
\author{M.~Adams} \affiliation{University of Illinois at Chicago, Chicago, Illinois 60607, USA}
\author{T.~Adams} \affiliation{Florida State University, Tallahassee, Florida 32306, USA}
\author{J.P.~Agnew} \affiliation{The University of Manchester, Manchester M13 9PL, United Kingdom}
\author{G.D.~Alexeev} \affiliation{Joint Institute for Nuclear Research, Dubna, Russia}
\author{G.~Alkhazov} \affiliation{Petersburg Nuclear Physics Institute, St. Petersburg, Russia}
\author{A.~Alton$^{a}$} \affiliation{University of Michigan, Ann Arbor, Michigan 48109, USA}
\author{A.~Askew} \affiliation{Florida State University, Tallahassee, Florida 32306, USA}
% ...
%
%
\collaboration{The D0 Collaboration\footnote{with visitors from
$^{a}$Augustana College, Sioux Falls, SD, USA,
$^{b}$The University of Liverpool, Liverpool, UK,
$^{c}$DESY, Hamburg, Germany,
$^{d}$Universidad Michoacana de San Nicolas de Hidalgo, Morelia, Mexico
$^{e}$SLAC, Menlo Park, CA, USA,
$^{f}$University College London, London, UK,
$^{g}$Centro de Investigacion en Computacion - IPN, Mexico City, Mexico,
$^{h}$Universidade Estadual Paulista, S\~ao Paulo, Brazil,
$^{i}$Karlsruher Institut f\"ur Technologie (KIT) - Steinbuch Centre for Computing (SCC)
and
$^{j}$Office of Science, U.S. Department of Energy, Washington, D.C. 20585, USA.
}} \noaffiliation
\vskip 0.25cm

\date{August 12, 2013}

\begin{abstract}
We study the process of associated photon and jet production, $p\bar{p}\rightarrow \gamma + \mathrm{jet} + X$, using 
8.7~fb$^{-1}$ of integrated luminosity collected by the D0 detector at the 
Fermilab Tevatron Collider at a center-of-mass energy $\sqrt{s}=1.96$~TeV. 
Photons are reconstructed with rapidity $|y^\gamma|\lt 1.0$ 
or $1.5<|y^\gamma|\lt 2.5$ 
and transverse momentum $\Ptg>20$ GeV.
The highest-$p_T$ jet is required  to be in one of four rapidity regions
up to $|y^\text{jet}|\leq 3.2$.
For each rapidity configuration  
we measure the differential cross sections in $\Ptg$ 
separately for events with the same sign ($y^{\gamma} y^{\mathrm{jet}}\gt0$) and opposite sign
($y^{\gamma} y^{\mathrm{jet}}\leq0$) of photon and jet rapidities.
We compare the measured triple differential cross sections, $\mathrm{d^3}\sigma / \mathrm{d}\Ptg\mathrm{d}y^{\gamma}\mathrm{d}y^{\mathrm{jet}} $, 
to next-to-leading order (NLO) perturbative QCD calculations 
using different sets of parton distribution functions and to predictions
from the {\sc sherpa} and {\sc pythia} Monte Carlo event generators.
The NLO calculations are found to be in general agreement with the data, 
but do not describe all kinematic regions.

\end{abstract}
\maketitle


\section{Introduction}
\label{Sec:intro}

In hadron-hadron collisions, high-energy photons ($\gamma$)
emerge unaltered from the hard scattering process of two partons and 
therefore provide a clean probe of the parton dynamics. 
The study of such photons (called {\em prompt}) produced in association with a jet 
can be used to extend inclusive photon production measurements 
\cite{UA2_phot,CDF_phot,D0Run1_phot,Photon_paper_erratum,Atlas_IncGam,CMS_IncGam} 
and provide information about the parton distribution functions (PDFs) of the incoming hadrons
\cite{PAurLindf80,JFOwens,Cont,Au2,Vo1,Mar,DBNS}. 
The term ``prompt'' means that these photons do not result from mesons, for example,
$\pi^0,\eta,\omega$, or $K^0_{\rm S}$ decays.
Such events are mostly produced in Quantum Chromodynamics (QCD) directly through the Compton-like scattering process
$gq\to \gamma q$ 
and through quark-antiquark annihilation 
$q\bar{q}\to \gamma g$. 
Inclusive \gpj production may also 
originate from partonic processes such as $gg \to q\bar{q}$, $qg \to qg$, or $qq\to qq$
where a final state quark or gluon produces a photon during fragmentation
(fragmentation photon) \cite{JFOwens,Ber}, and another parton fragments into a jet. 
Photon isolation requirements substantially reduce the rates of these events.
However, their contribution is still noticeable in some regions of phase space,
for example, at low photon transverse momentum, $\Ptg$.

By selecting events with different angular configurations between the photon and the jets,
the data probe different ranges of parton momentum fraction $x$ and hard-scattering scales $Q^2$,
as well as providing some differentiation between contributing partonic subprocesses.

In this article, we present an analysis of \gpj production in $p\bar{p}$ 
collisions at a center-of-mass energy $\sqrt{s}=1.96$ TeV in 
which the highest-$p_T$ (leading) photon is produced either centrally with 
a rapidity $|y^{\gamma}|<1.0$ or in the forward rapidity region with $1.5<|y^{\gamma}|<2.5$~\cite{d0_coordinate}. 
The leading jet is required to be in one of the four rapidity regions,
$|y^\text{jet}|\leq 0.8$, $0.8<|y^\text{jet}|\leq 1.6$, $1.6<|y^\text{jet}|\leq 2.4$, or $2.4<|y^\text{jet}|\leq 3.2$,
and to satisfy the minimum transverse momentum requirement $p_T^{\rm jet}>15$ GeV. 
The cross section as a function of $\Ptg$ is measured differentially for sixteen angular 
configurations of the leading jet and the photon rapidities. 
These configurations are obtained by combining the two photon and four jet rapidity regions,
considered separately for events having the same sign and opposite sign of photon and jet rapidities, 
i.e. $y^{\gamma}y^{\mathrm{jet}}\gt 0$ and $y^{\gamma}y^{\mathrm{jet}}\leq0$.

\begin{figure}[h]
\includegraphics[scale=0.4]{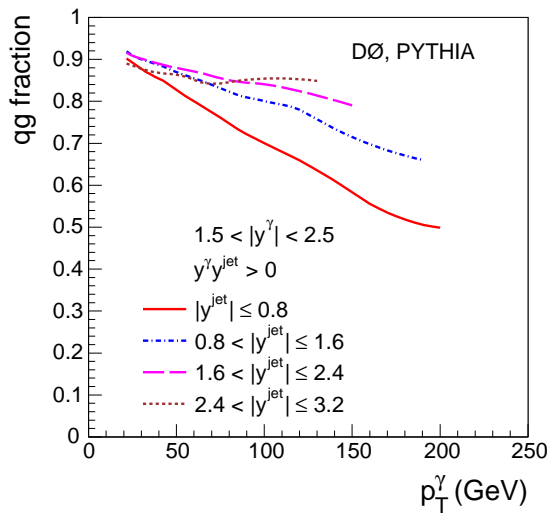}
\vspace*{-3mm}
\caption{(color online)
The fraction of events, estimated using the {\sc pythia} event 
generator \cite{pythia} with {\sc cteq6L} PDF set \cite{CTEQ}, 
produced via the $qg \to q\gamma$ subprocess relative to the total cross section of
associated production of a direct photon in the forward rapidity region, $1.5<|y^\gamma|<2.5$, 
and a leading jet in one of the four rapidity intervals satisfying $y^{\gamma} y^{\mathrm{jet}}\gt 0$.}
\label{fig:subproc}
\end{figure}

The primary motivation of this measurement is to constrain the gluon PDF that directly affects
the rate of Compton-like $qg \rightarrow q\gamma $ parton scattering.
The rate of this process varies for different photon-jet rapidity configurations and drops with increasing $\Ptg$.
Estimates using the {\sc pythia} \cite{pythia} Monte Carlo (MC) event generator and {\sc cteq6L} 
PDF set~\cite{CTEQ} show that the highest fraction of $qg$ events is observed in same-sign events with forward photons
($y^{\gamma}y^{\mathrm{jet}}\gt 0$ and $1.5<|y^{\gamma}|<2.5$).
Figure~\ref{fig:subproc} shows the expected contributions
of the Compton-like process to the total associated production cross section of a photon and a 
jet for the four jet rapidity intervals in same-sign events with forward photons. 
In these events the $qg$ fraction increases with jet rapidity.

The PDFs entering the theoretical predictions have substantial uncertainties, 
particularly for the gluon contributions at small $x$, or large $x$ and large $Q^{2}$ \cite{CTEQ}.
The \gpj cross sections probe different regions of parton momentum fraction 
$x_{1}$ and $x_{2}$  of the two initial interacting partons. For example, at 
$\Ptg \approx 20-25$ GeV, events with a central photon and central jet cover 
the interval in $0.01 \lt x \lt 0.06$, while same-sign events with a forward photon and very forward jet 
($2.4<|y^\text{jet}|\leq 3.2$) cover the regions within $0.001 \lt x \lt 0.004$ and $0.2 \lt x \lt 0.5$.
Here, $x$ is defined using the leading-order approximation 
$x_{1,2} = (\Ptg/\sqrt{s})(\exp({\pm y^\gamma}) + \exp({\pm y^{\text {jet}}}))$ 
\cite{JFOwens}.
The total $x$ and $Q^{2}$  region (with $Q^2$ taken as $(p_T^\gamma)^2$)
covered by the measurement is $0.001\leq x \leq 1$ and 
$400 \leq Q^2 \leq 1.6 \times 10^{5} ~\rm GeV^2 $, 
extending the kinematic reach of previous \gpj 
measurements \cite{ISR,UA2_g,CDF2,H1_gp,H1_g,ZEUS_g,ZEUS_gj,D0_tgj_2008,Atlas_gj}. 

\begin{figure}[h]
\includegraphics[scale=0.4]{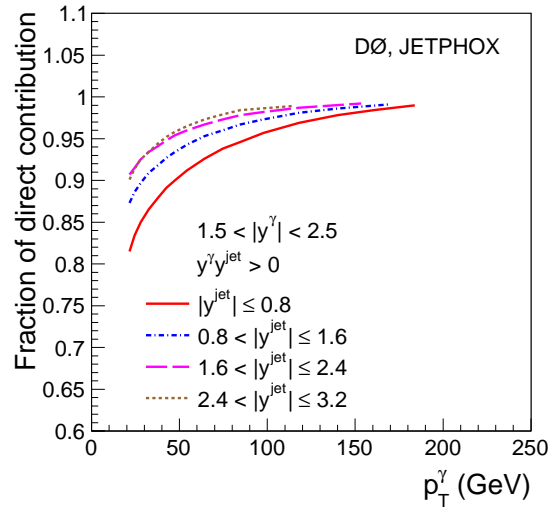}
\vspace*{-3mm}
\caption{(color online)
For \gpj events, the fraction of the direct (non-fragmentation) photon contribution 
of the total (direct+fragmentation) cross section,
estimated with {\sc jetphox} \cite{JETPHOX} for events with forward photons.}
\label{fig:dirfrac}
\end{figure}

The expected ratio of the direct photon contribution to the sum of direct and 
fragmentation contributions of the \gpj ~cross section is shown in Fig.~\ref{fig:dirfrac},
for the chosen photon isolation criteria (see Sec.~\ref{sec:evsel}), in the four studied
regions. The fragmentation contribution
decreases with increasing $\Ptg$ for all regions \cite{CFGP,Ber,PatrickPR06}.

Compared to the latest \gpj cross sections published by the D0 \cite{D0_tgj_2008}
and ATLAS \cite{Atlas_gj} Collaborations, this measurement 
considers not only central but also forward photon rapidities, four jet rapidity intervals, 
and uses a significantly larger data set. 

This paper is organized as follows.
In Sec.~\ref{sec:data}, we briefly describe the D0 detector and \gpj events selection criteria.
  In Sec.~\ref{sec:mc}, we describe the MC signal and background samples used in the analysis. 
  In Sec.~\ref{sec:analysis}, we assess the main corrections applied to the data needed to measure the cross sections
  and discuss related uncertainties in Sec.~\ref{sec:syst}.
  Measured cross sections and comparisons with theoretical predictions are presented in Sec.~\ref{sec:results}.
  Finally, Sec.~\ref{sec:summary} summarizes the results.

\section{D0 detector and data set}
\label{sec:data}

\subsection{D0 detector}
\label{sec:det}

The D0 detector is a general purpose detector described in detail elsewhere~\cite{d0det, l0, l1cal}. 
The subdetectors most relevant in this analysis 
are the calorimeter, the central tracking system, and the central preshower. 
The muon detection system is used
for selecting a clean $Z \rightarrow \mu^+ \mu^- \gamma$ sample
to obtain data-to-MC correction factors for the photon reconstruction efficiency. 
The central tracking system, used to reconstruct tracks of 
charged particles, consists of a silicon micro-strip detector (SMT) and a central fiber track detector (CFT), 
both 
inside a 2~T solenoidal magnetic field. 
While the amount of material traversed by a charged particle depends on its trajectory, 
it is typically on the order of 0.1 radiation lengths in the tracking system.
The tracking system provides
a $35\thinspace\mu$m vertex resolution along the beam line and $15\thinspace\mu$m
resolution in the transverse plane near the beam line for 
charged particles with $p_T \approx 10$~ GeV.
The solenoid magnet is surrounded by the central preshower (CPS) 
detector located immediately before the inner layer of the electromagnetic calorimeter. The CPS consists of 
approximately one radiation length of lead absorber surrounded by three layers of scintillating strips. 
The preshower detectors are in turn surrounded by sampling calorimeters constructed of depleted uranium
absorbers in an active liquid argon volume.
The calorimeter is composed of three sections: a central calorimeter (CC) covering the range of pseudorapidities 
$|\eta_{\rm det}|<1.1$~\cite{d0_coordinate} and two end calorimeters (EC) with coverage 
extending to $|\eta_{\rm det}|\approx4.2$, with all three housed in separate cryostats.
The electromagnetic (EM) section of the central calorimeter contains four 
longitudinal layers of approximately 2, 2, 7, and 10 radiation lengths, and is 
finely-segmented transversely into cells of size
$\Delta \eta_{\rm det} \times \Delta \phi_{\rm det}=0.1 \times 0.1$, with the exception of layer three 
with $0.05 \times 0.05$ granularity. The calorimeter resolution for measurements
 of the electron/photon energy at 50 GeV is about 3.6\%. 
The luminosity is measured using plastic scintillator arrays placed in front of the EC cryostats
at $2.7 <|\eta_{\rm det}| < 4.4$.

\subsection{Event selection}
\label{sec:evsel}

Triggers for the events used for this analysis are based on at least one
cluster of energy found in the EM calorimeter with loose shower shape requirement and
various $\Ptg$ thresholds.
The data set with photon candidates covering the interval of $20<\Ptg \leq35$ GeV is 
selected using prescaled EM triggers
with a $p_T$ threshold of 17 GeV and corresponds to a total integrated luminosity of $7.00 \pm 0.43$~pb$^{-1}$.
The selection efficiency of photons with respect to this trigger condition exceeds $96\%$.
As a cross check, the cross sections in this $\Ptg$ region are also measured using events
that are heavily prescaled with trigger thresholds of 
$p_{T}=13$~GeV or 9 GeV  corresponding to total luminosities of 
$2.63 \pm 0.16$~pb$^{-1}$  and $0.65 \pm 0.04$~pb$^{-1}$, respectively~\cite{d0lumi}. 

Photon candidates with $p_T>35$ GeV are selected using a set of unprescaled
EM triggers with $p_T$ thresholds between 20 GeV and 70 GeV, with a signal selection efficiency
with respect to the trigger requirements close to 100\%.
This data set corresponds to an integrated luminosity of $8.7 \pm 0.5$ fb$^{-1}$ \cite{d0lumi}
after relevant data quality cuts.

The D0 tracking system is used to select events 
containing at least one $p\bar{p}$ collision vertex reconstructed with at least three tracks
and within 60~cm of the center of the 
detector along the beam axis. The efficiency of the vertex requirements above
varies as a function of instantaneous luminosity within $95\%-97\%$. 

The longitudinal segmentation of the EM calorimeter and CPS detector 
allows the estimation of the direction of the central photon candidate and 
the coordinate of its origin along the beam axis (``photon vertex pointing'').
This 
position is required to be within $10$ cm (3 standard deviations) 
of the $p\bar{p}$ collision vertex 
if there is a CPS cluster matched to the photon EM cluster ($\sim80\%$ of events) 
or within 32~cm otherwise (about 1.5 standard deviation for such events).
Forward photons are assumed to originate from the default $p\bar{p}$ collision vertex.
A systematic uncertainty is assigned to account for this assumption.

\subsection{Photon and jet selections}
\label{sec:gamjetsel}

EM clusters for photon candidates are formed from calorimeter towers in a cone of radius 
$\mathcal R = \sqrt{(\Delta \eta)^2+(\Delta \phi)^2}=0.4$ 
around a seed tower~\cite{d0det}. A stable cone is found iteratively, 
and the final cluster energy is recalculated from an inner cone 
within $\mathcal R = 0.2$. The photon candidates are required (i) to have $\geq 97\%$ of the cluster energy 
deposited in the EM calorimeter layers;
(ii) to be isolated in the calorimeter with ${\cal I}=[E_{\rm tot}(0.4)-E_{\rm EM}(0.2)]/E_{\rm EM}(0.2)<0.07$,
where $E_{\rm tot}(\mathcal R)$ $[E_{\rm EM}(\mathcal R)]$ is the total [EM only] energy in a cone of radius $\mathcal R$;
(iii) to have a scalar sum of the $p_T$ of all charged particles originating from the vertex in an annulus of $0.05<\mathcal{R}<0.4$ 
around the EM cluster to be less than $1.5$~GeV;
and (iv) to have an energy-weighted EM shower width consistent with that expected for a photon. 
To suppress electrons misidentified as photons, the EM clusters are required to have no spatial match 
to a charged particle track or any hit configuration in the SMT and CFT detectors consistent with an electron trajectory~\cite{track_veto}. 
This requirement is referred to as a ``track-match veto''.

An additional group of variables exploiting the differences between the photon- and 
jet-initiated activity in the EM calorimeter, CPS
(for central photons), and the tracker is combined into an artificial neural network (NN) to further reject jet background~\cite{nn}. 
In these background events, photons are mainly produced from decays of energetic $\pi^0$ and $\eta$ mesons.
The NN is trained on a {\sc pythia}~\cite{pythia} MC sample of photon and jets events. 
The generated MC events are processed through 
a {\sc geant}-based simulation of the D0 detector~\cite{geant}. 
Simulated events are overlaid with data events from random $p\bar{p}$ crossings 
to properly model the effects of multiple $p \bar p$ interactions and detector noise in data.
Care is taken to ensure that the instantaneous luminosity distribution 
in the overlay events is similar to the data used in the analysis. 
MC events are then processed through the same reconstruction procedure as the data. 
They are weighted to take into account the trigger efficiency in data, and
small observed differences in the distributions of the instantaneous luminosity 
and of the $z$ coordinate of the $p\bar{p}$ collision vertex.
Photons radiated from charged leptons 
in $Z$ boson decays ($Z \rightarrow \ell^+ \ell^- \gamma, \mbox{\space}\ell=e,\mu$) are used 
to validate the NN performance~\cite{nnvalid,d0_dpp,hgg}. 
The shape of the NN output ($O_{\rm NN}$) distribution in the MC simulation describes the data well and gives 
an additional discrimination against jets. 
The $O_{\rm NN}$ distribution for jets is validated using dijet MC and data samples enriched in jets misidentified as photons.
For this purpose, the jets are required to pass all photon identification criteria listed above,
but with an inverted calorimeter isolation requirement of ${\cal I}>0.1$ or 
by requiring at least one track in a cone of $\mathcal R < 0.05$ around the photon candidate.
The photon candidates are selected with a requirement 
$O_{\rm NN}>0.3$ to retain $97\%-98\%$  
of photons and to reject $\approx\!40\%$ ($\approx\!15\%$) of jets remaining after 
the other selections described above for central (forward) photons have been applied.

Background contributions 
from cosmic rays and 
from isolated electrons, originating from the leptonic decays of 
$W$ bosons,
are suppressed by requiring the missing transverse energy {\mbox{$\not\!\!E_T$}}, calculated 
as a vector sum of the transverse energies of all calorimeter cells
and corrected for reconstructed objects (photon and jet energy scale corrections), 
to satisfy the condition ${\mbox{$\not\!\!E_T$}}<0.7~\Ptg$.

The measured energy of a photon EM cluster is calibrated in two steps.
First, the absolute energy calibration of the EM cluster is obtained using 
electrons from $Z \rightarrow e^+e^-$ decays
as a function of $\eta_{\rm det}$ and $p_T$. However, photons interact less with the material 
in front of the calorimeter than electrons.
As a result the electron energy scale correction overestimates the
photon $p_T$ relative to the particle (true) level. 
The relative photon energy correction as a function of $\eta$ is derived using a
detailed {\sc geant}-based \cite{geant} simulation of the D0 detector response. 
It is particularly
sizable at low $p_T$ ($\Ptg \approx 20$ GeV), where
the photon energy overcorrection is found to be $\approx 3\%$. 
The difference
between electron and photon calibrations becomes smaller at higher energies. 
A systematic uncertainty of $0.60\%-0.75\%$ on this correction 
is due to the electron energy calibration
and uncertainties in the description of the amount of material in front of the calorimeter.
Combined with the steeply falling $\Ptg$ spectrum this results in a 
$3\%-5\%$ uncertainty on the measured cross sections (see Section \ref{sec:syst}).

Selected events should contain at least one hadronic jet. Jets are reconstructed using the 
D0 Run II Midpoint Cone jet-finding algorithm with a cone of ${\cal R}=0.7$~\cite{JetAlgo}, 
and are required to satisfy quality criteria that suppress backgrounds from leptons, photons, 
and detector noise effects. 
Jet energies are corrected to the particle level using a jet energy scale correction procedure ~\cite{jesnim}.
The leading jet must satisfy two requirements: 
$p_{T}^{\mathrm{jet}}\gt 15$ GeV and $p_{T}^{\mathrm{jet}}\gt0.3\Ptg$,
where the first is related with the jet $p_T$ reconstruction threshold of 6 GeV for the uncorrected jet $p_T$.
The second requirement reflects the correlation between photon and leading jet $p_T$, and
is optimized at the reconstruction level to account for jet $p_T$ resolution.
At the particle level, this selection reduces the fraction of events with strong radiation in the
initial or/and final state which potentially may lead to higher order corrections in theory, 
i.e., uncertainty to the current {\it NLO} QCD predictions.
The jet $p_T$ selections above have about $90\%-95\%$ efficiency for the signal.
The leading photon candidate and the leading jet are also required to be 
separated in $\eta$-$\phi$ space by $\Delta{\cal{R}}(\gamma,\mathrm{jet})>0.9$. 

In total, approximately 7.2 (8.3) million \gpj candidate events 
with central (forward) photons are selected after application 
of all selection criteria.

\section{Signal and background models}
\label{sec:mc}

To study the characteristics of signal 
events, MC samples are generated using {\sc pythia} \cite{pythia} and {\sc sherpa} 
\cite{sherpa} event generators,
with  {\sc cteq}6.1L and  {\sc cteq}6.6M PDF sets \cite{CTEQ}, respectively.
In {\sc pythia}, the signal events are included via $2\to 2$ matrix elements (ME) with 
$gq\to \gamma q$ and $q\bar{q}\to \gamma g$ hard scatterings (defined at the leading order)
followed by the leading-logarithm approximation of the partonic shower.
The soft underlying events, as well as fragmentation, are based on an empirical model (``Tune A''),
tuned to Tevatron data~\cite{tuneA}.

In {\sc sherpa}, up to two extra partons (and thus jets) are allowed at the ME level
in the $2\to \{2,3,4\}$ scattering, but jets can also be produced in parton showers (PS). 
Matching between partons coming from real emissions in the ME and jets from PS
is done at an energy scale $Q_{\rm cut}$ defined following the prescriptions given in Ref.~\cite{sherpa_gam}.
Compared with Tune A the multiple parton interaction (MPI) model implemented in {\sc sherpa}  is characterized by 
(a) showering effects in the second interaction, which makes it closer to the $p_T$-ordered showers ~\cite{pT_order} 
in the Perugia tunes~\cite{Perugia}, 
and (b) a combination of the CKKW merging approach with the MPI modeling \cite{sherpa,CKKW}.
Another distinctive feature of {\sc sherpa} is the modeling of the parton-to-photon fragmentation
contributions through the incorporation of QED effects into the parton shower \cite{sherpa_gam}.
This contribution is available in {\sc sherpa} with default settings for \gpj events.

Since we measure the cross section of {\it isolated} prompt photons, the isolation criterion
should be defined in the MC sample as well to allow a comparison of data to expectations.
In the {\sc pythia} and {\sc sherpa} samples, the photon is required to be isolated at the particle level
by $p_T^{\rm iso} = p_T^{\rm tot}(0.4) - p_T^\gamma < 2.5$~GeV,
where $p_T^{\rm tot}(0.4)$ is the total transverse energy of particles within a cone of radius ${\cal R} = 0.4$
centered on the photon.
Here, the particle level includes all stable particles as defined in Ref.~\cite{particle}.
The photon isolation at the particle level differs from that at the reconstruction level
(see Sec.~\ref{sec:gamjetsel}), and includes specific requirements on the calorimeter isolation 
(defined around the EM cluster) and track isolation.

To estimate backgrounds to \gpj production, we also consider dijet events simulated in {\sc pythia}.
In the latter, constraints are placed at the generator level to increase the number of jet events fluctuating 
into photon-like objects~\cite{nn} after applying photon selection criteria. 
The signal events may contain photons originating from the parton-to-photon fragmentation process.
For this reason, the background events, produced with QCD processes in {\sc pythia}, 
were preselected to exclude bremsstrahlung photons produced from partons. 
Finally, to estimate other possible backgrounds, we have also used $W$+jet and $Z$+jet samples simulated with {\sc alpgen+pythia} \cite{alpgen},
and diphoton events simulated with {\sc sherpa}.
Signal and background events are 
processed through a {\sc geant}-based~\cite{geant} simulation and event reconstruction as described in the previous section.

\section{Data analysis and corrections}
\label{sec:analysis}

\subsection{Estimating signal fraction}
\label{sec:purity}

Two types of instrumental background contaminate the
$\gamma$+jet sample: electroweak interactions resulting in
one or more electromagnetic clusters (from electrons or photons),
and strong interactions producing a jet misidentified as a photon.

The first type of background includes
$W(\to e\nu)$+jet, $Z/\gamma^*(\to e^+e^-)$+jet, and diphoton production.
The contributions from these backgrounds are estimated from MC simulation. 
In the case of $W(\to e\nu)$+jet events, with the electron misidentified as a photon, 
the neutrino will contribute additional {\mbox{$\not\!\!E_T$}}.
The combination of the track-match veto (part of the photon identification criteria), and {\mbox{$\not\!\!E_T$}} requirement reduces
the contribution from this process to a negligible level, less than 0.5\% for events with central photons,
and less than 1.5\% for events with forward photons.
Contributions from $Z$+jet and diphoton events, in which either $e^\pm$ from $Z$ decay 
is misidentified as a photon, or one of the photons in the diphoton events is misidentified as a jet,
are found to be even smaller.
These backgrounds are subtracted from the selected data sample.

\begin{figure}
\includegraphics[scale=0.45]{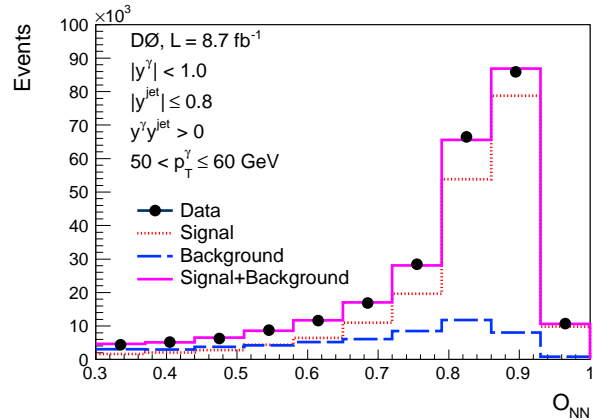}
\vspace*{-1mm}
\caption{(color online)
Distribution of observed events for $O_{\rm NN}$ after all selection criteria 
for the representative bin $50 <\Ptg < 60$ GeV ($|y^\gamma| < 1.0$). 
The distributions for the signal and background templates are shown normalized to their respective
fitted fractions. 
Fits in the other $\Ptg$ bins are of similar quality.
}
\label{fig:onn}
\end{figure}

\begin{figure*}
\includegraphics[scale=0.39]{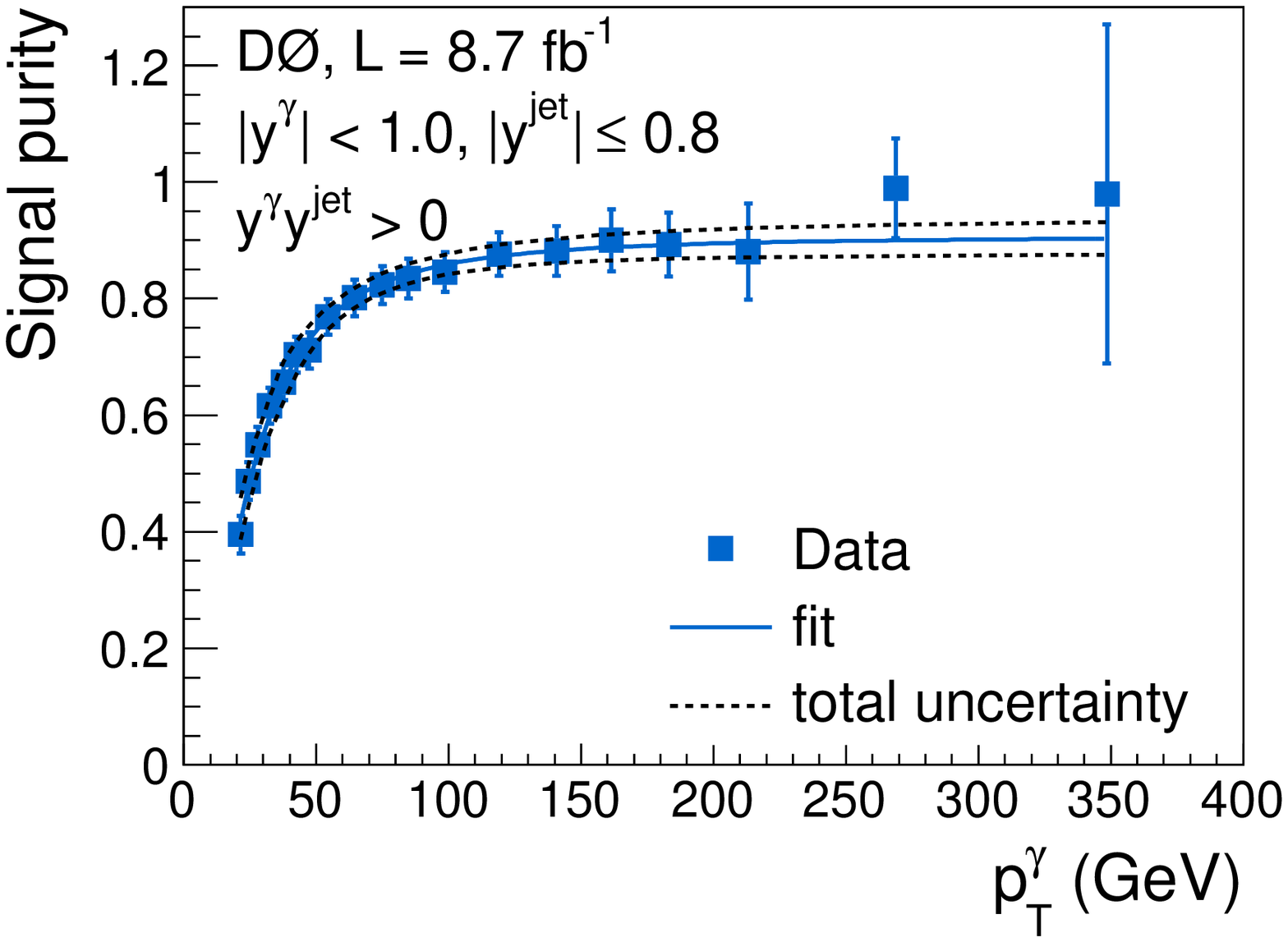}
\includegraphics[scale=0.39]{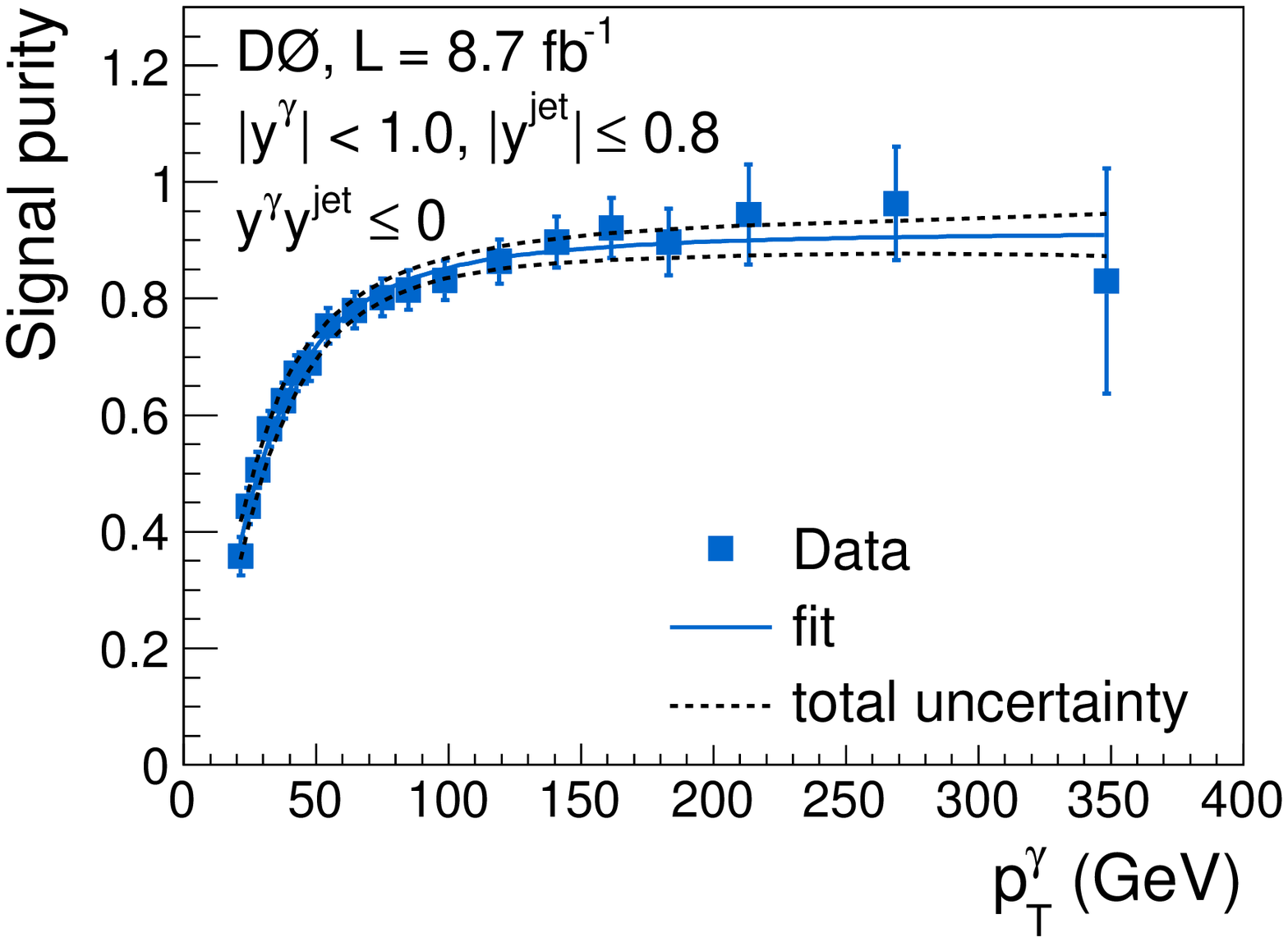}
\includegraphics[scale=0.39]{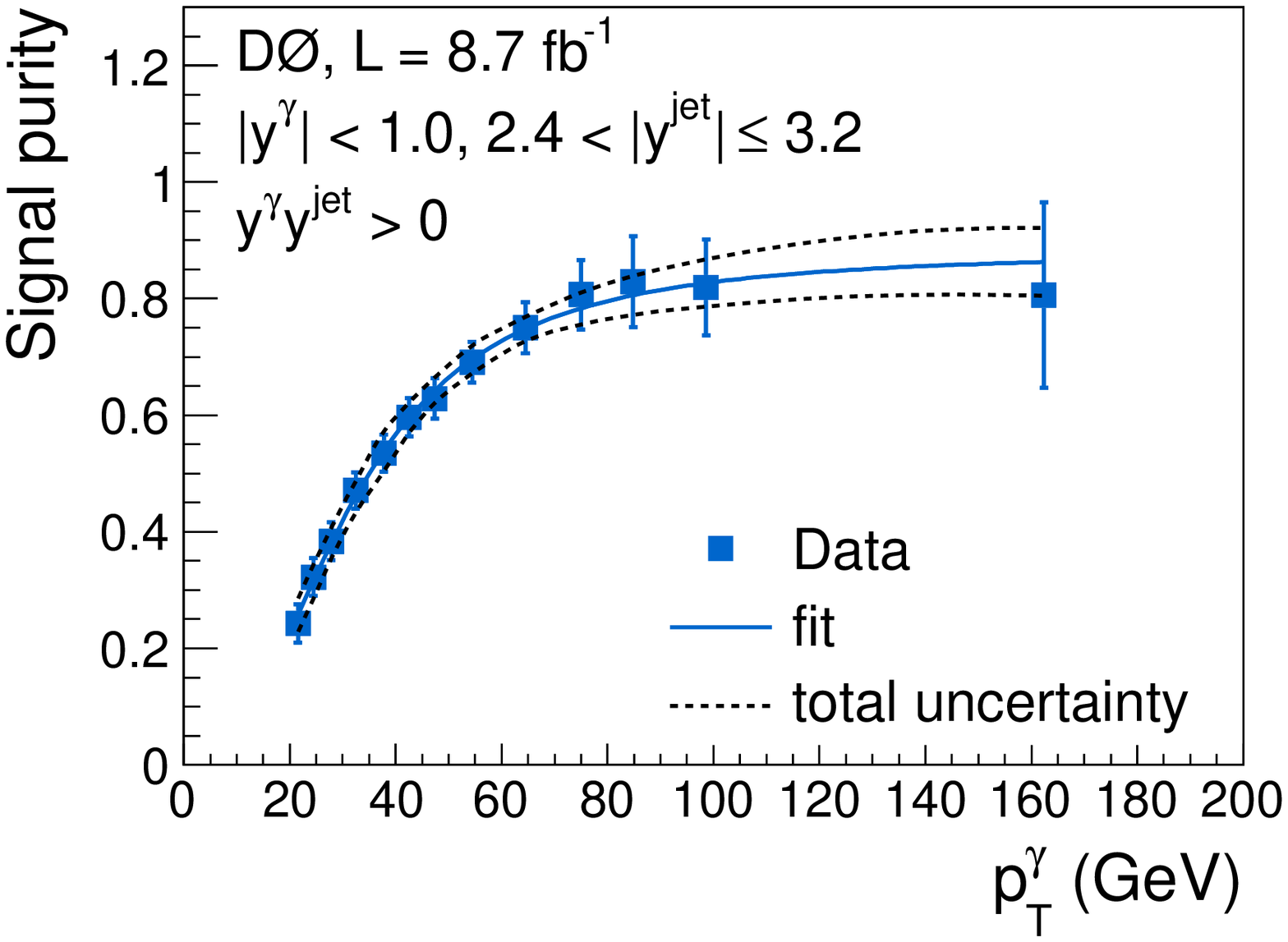}
\includegraphics[scale=0.39]{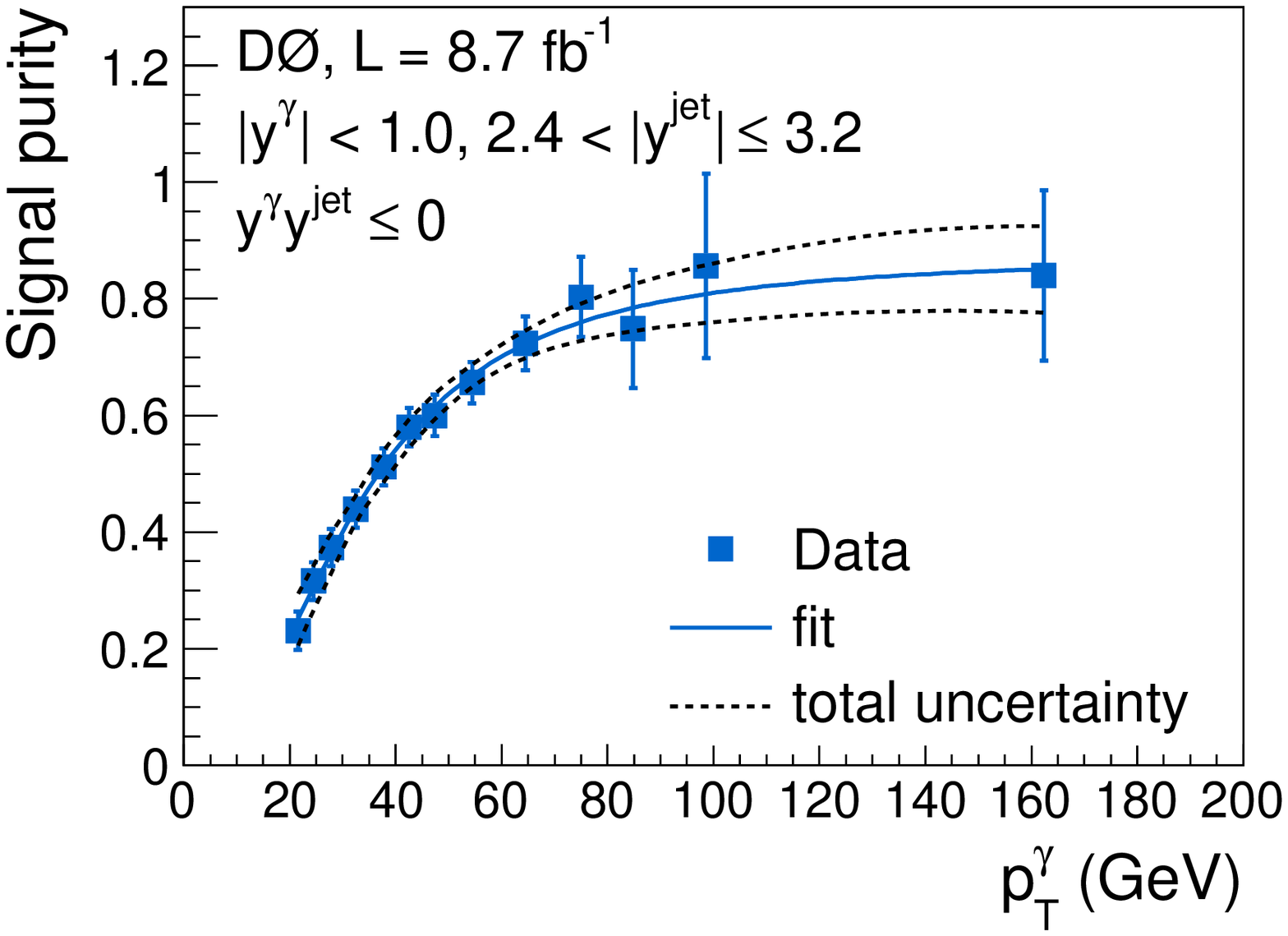}
\vspace*{-4mm}
\caption{(color online)
Purity of the selected \gpj ~sample as a function of $\Ptg$, shown for 
central photons, very central and very forward jet rapidities, same-sign, and opposite-sign rapidity events.
The solid line shows the fit and the dashed lines show the total fit uncertainty.
}
\label{fig:pur_CC}
\end{figure*}

\begin{figure*}
\includegraphics[scale=0.39]{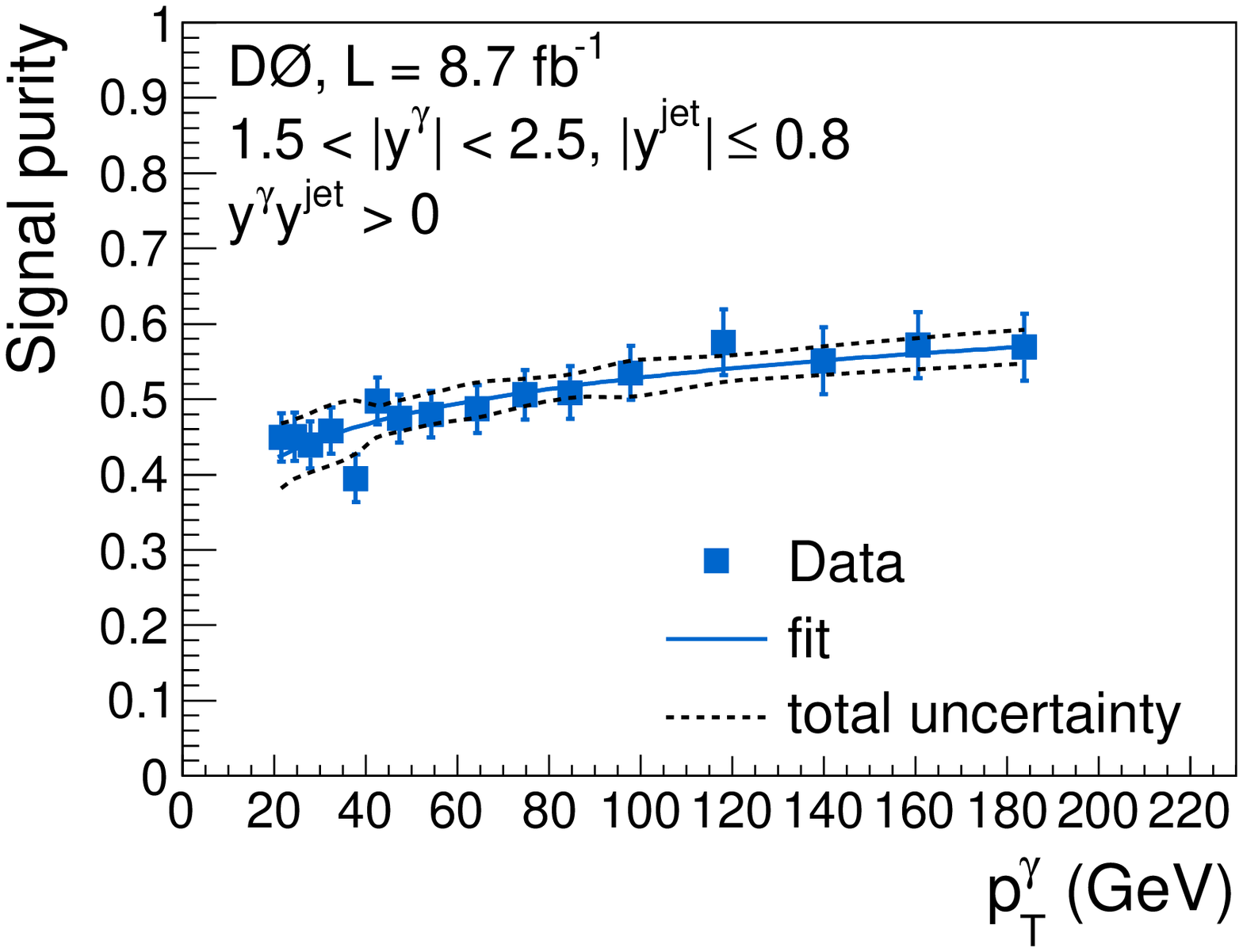}
\includegraphics[scale=0.39]{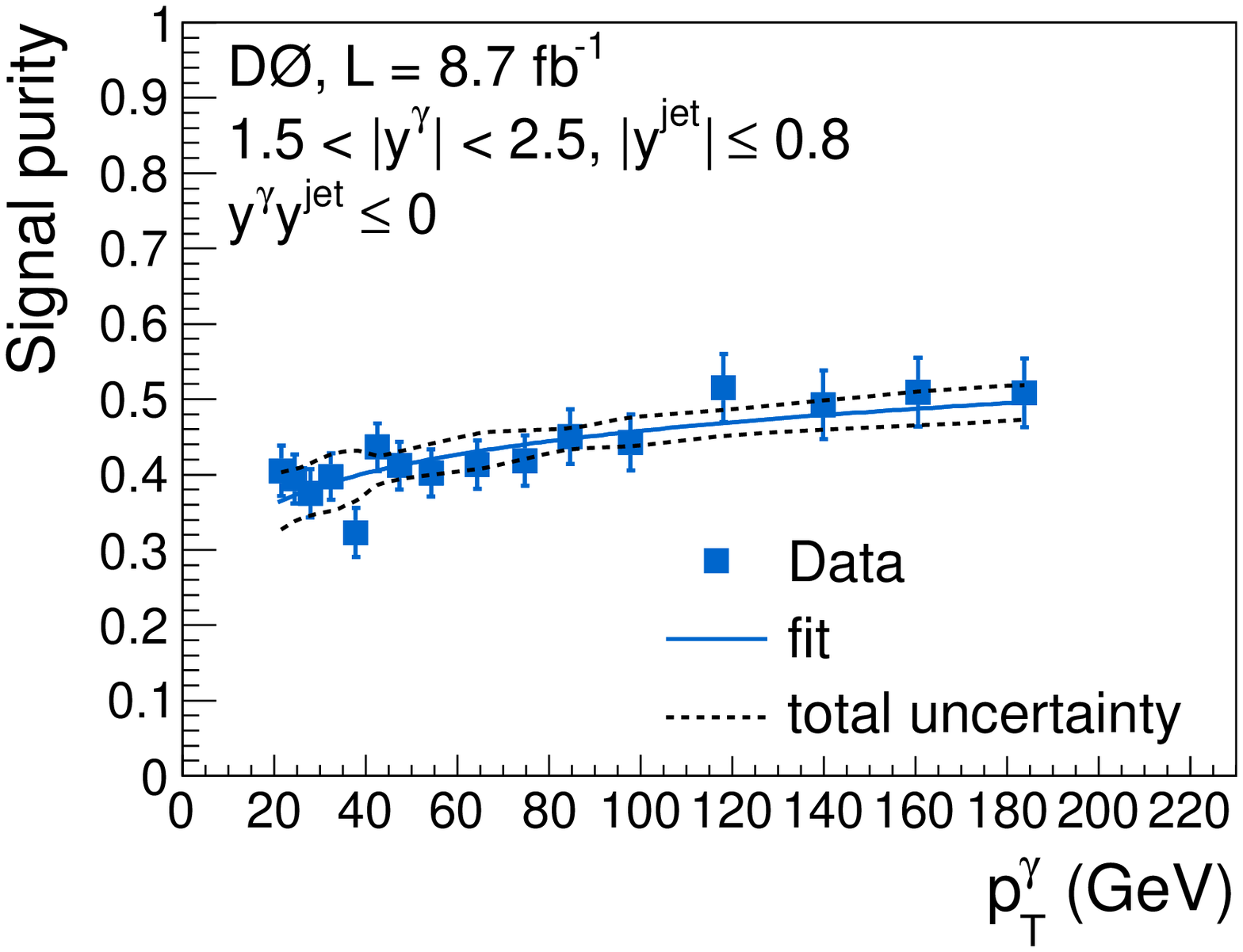}
\includegraphics[scale=0.39]{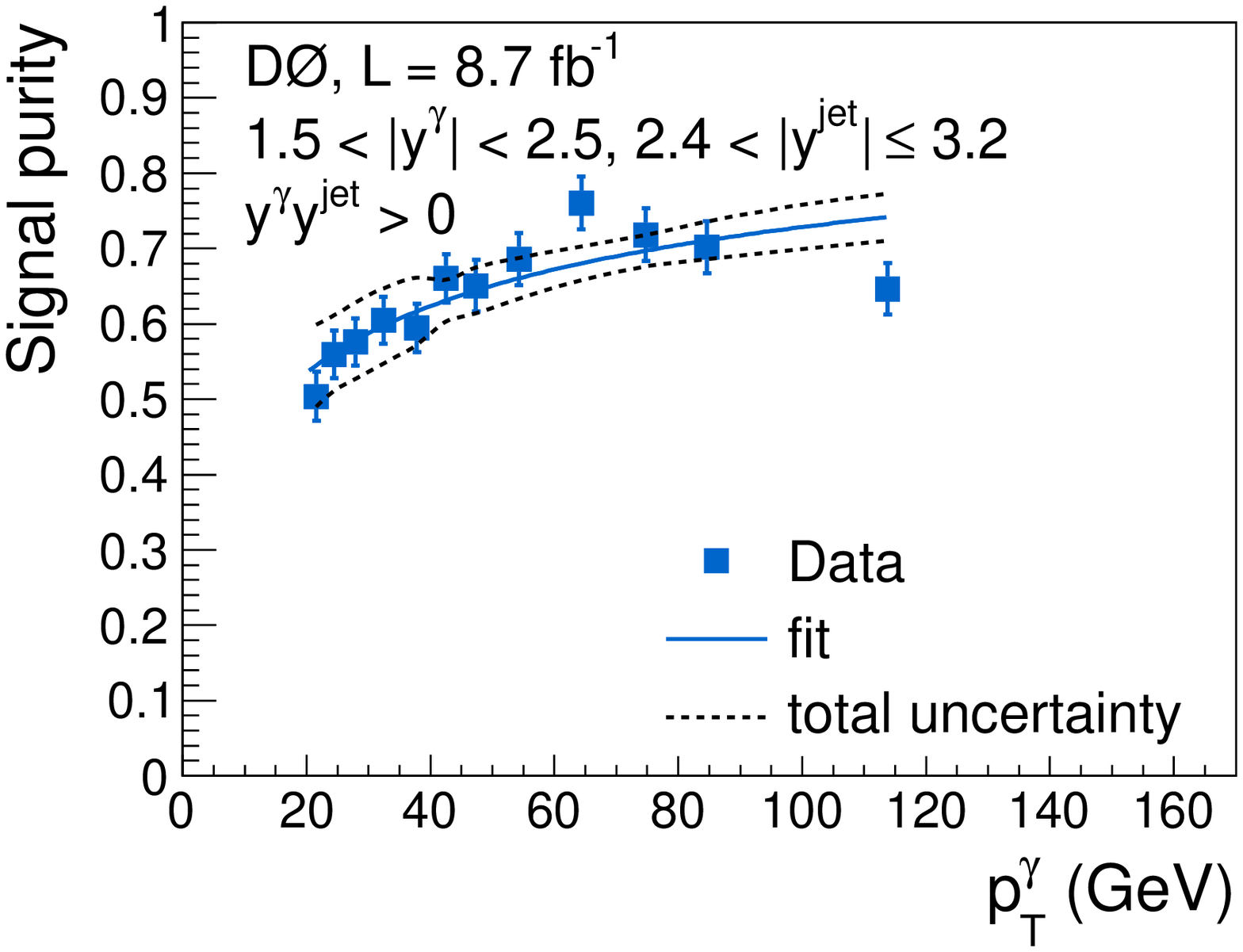}
\includegraphics[scale=0.39]{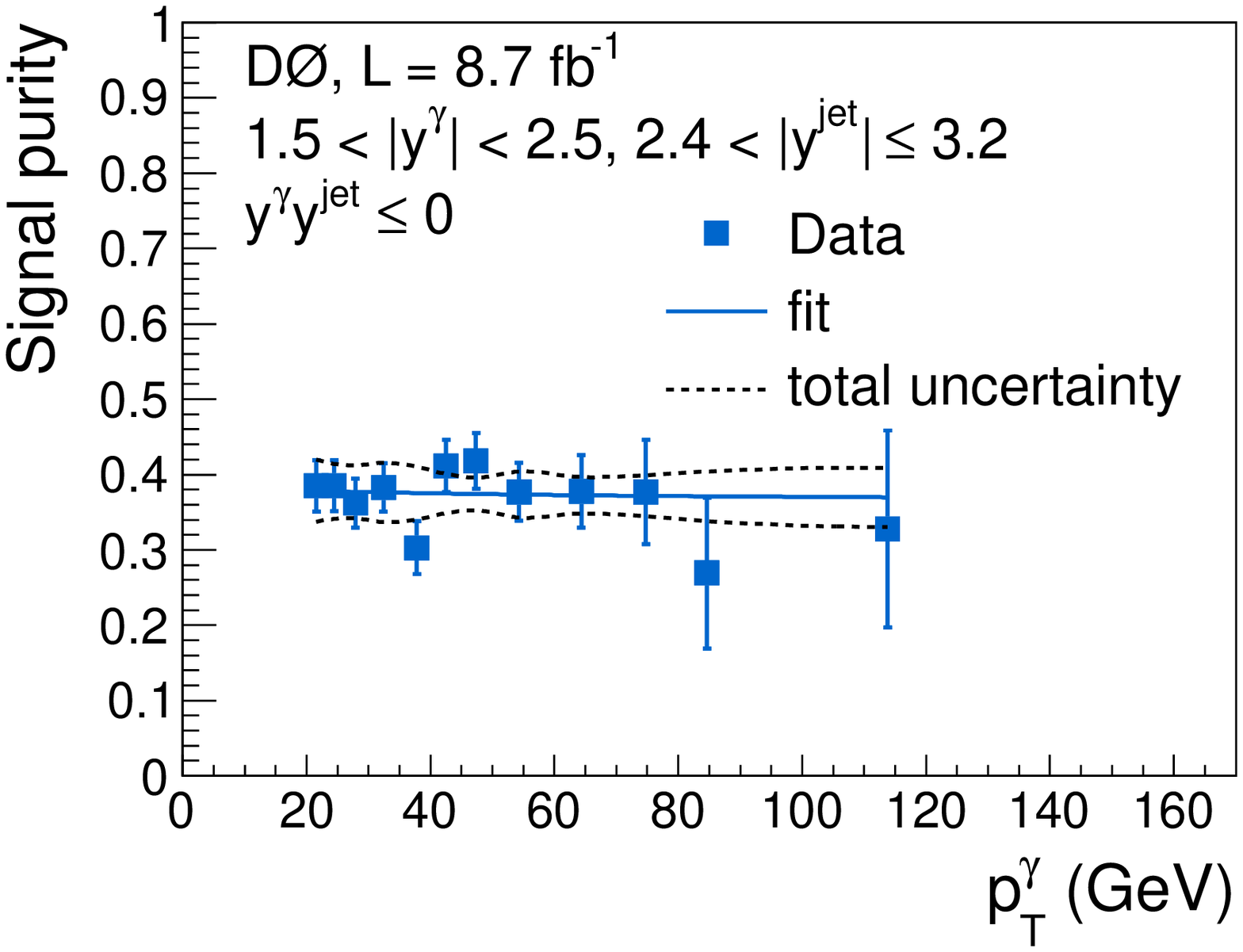}
\vspace*{-4mm}
\caption{(color online)
Same as Fig.~\ref{fig:pur_CC} but for events with forward photons.}
\label{fig:pur_EC}
\end{figure*}

To estimate the remaining background contribution from dijet events, we consider photon candidates
in the region $0.3<O_{\rm NN}\leq 1$ (i.e. the region used for data analysis).
The distributions for the simulated  photon signal and dijet background samples are fitted to 
the data for each $\Ptg$ bin using a maximum likelihood fit \cite{HMCMLL} 
to obtain the fractions of signal and background components in data.
The result of this fit to $O_{\rm NN}$ templates,
normalized to the number of events in data, is
shown in Fig.~\ref{fig:onn} for central photons with $50 <\Ptg < 60$ GeV, as an example.
The  $\Ptg$ dependence of the signal fraction (purity) is fitted in each region using a three-parameter function, ${\cal P}=a/(1+b(\Ptg)^c)$.
Two alternative fitting functions have also been considered.
Figure~\ref{fig:pur_CC} shows the resulting purities for events with central photons, very central and very forward jet rapidities,
for same-sign and opposite-sign rapidities. 
Figure~\ref{fig:pur_EC} shows similar results for events with forward photons.
The signal fractions, typically, grow with $\Ptg$, while the growth is not as significant
for the events with forward photons. 
The signal fractions are somewhat greater for the same-sign rapidity 
events than for the opposite-sign,
and also greater for events with central jets as opposed to forward jet events.

The measured fractions of signal events have to be corrected for events with prompt photons 
with the isolation parameter value at the particle level $p_T^{\rm iso}\geq2.5$ GeV. 
Such events can migrate into our data sample 
even after applying the photon selections described in Sec.~\ref{sec:gamjetsel}.
The fractions of such events are estimated in two ways.
First, we use the signal models in {\sc sherpa} and {\sc pythia} MC generators to determine the fraction
of events with $p_T^{\rm iso}\geq2.5$ GeV after all selections.
The fraction of such events is $1\%-3\%$ for events with central photons
and $1\%-2\%$ for events with forward photons. 
This procedure gives consistent results for both MC generators.
In the second method,
we calculate signal purities for the signal events in which we keep {\it all} photons, i.e., including those 
with isolation $p_T^{\rm iso}\geq2.5$ GeV, and compare them with the 
default case where photons satisfy the isolation cut $p_T^{\rm iso}<2.5$. 
The difference of $1\%-3\%$ 
is in good agreement with the direct MC estimates.
We subtract this fraction from data and assign 
an additional systematic uncertainty on the signal purity of $1\%-1.5\%$.

Other systematic uncertainties on the signal purity are caused by the $O_{\rm NN}$ template fitting uncertainties 
derived from the error matrix, 
the choice of fit functions, and the signal model dependence estimated 
by a comparison of signal purities obtained with the photon templates taken from {\sc pythia} and {\sc sherpa}.
An additional systematic uncertainty on the background template due to the fragmentation model 
implemented in {\sc pythia} is also taken into account.
It is found to be about $5\%$ at $\Ptg\simeq 30$~GeV, $2\%$ at $\Ptg\simeq 50$~GeV, and $1\%$ at $\Ptg\gtrsim 70$~GeV 
and is estimated using the method described in Ref.~\cite{Photon_paper_erratum}.

\subsection{Acceptance and efficiency corrections}
\label{sec:eff}

We calculate corrections to the observed rate of \gpj candidates 
to account for the photon and jet detection efficiencies 
(and for the geometric and kinematic acceptances)
using samples of simulated $\gamma$+jet events in which the photon is required to be isolated at 
the particle level by applying $p_T^{\rm iso}<2.5$ GeV.

The bin size is chosen to be larger than the resolution on $p_T^\gamma$, 
yielding 
more than $80\%$ of the particle-level events  located in the same 
$\Ptg$ bins at the reconstruction level.
The acceptance is dominated by the EM cluster quality selection requirements on $\eta_{\rm det}$, 
applied to avoid edge effects in the calorimeter regions used for the measurement, 
and on $\phi_{\rm det}$ in the central rapidity region, 
applied to avoid periodic calorimeter module boundaries~\cite{d0det} 
that bias the EM cluster energy and position measurements. 
The acceptance typically varies within about $1.4-0.8$ with a relative systematic uncertainty of $3\%-12\%$,
and takes into account correlation between the same-sign and opposite-sign events.
The acceptance greater than unity corresponds to opposite-sign rapidity events with forward jets and low $\Ptg$ central photons,
and are caused by a migration of the (particle-level) same-sign events into the other category. 
Migration significantly increases the number of reconstructed
opposite-sign events due to a much larger cross section for same-sign events at small $\Ptg$
(see Sec.\ref{sec:results}).
Correction factors to account for 
differences between jet-$p_T$ and rapidity spectra in data and simulation are estimated with
{\sc pythia}, and used as weights to create a data-like MC sample.
The differences between acceptance corrections obtained with standard and data-like MC samples are taken
as a systematic uncertainty of up to 10\% at small $\Ptg$. An additional systematic uncertainty of up to 7\%
is assigned from a comparison of the photon selection efficiency calculated 
with {\sc pythia} and {\sc sherpa}.

Small differences between data and MC in the photon selection efficiencies are corrected using factors
derived from $Z \rightarrow e^+e^-$ control samples, as well as photons from
radiative $Z$ boson decays~\cite{nnvalid}.
The total efficiency of the photon selection criteria is $68\%-80\%$, depending on the ~\ptg~ and $y^\gamma$ region.  
The systematic uncertainties caused by these correction factors are 3\% for $|y^\gamma|<1.0$ 
and $7.3\%$ for $1.5<|y^\gamma|<2.5$ and are mainly due to uncertainties 
caused by the track-match veto, isolation, and the photon NN requirements.

\section{Summary of systematic uncertainties}
\label{sec:syst}

\begin{figure*}
\includegraphics[scale=0.4]{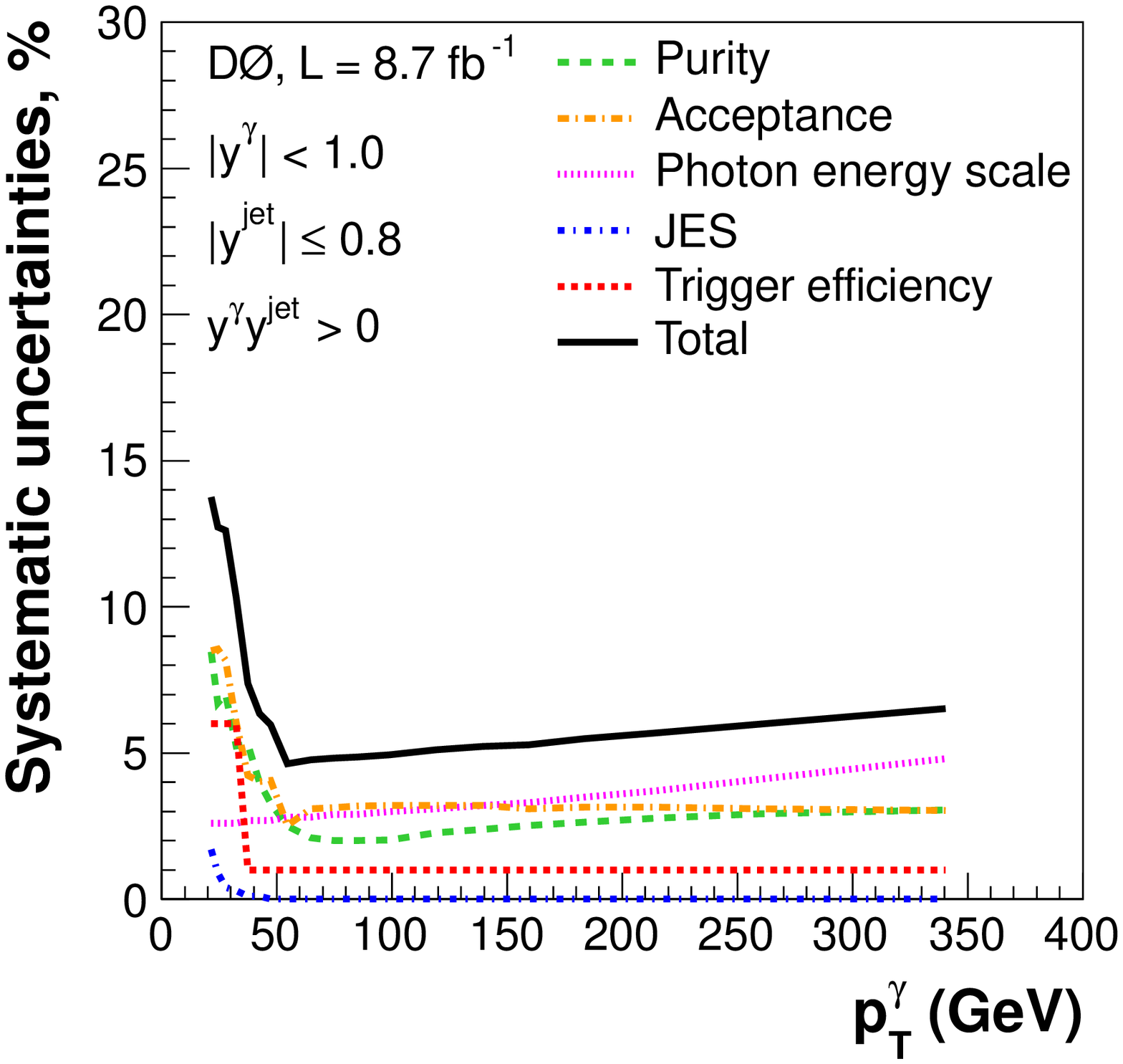}
\includegraphics[scale=0.4]{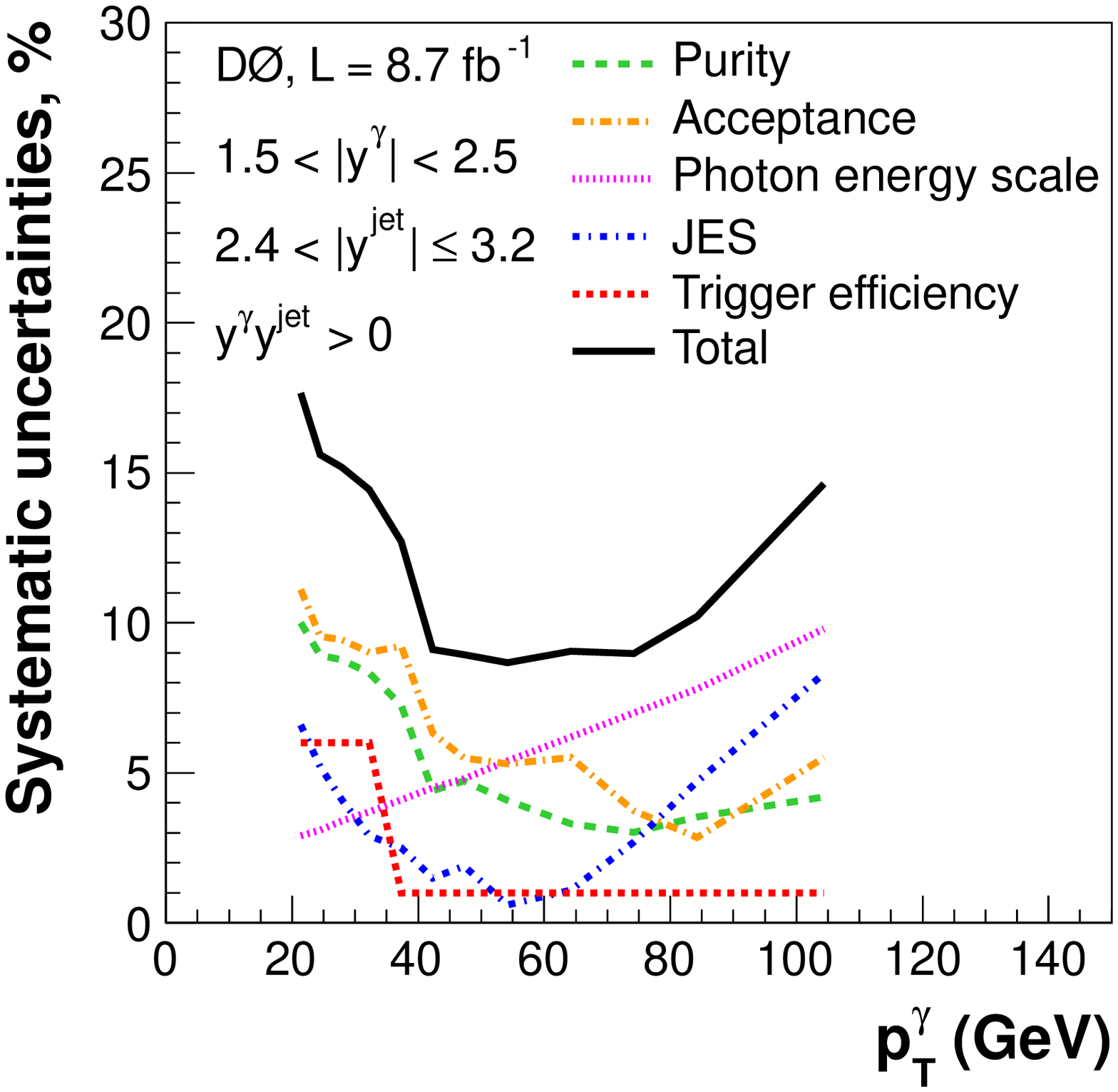}
\vspace*{-3mm}
\caption{(color online)
Systematic uncertainties on the prompt \gpj production cross sections 
for events with central and forward photons 
(same-sign events with $|y^{\rm jet}| \leq 0.8$ and $2.4<|y^{\rm jet}| \leq 3.2$ are shown as an example).
A common normalization uncertainty of 6.8\% for events with central photons and 11.2\% for forward photons
resulting from integrated luminosity, photon selection efficiency, and photon production vertex are not included in the figures. }
\label{fig:syst}
\end{figure*}

The main sources of experimental systematic 
uncertainty on the prompt \gpj production cross section
in two kinematic regions, 
$|y^\gamma|<1.0$, $|y^{\mathrm{jet}}|\leq0.8$, $y^{\mathrm{\gamma}}y^{\mathrm{jet}}>0$ and,
$1.5<|y^\gamma|<2.5$, $2.4<|y^{\mathrm{jet}}|\leq3.2$, $y^{\mathrm{\gamma}}y^{\mathrm{jet}}>0$,
are shown, as an example, in Fig.~\ref{fig:syst}. 
Similar uncertainties are found for the other kinematic regions.
The largest uncertainties are assigned to the signal purity estimation $(11\%-3\%)$, 
photon and jet selections $(3\%-10\%)$, jet energy scale $(7\%-1\%)$, photon energy scale $(3\%-8\%)$, 
EM trigger selection (6\% for $20<\Ptg<35$ GeV and 1\% for $\Ptg\geq35$ GeV)
and the integrated luminosity (6.1\%). 
The uncertainty ranges cover the intervals from low $\Ptg$ to high $\Ptg$.
The systematic uncertainty on the photon selection is due to 
the correction determined by comparing the observed data/MC 
difference in the efficiency to pass the photon selection criteria, 
and a reconstruction of the photon production vertex $z$-position 
(2\% for events with central photons and 6\% for forward photons).
The total experimental systematic uncertainty for each data point is 
obtained by adding the individual contributions in quadrature. 
A common normalization uncertainty of 6.8\% for events with central photons and 11.2\% for forward photons
resulting from uncertainties on integrated luminosity, photon selection efficiency, 
and photon production vertex selection (see above) is not included in the figures,
but is included in the tables.
Correlations between systematic uncertainties are given in Ref.~\cite{epaps} 
to increase the value of these data in future PDF fits. 
Bin-by-bin correlations in $\Ptg$ are provided for the seven sources of 
systematic uncertainty. The normalization uncertainties are not included in
those tables.

\section{Differential cross section and comparison with theory}
\label{sec:results}

The differential cross section $\mathrm{d^3}\sigma / \mathrm{d}\Ptg\mathrm{d}y^{\gamma}\mathrm{d}y^{\mathrm{jet}} $ for 
\gpj production is obtained from the number of data events in each interval after applying corrections 
for signal purity, acceptance and efficiency, divided by the integrated luminosity and the widths 
of the interval in the photon transverse momentum, photon rapidity, and jet rapidity.
For all regions we choose intervals of $\mathrm{d}y^{\gamma}=2.0$ and $\mathrm{d}y^{\rm jet}=1.6$.

\begin{figure*}[htbp]
\vspace*{-5mm}
\includegraphics[scale=0.34]{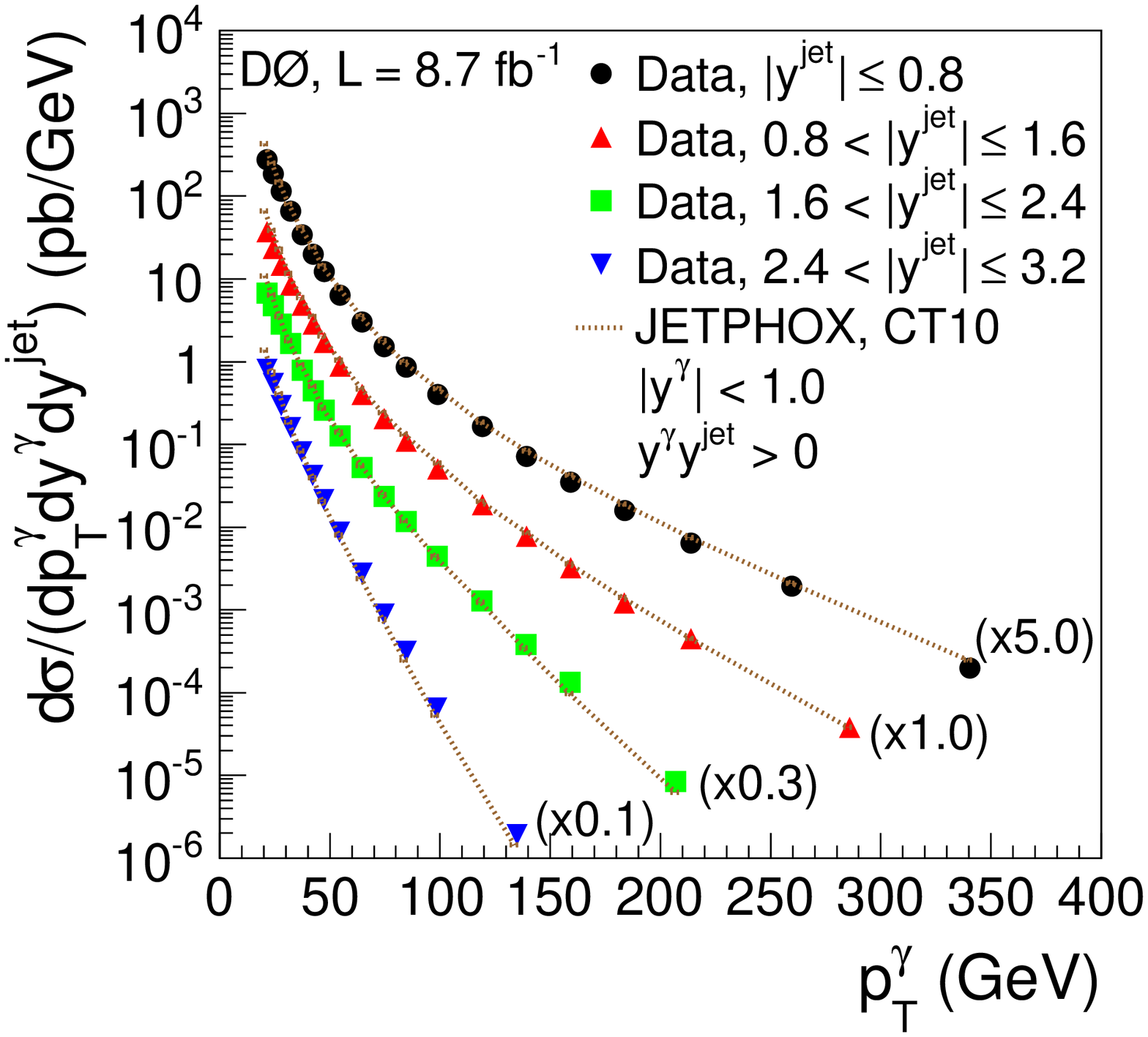}
\includegraphics[scale=0.34]{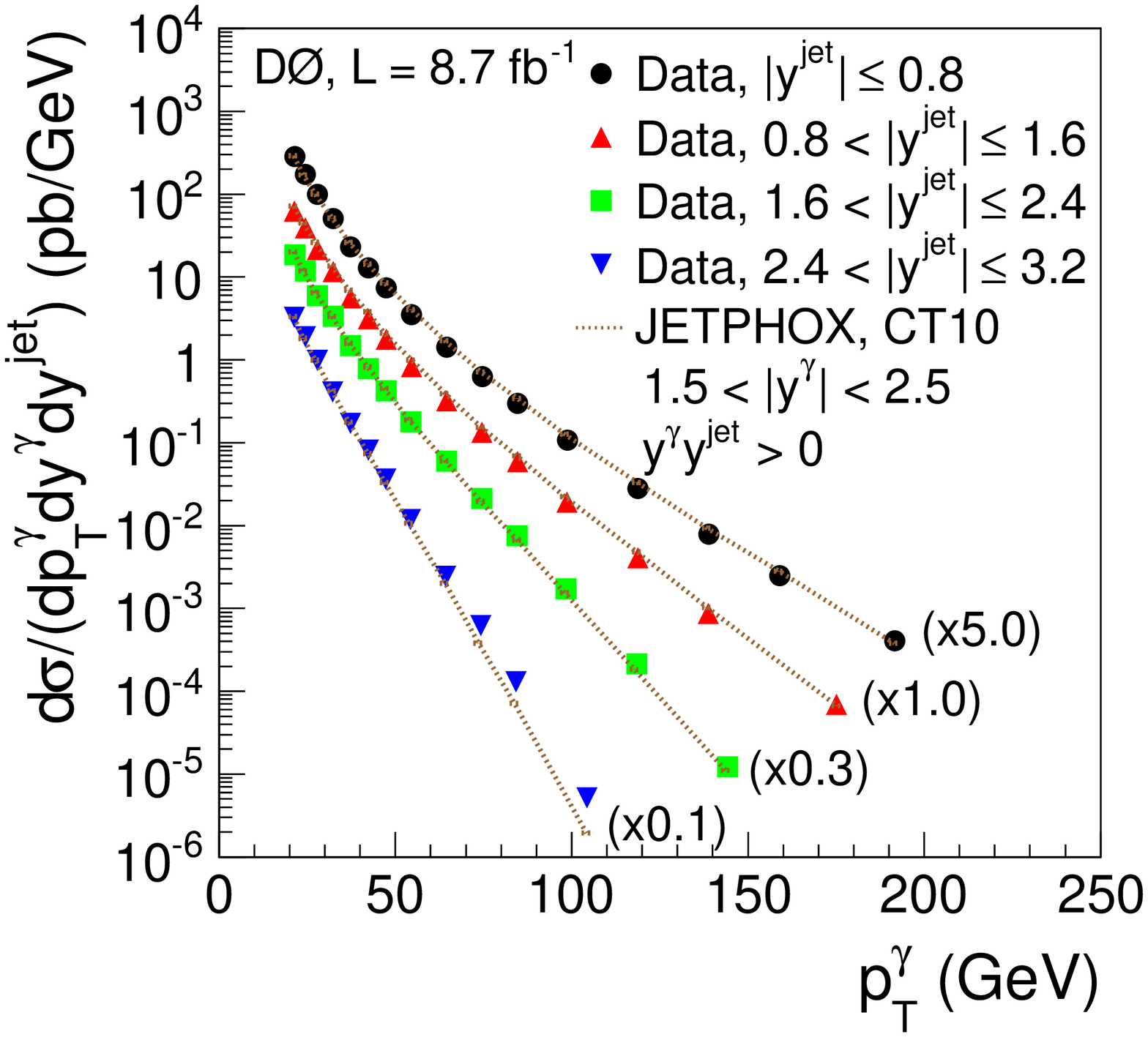}
\includegraphics[scale=0.34]{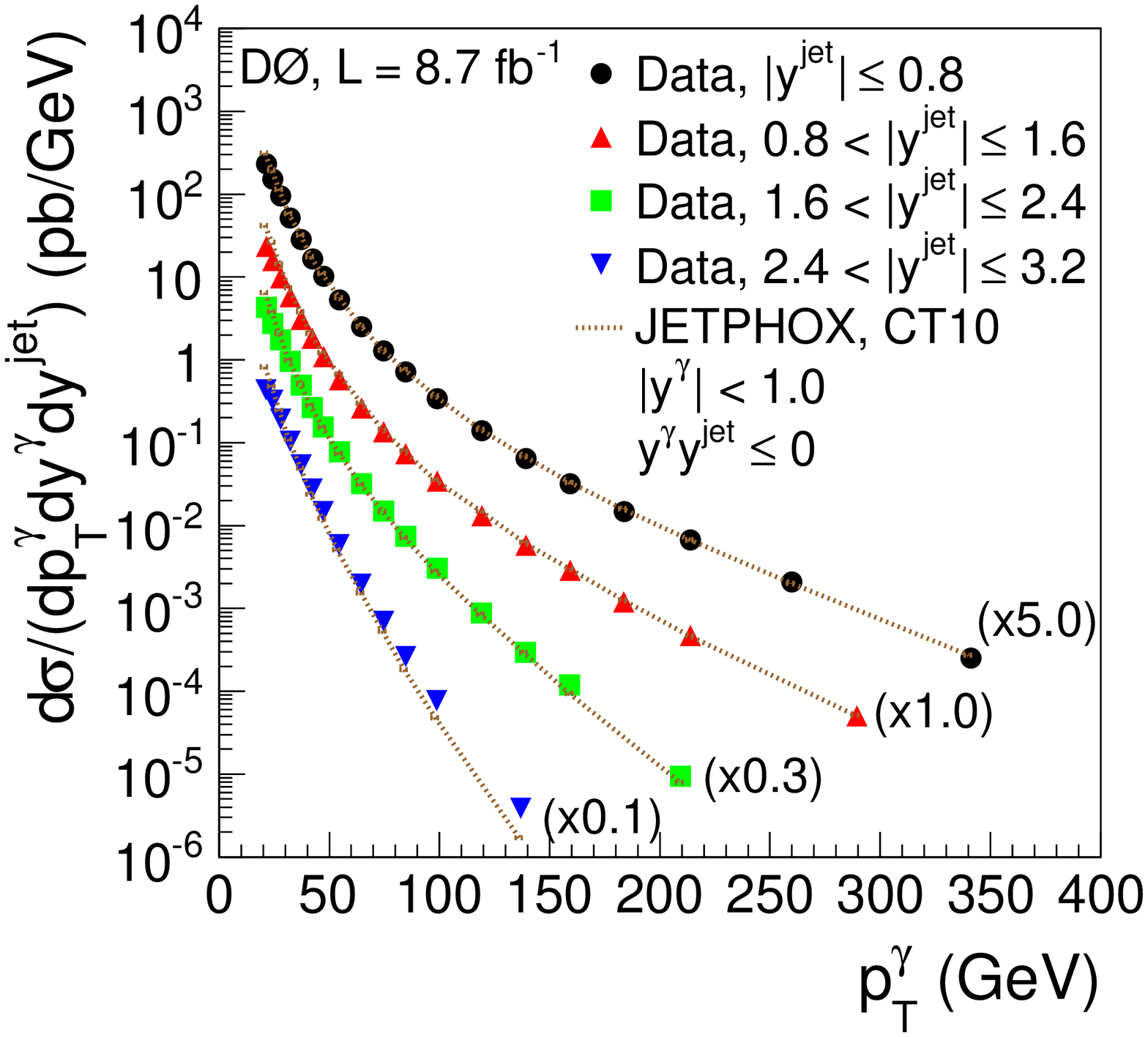}
\includegraphics[scale=0.34]{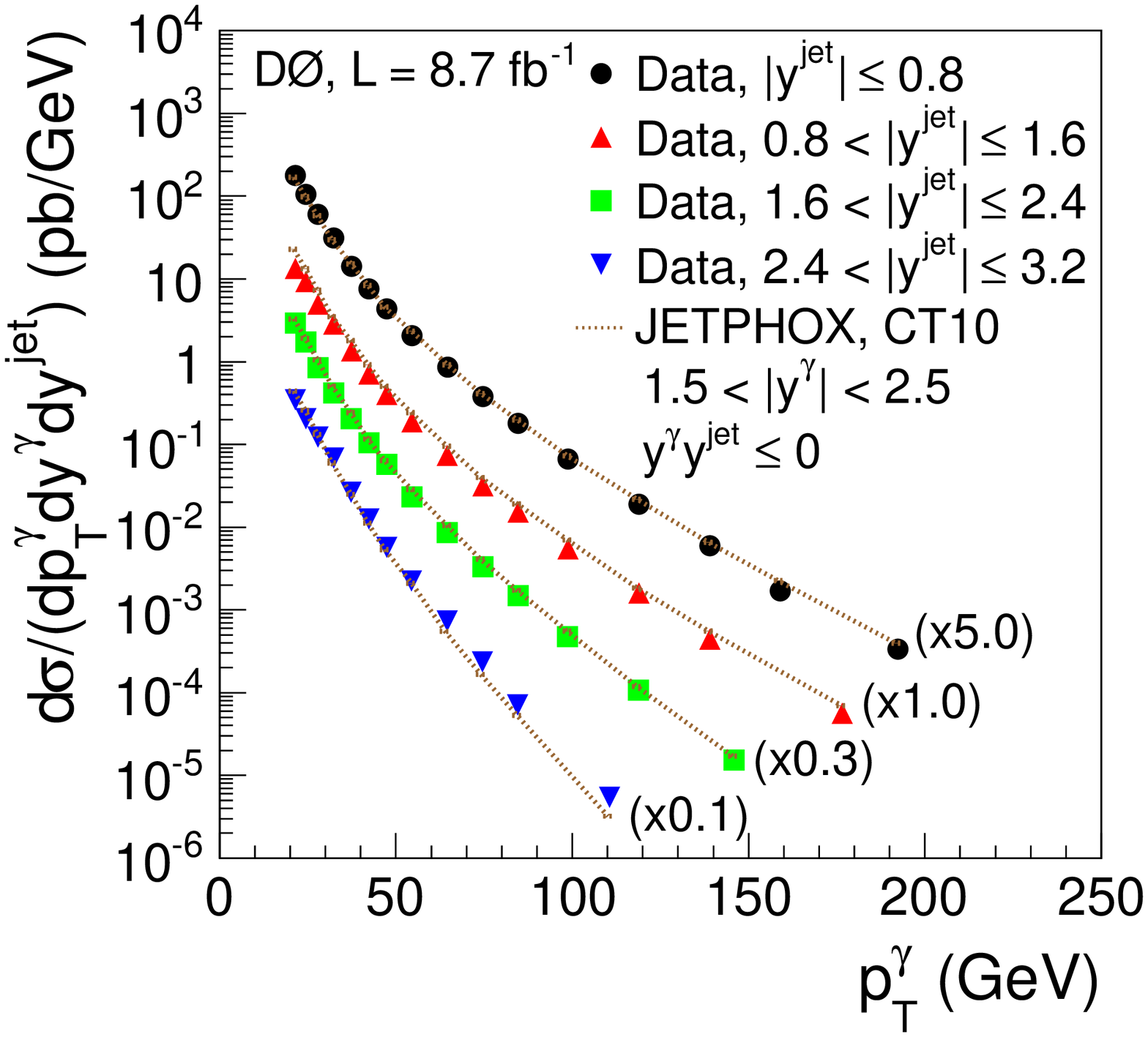}
\vspace*{-5mm}
\caption{(color online)
The measured differential \gpj ~cross section as a function of
$\Ptg$ for the four measured jet rapidity intervals, with central photons, $|y^\gamma|<1.0$, 
and forward photons, $1.5\lt |y^\gamma|\lt 2.5$,
for same-sign and opposite-sign of photon and jet rapidities.
For presentation purposes, cross sections for $|y^{\mathrm{jet}}|\leq0.8$, $0.8<|y^{\mathrm{jet}}|\leq1.6$, $1.6<|y^{\mathrm{jet}}|\leq2.4$
and $2.4<|y^{\mathrm{jet}}|\leq3.2$ are scaled by factors of 5, 1, 0.3 and 0.1, respectively.
The data are compared to the NLO QCD predictions using the
{\sc jetphox} package \cite{JETPHOX} with the CT10 PDF set \cite{CT10}
and $\mu_{R}=\mu_{F}=\mu_f=\Ptg$.}
\label{fig:cross1}
\end{figure*}

\begin{figure*}[htbp]
\vspace*{-5mm}
\hspace*{-9mm} \includegraphics[scale=0.345]{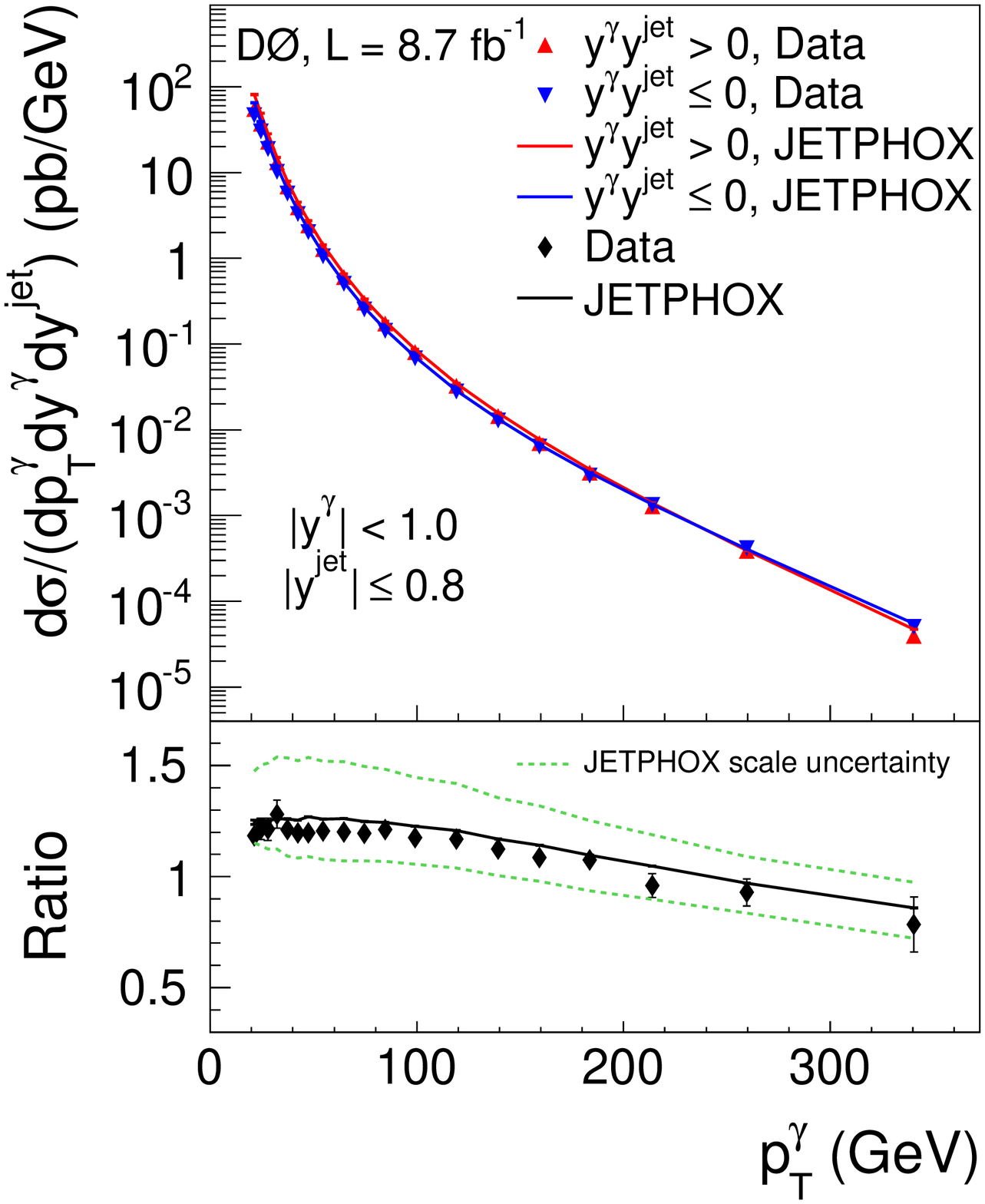}
\hspace*{-1mm} \includegraphics[scale=0.345]{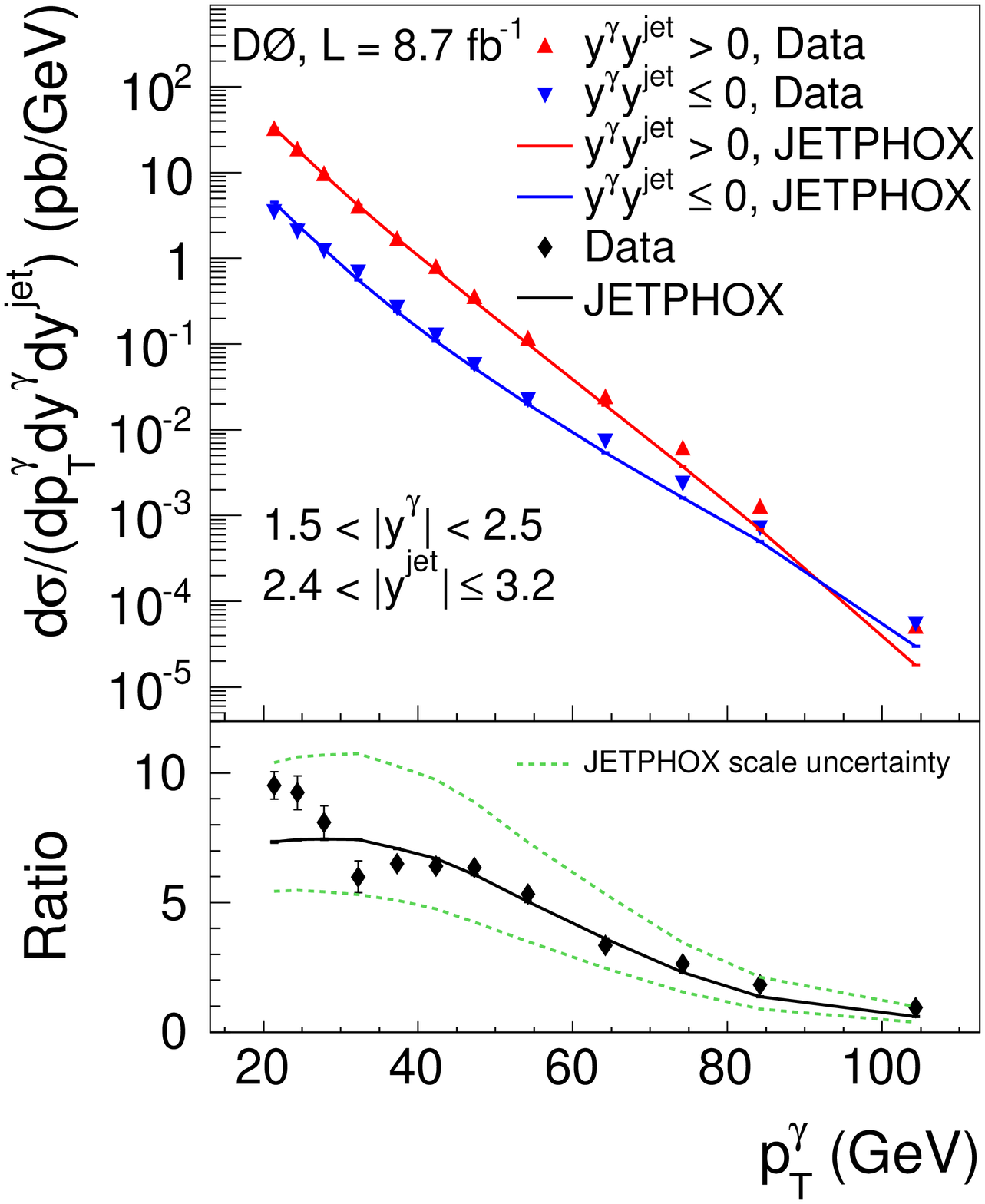}
\vspace*{-5mm}
\caption{(color online)
Comparison of the same-sign to opposite-sign cross section ratios for events 
with a central photon and central jet and those with a
forward photon and very forward jet.
}
\label{fig:SStoOS}
\end{figure*}

The cross sections for each region are presented as a function of $\Ptg$ in Fig.~\ref{fig:cross1}. 
The data points are shown at the value $\langle \Ptg \rangle$ 
for which a value of a smooth function describing the cross section dependence 
equals the average cross section in that bin \cite{TW}.
The cross sections cover 5--6 orders of magnitude in each rapidity range,
and fall more rapidly for events with larger jet and/or photon rapidities. 
The cross section of events with same-sign rapidities
has a steeper $\Ptg$ spectrum than for the opposite-sign events.
As an example, in Fig.~\ref{fig:SStoOS} we show ratios of the same-sign to opposite-sign cross sections 
for two extreme cases,
central photon and central jet, and forward photon and very forward jet.
The ratio reaches about a factor of 1.2 at low $\Ptg$ at central photon and jet rapidities, 
while for the forward rapidities it varies by up to a factor of 10.
In both cases the ratio drops to about unity at high $\Ptg$.

The data are compared to next-to-leading order (NLO) perturbative QCD (pQCD)
predictions obtained using {\sc jetphox} \cite{CFGP,JETPHOX}, with 
CT10 PDF \cite{CT10} and BFG fragmentation functions of partons to photons \cite{BFG}. 
The renormalization, factorization, and fragmentation scales ($\mu_{R}$, $\mu_{F}$, and $\mu_f$) 
are set equal to $\Ptg$.
The uncertainty due to the scale choice is estimated via a simultaneous 
variation, up and down, by a factor of two of all three scales relative to the central value
($\mu_{R}=\mu_{F}=\mu_f=\Ptg$).
The CT10 PDF uncertainties are estimated using 26 pairs of
eigenvectors following the prescription of Ref.~\cite{CTEQ_asym}.

To compare data to the {\sc jetphox}  predictions at the particle level,
the latter are corrected for non-perturbative effects caused by (a) parton-to-hadron fragmentation and (b) MPI. 
These corrections are evaluated using {\sc pythia} MC samples in two steps:
(a) as a ratio of $\gamma$+jet cross section after fragmentation to that before fragmentation (i.e., at the parton level)
with the MPI effect switched off, and (b) as the ratio of $\gamma$+jet cross section after
switching on the MPI effect to that without it. The typical size of the correction for the fragmentation effect is about $0.98-1.02$
with 1\% uncertainty. As the default MPI tune we choose
Perugia-0 (P0) \cite{Perugia} since it 
shows the best description of the azimuthal distributions 
in $\gamma$+2-jet and $\gamma$+3-jet events \cite{g3j_PRD}.
To estimate a systematic uncertainty due to the MPI effect, other tunes have been considered as well:
P-hard and P-soft \cite{Perugia}, that explore the dependence on the strength of initial- and final-state radiation effects,
while maintaining a roughly consistent MPI model as implemented in the P0 tune; P-nocr, which excludes
any color reconnections in the final state; DW \cite{tuneDWT} with $Q^2$-ordered showers
as an alternative to the P0 tune with $p_T$-ordered showers.
We take asymmetric systematic uncertainties 
defined as maximal deviations up and down from the central prediction with P0. 
Generally, they correspond to P-hard and P-soft tunes.
The typical size of the correction for the MPI effect is $0.96-0.98$ with an uncertainty of $2\%-5\%$.
The overall correction for the non-perturbative effects is applied to the {\sc jetphox}  predictions with uncertainties added to the theory scale uncertainty.
Tables \ref{tab:cross1}--\ref{tab:cross16} show measured and predicted NLO cross sections with their uncertainties
for all sixteen studied regions.

\begin{figure*}
\vskip-10mm
\includegraphics[scale=0.33]{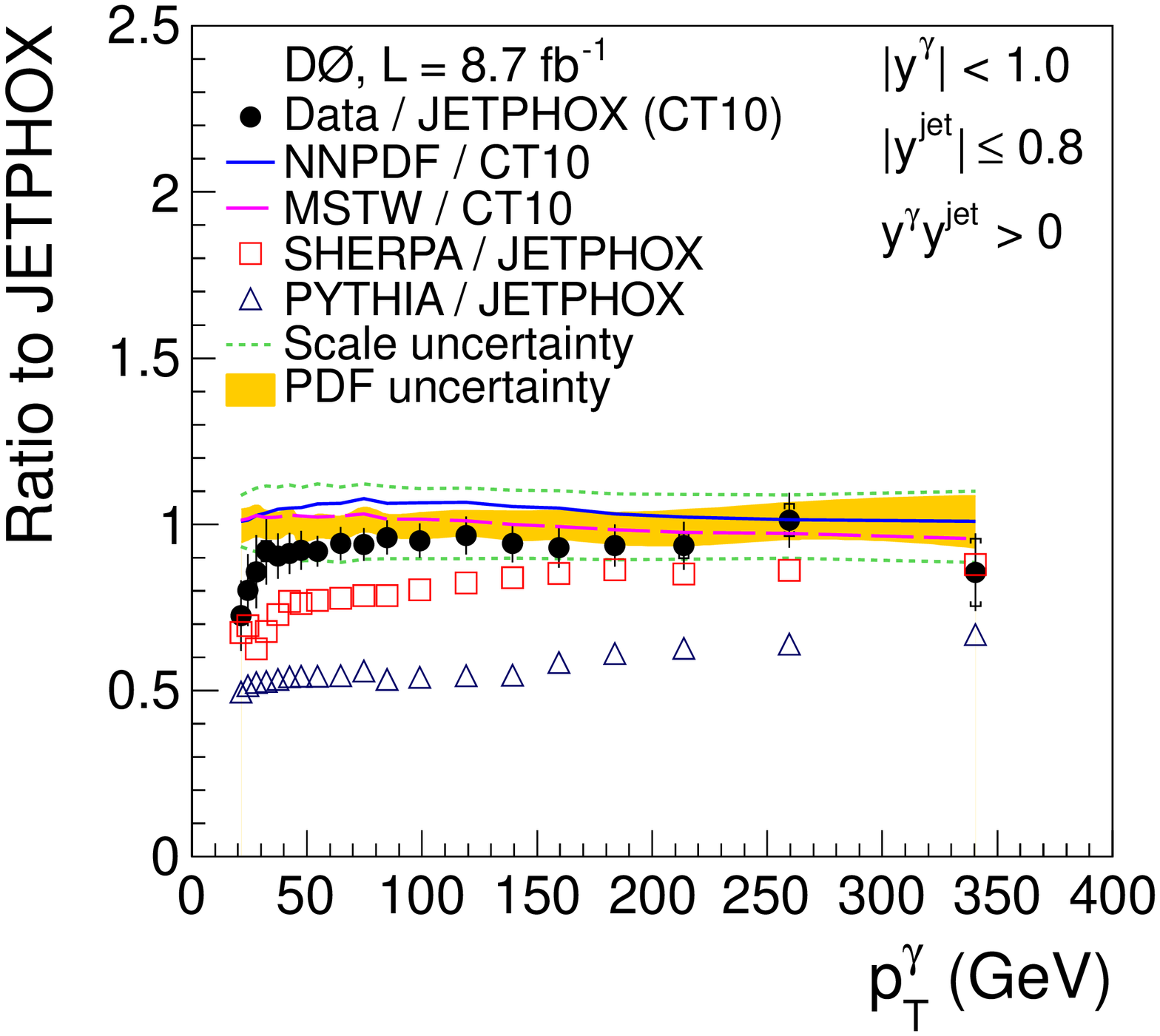}
\includegraphics[scale=0.33]{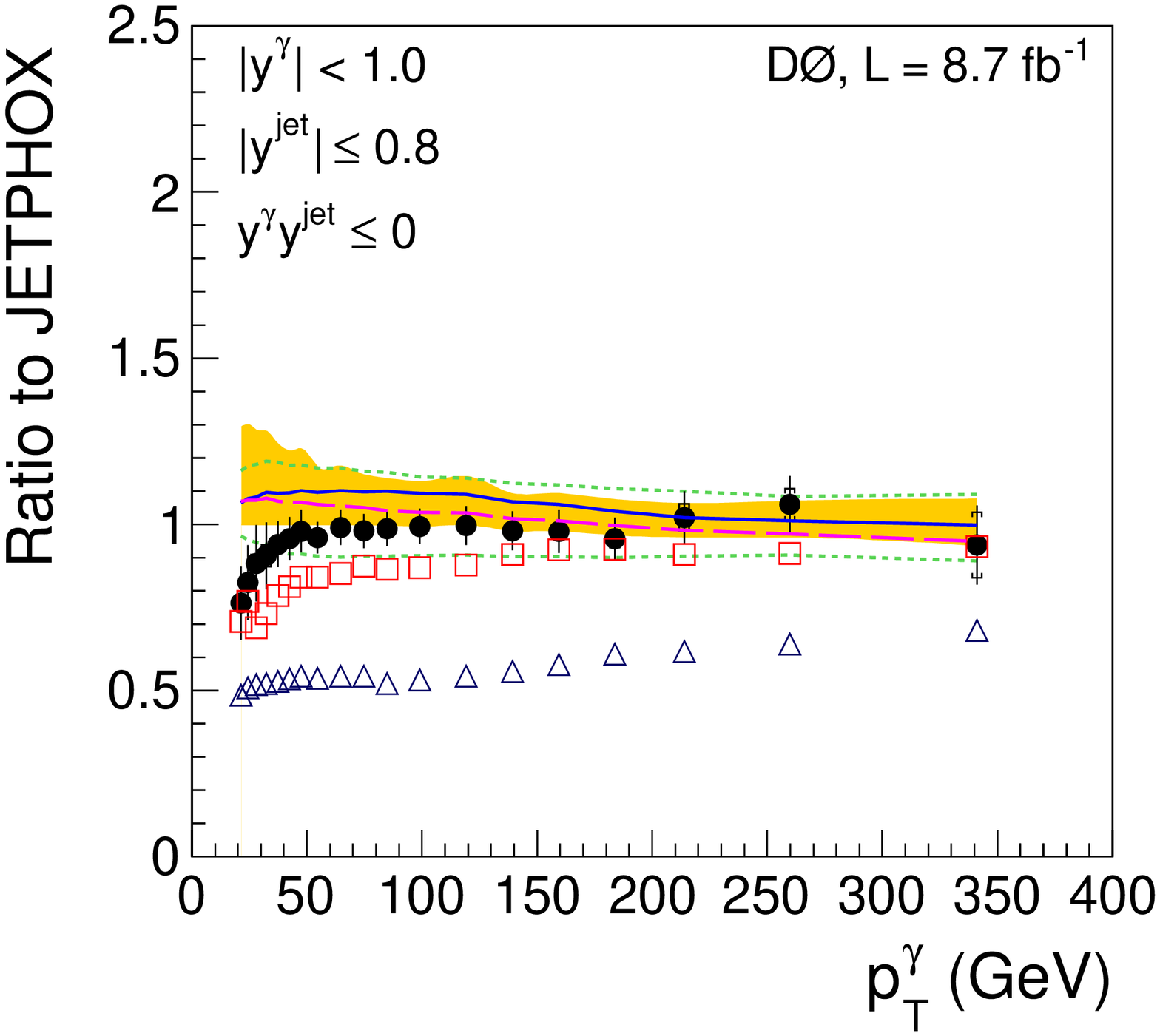}
\vskip-7mm
\includegraphics[scale=0.33]{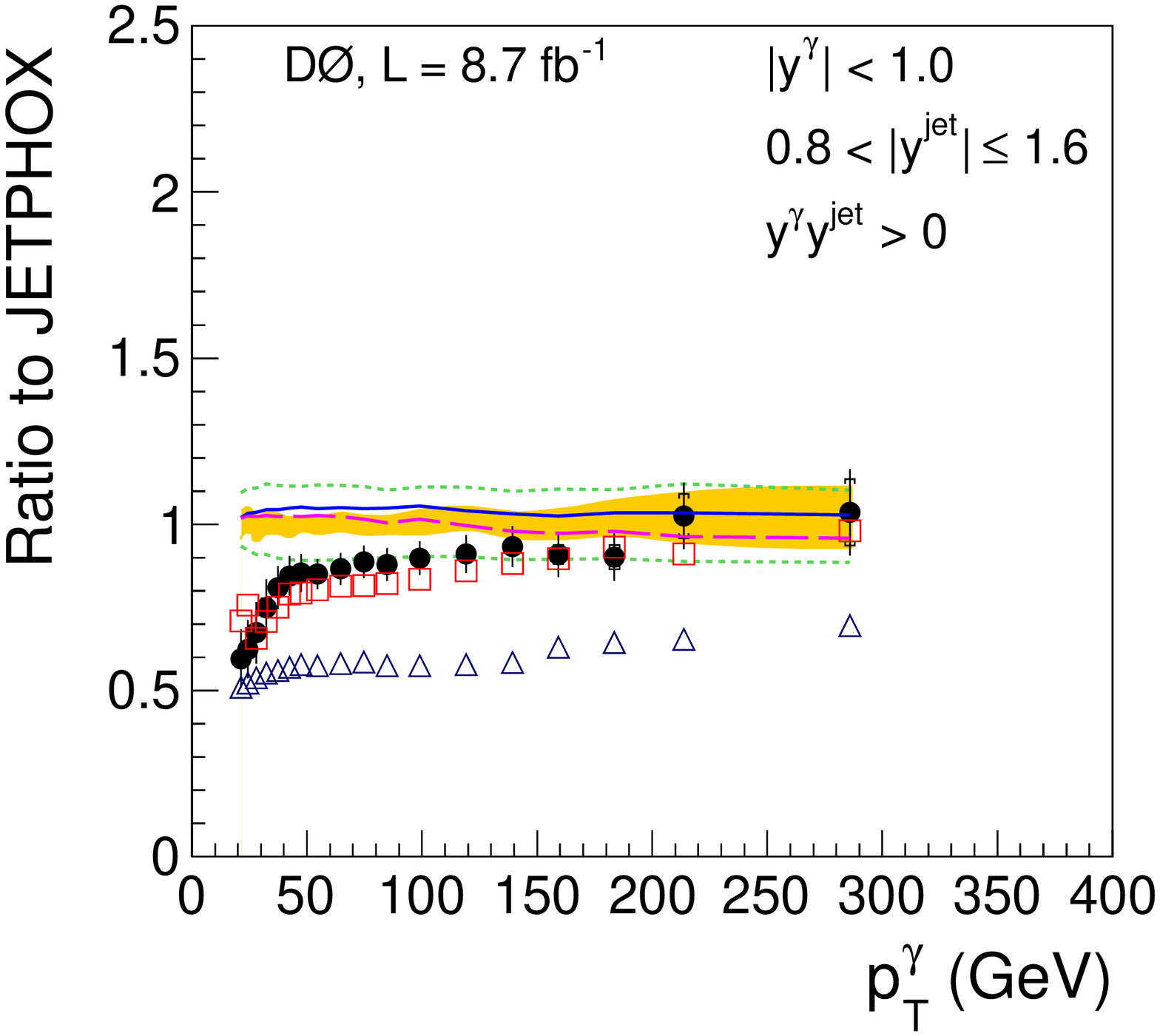}
\includegraphics[scale=0.33]{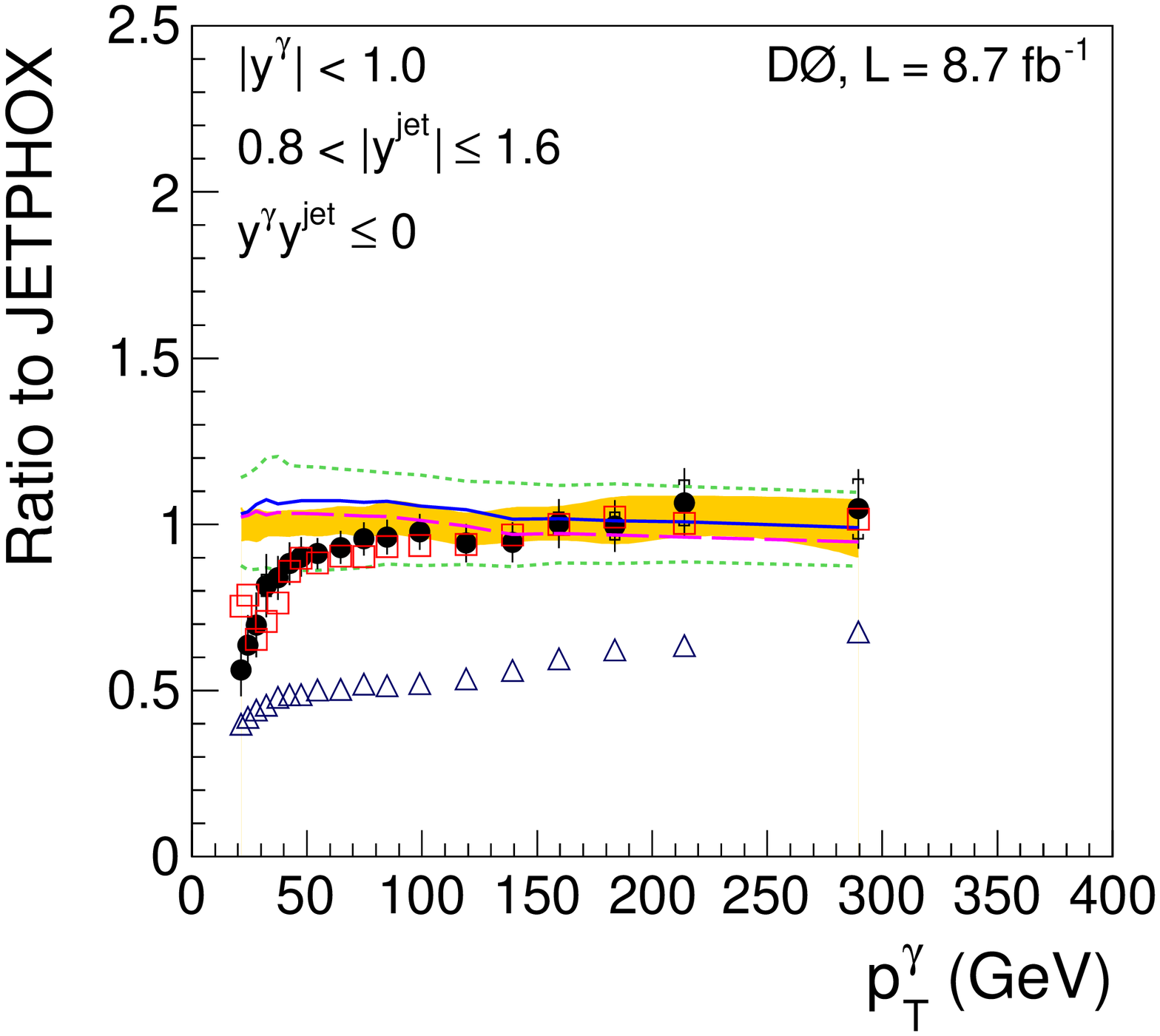}
\vskip-7mm
\includegraphics[scale=0.33]{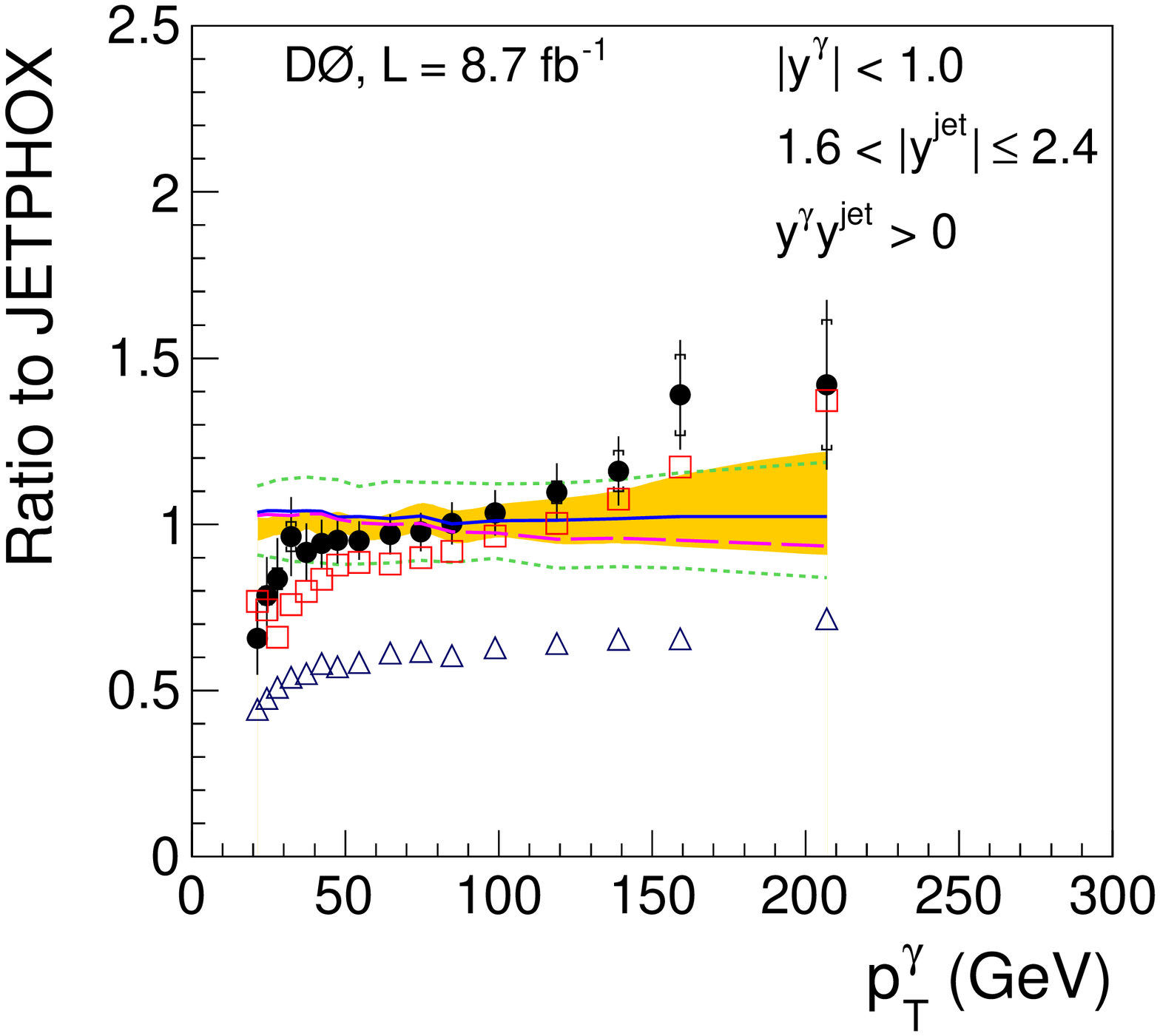}
\includegraphics[scale=0.33]{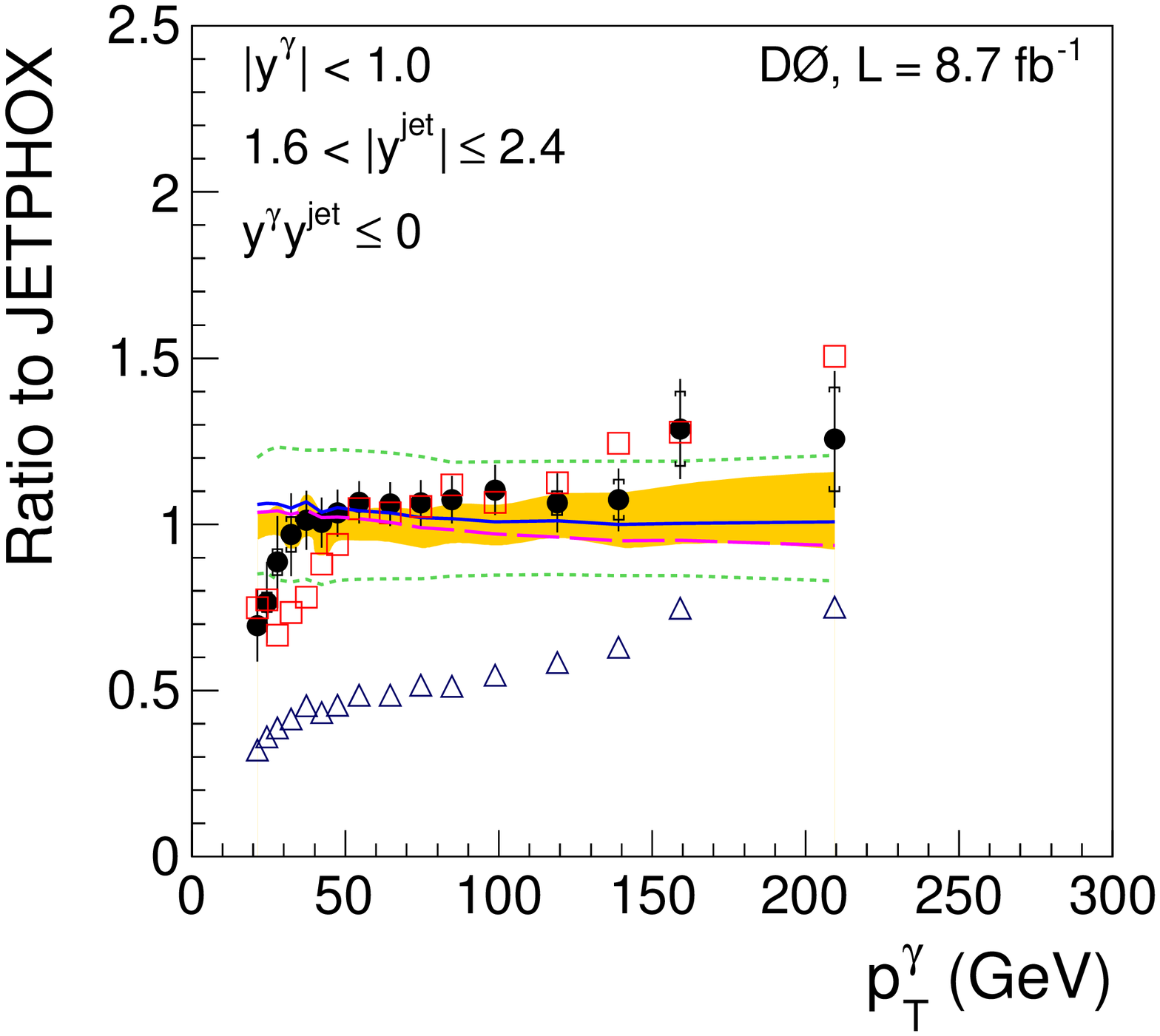}
\vskip-7mm
\includegraphics[scale=0.33]{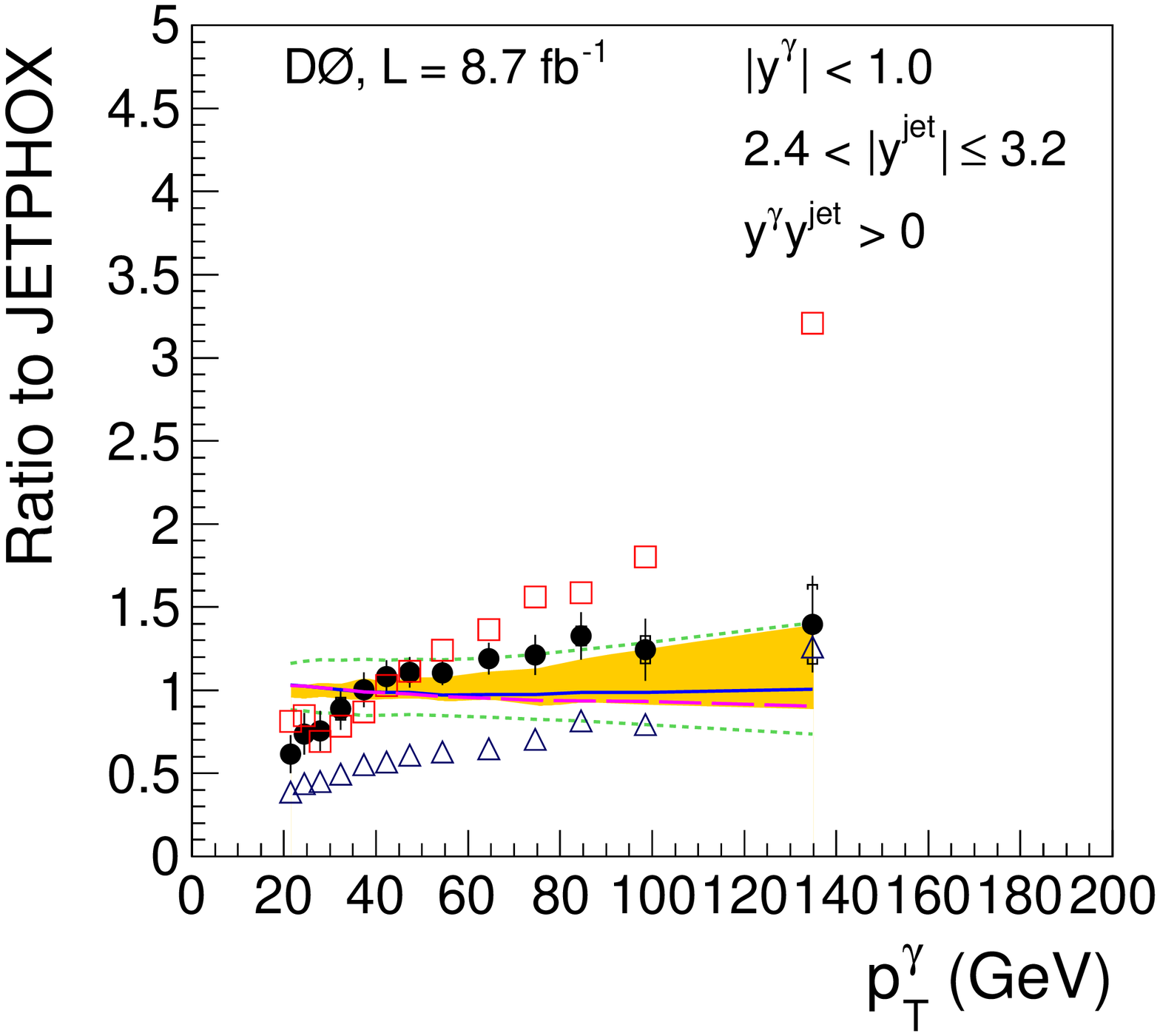}
\includegraphics[scale=0.33]{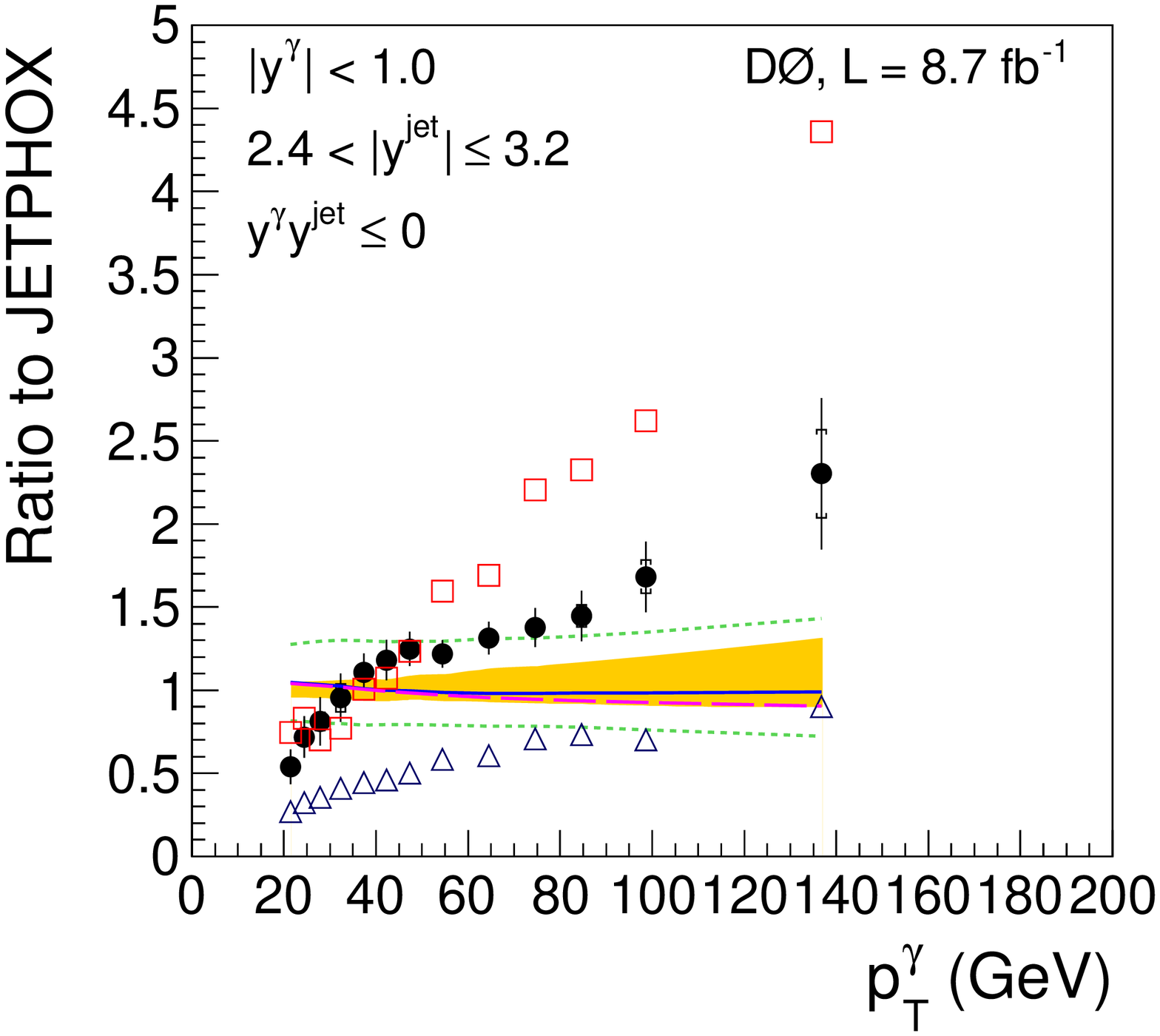}
\vskip-5mm
\caption{(color online)
Ratios of the measured differential cross sections with central photons in each of 
the four measured jet rapidity intervals
to the pQCD NLO prediction using {\sc jetphox} \cite{JETPHOX} with the CT10 PDF set and $\mu_{R}=\mu_{F}=\mu_f=\Ptg$.
The solid vertical line on the points shows the statistical and $p_T$-dependent systematic uncertainties added in quadrature,
while the internal line shows the statistical uncertainty.
A common $6.8\%$ normalization uncertainty on the data points is not shown.
The two dotted lines represent the effect of varying the
theoretical scales of {\sc jetphox} by a factor of two. The shaded region is the CT10 \cite{CT10} PDF uncertainty.
The dashed and dash-dotted lines show ratios of the {\sc jetphox} predictions with MSTW2008NLO \cite{MSTW}
and NNPDFv2.1 \cite{NNPDF} to CT10 PDF sets.
The predictions from {\sc sherpa} and {\sc pythia} are shown by the open squares and triangles, respectively.
}
\label{fig:DT_reg1}
\end{figure*}

\begin{figure*}
\vskip-10mm
\includegraphics[scale=0.33]{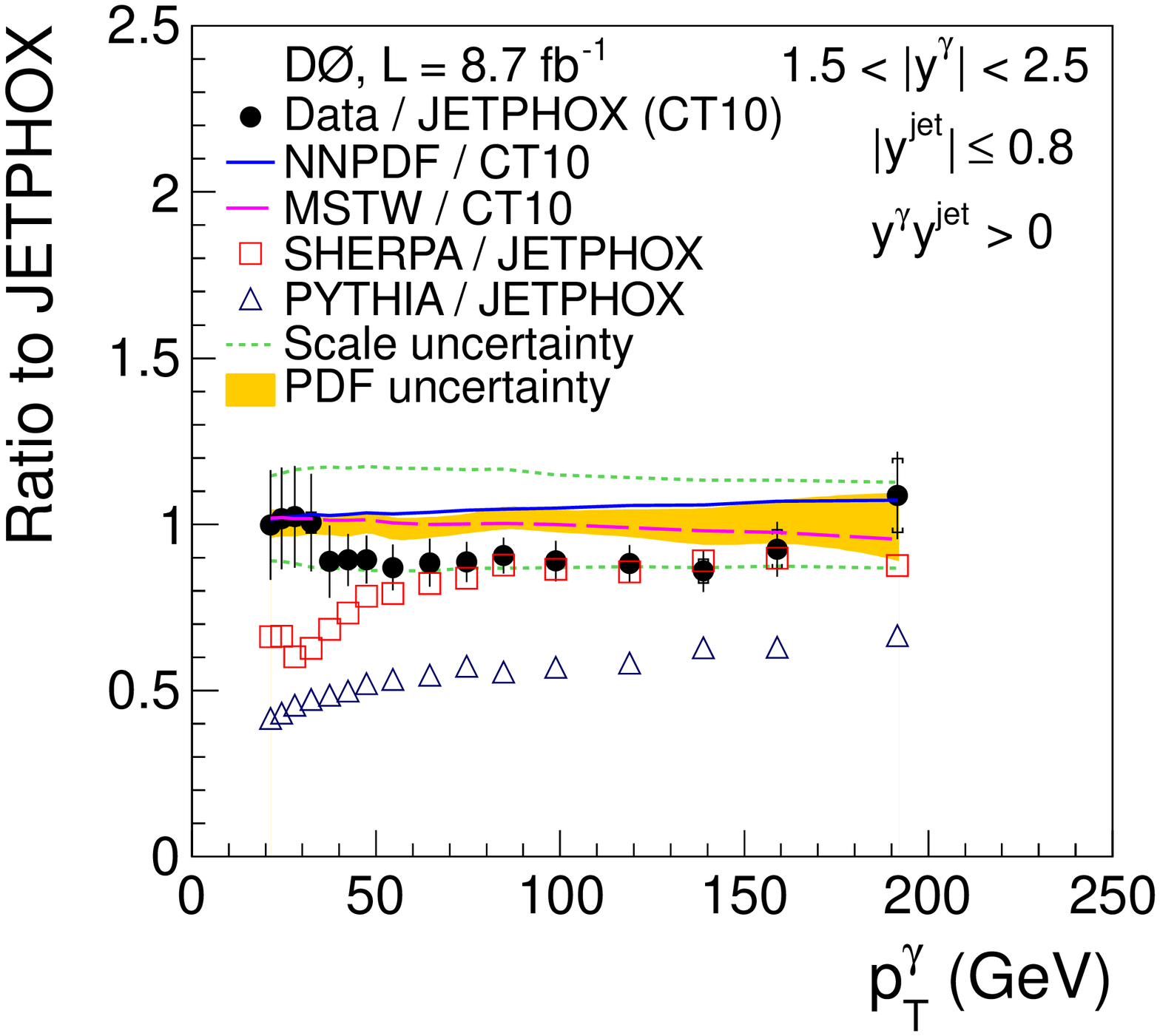}
\includegraphics[scale=0.33]{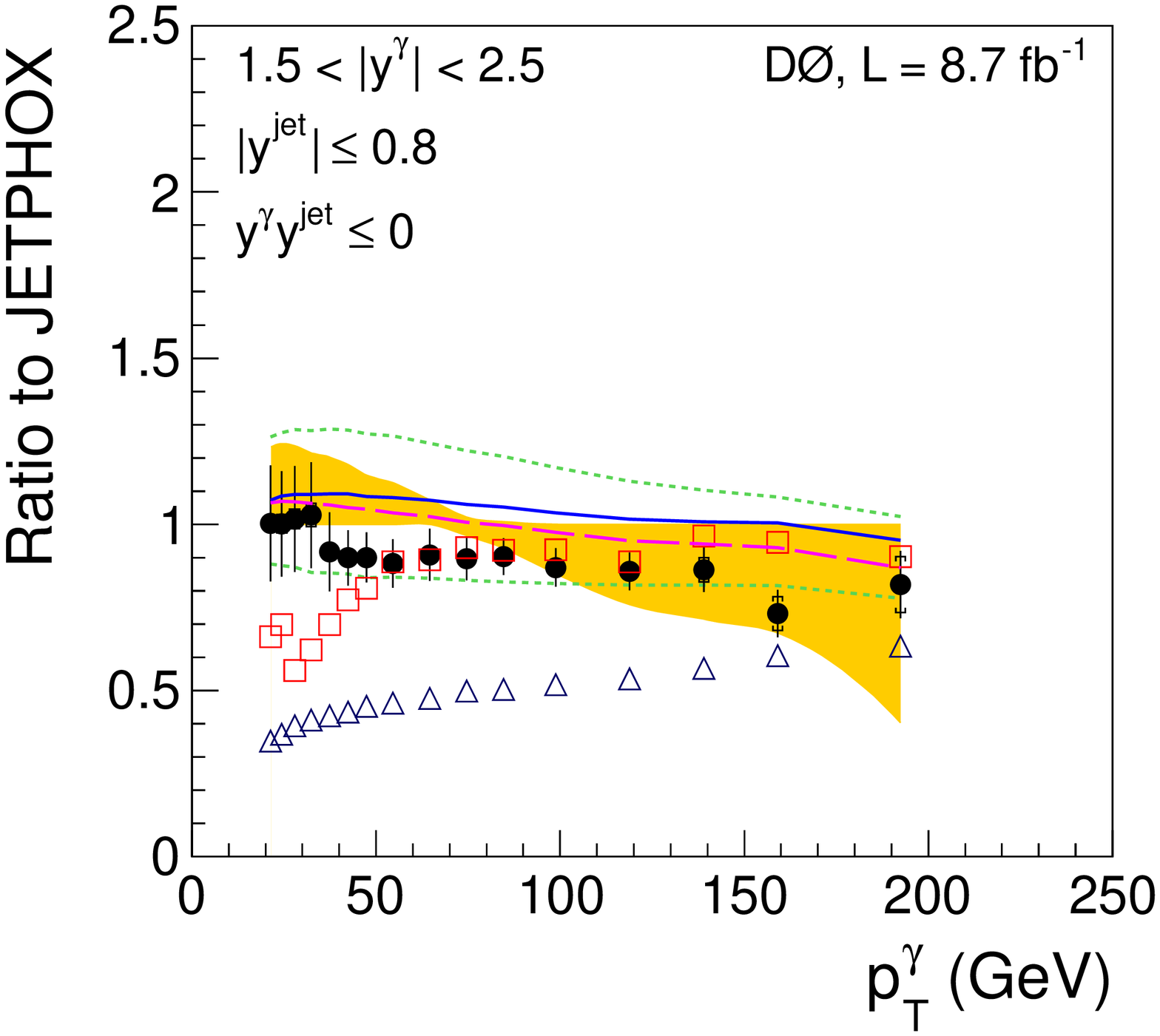}
\vskip-7mm
\includegraphics[scale=0.33]{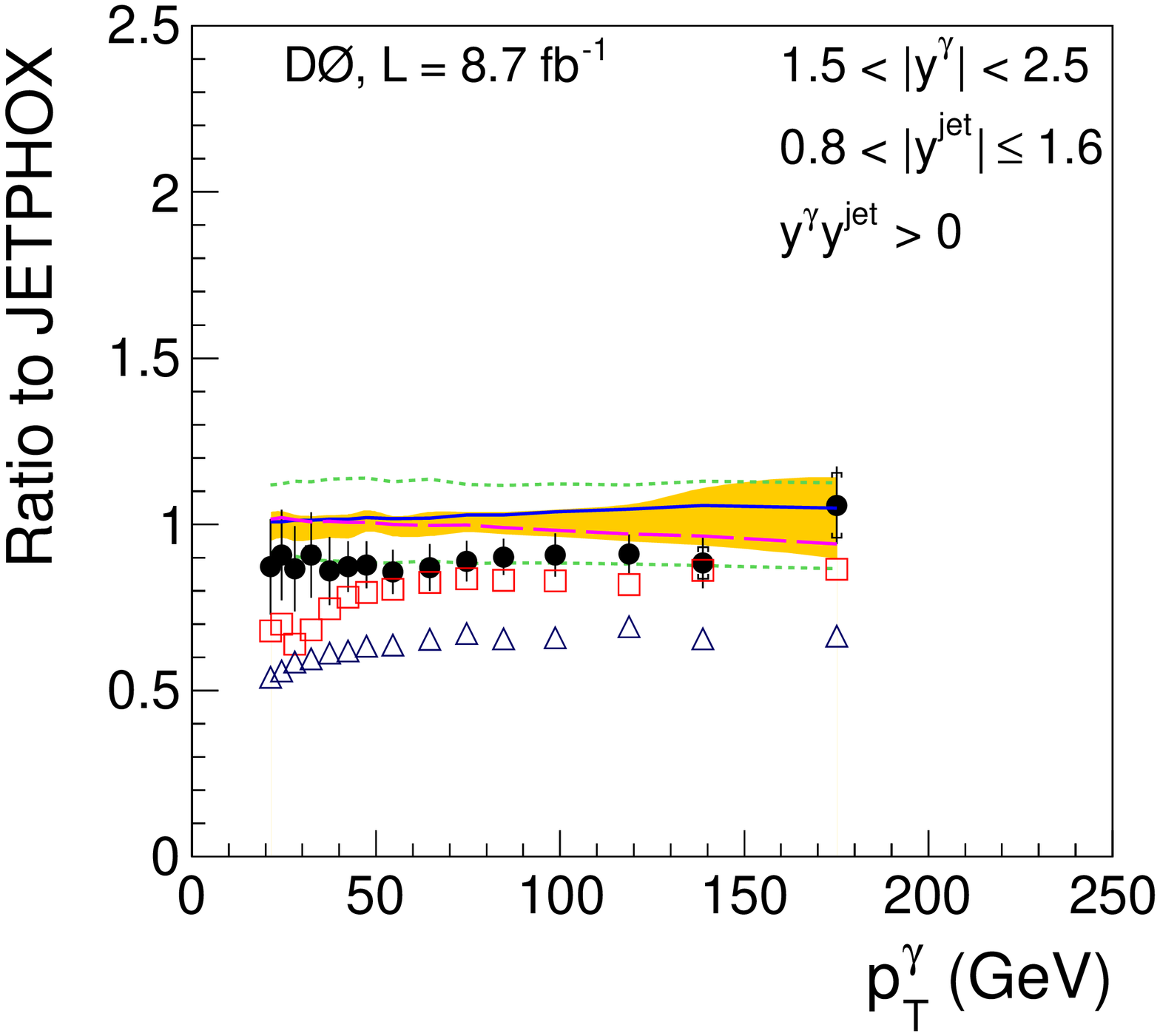}
\includegraphics[scale=0.33]{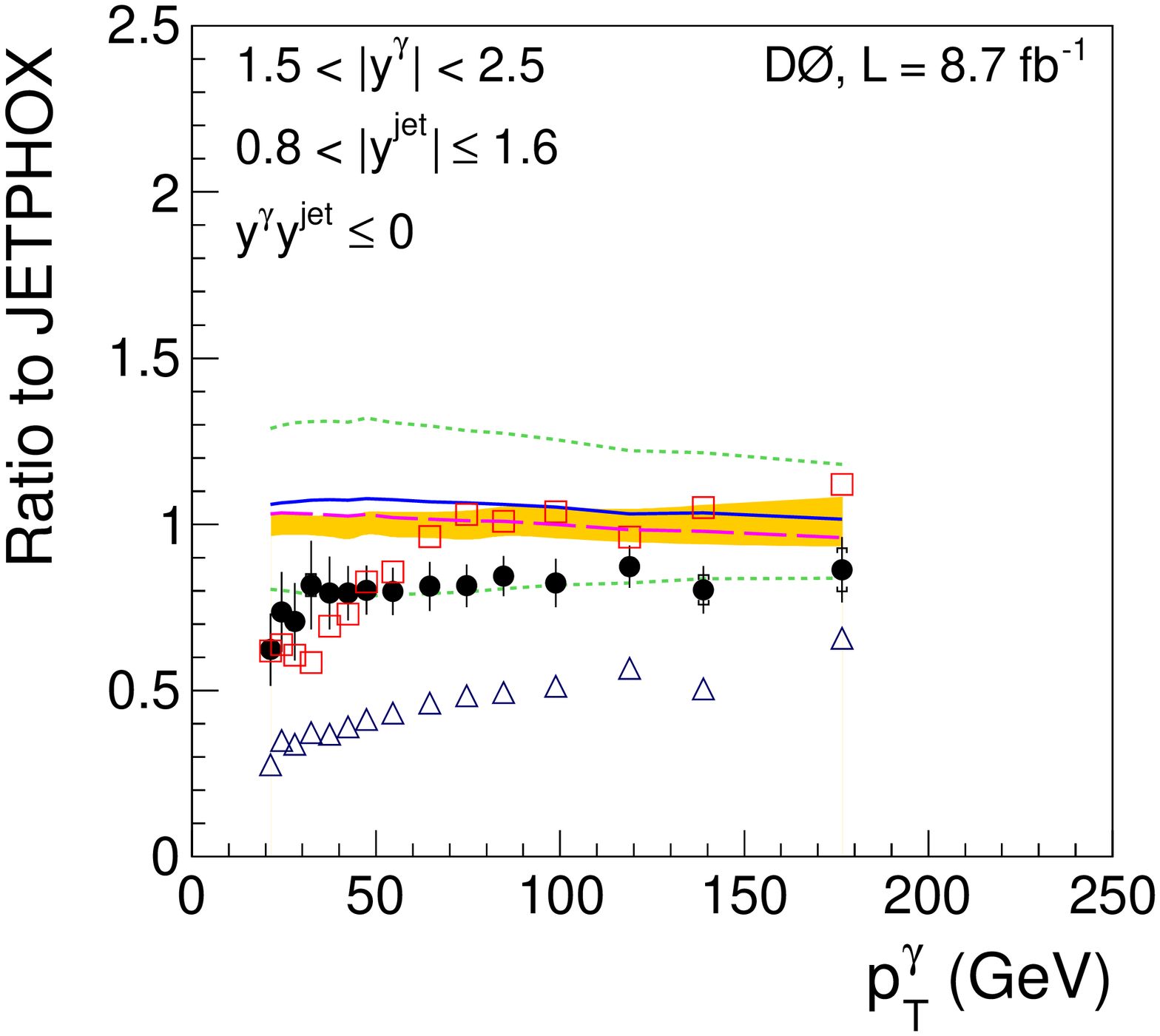}
\vskip-7mm
\includegraphics[scale=0.33]{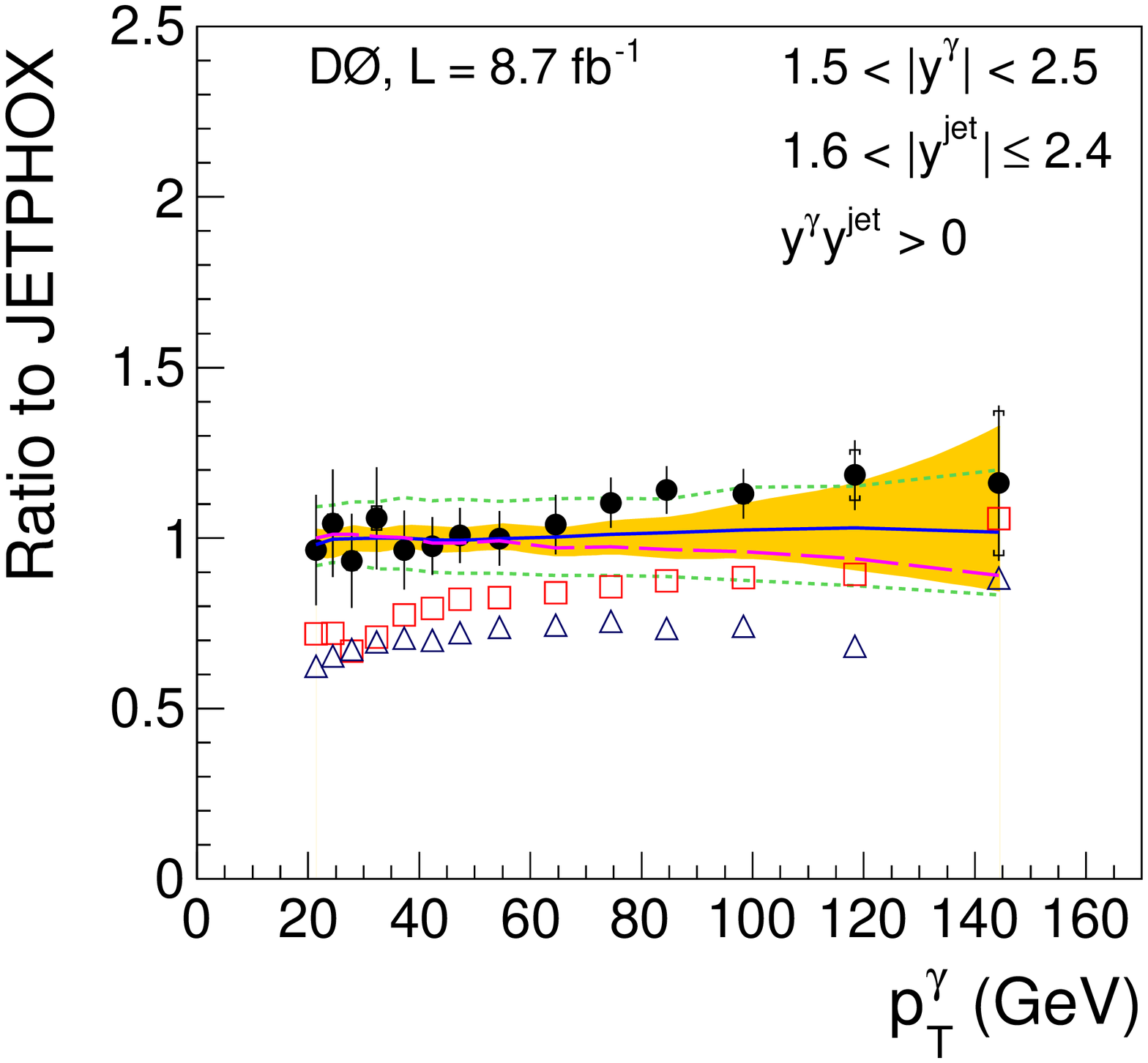}
\includegraphics[scale=0.33]{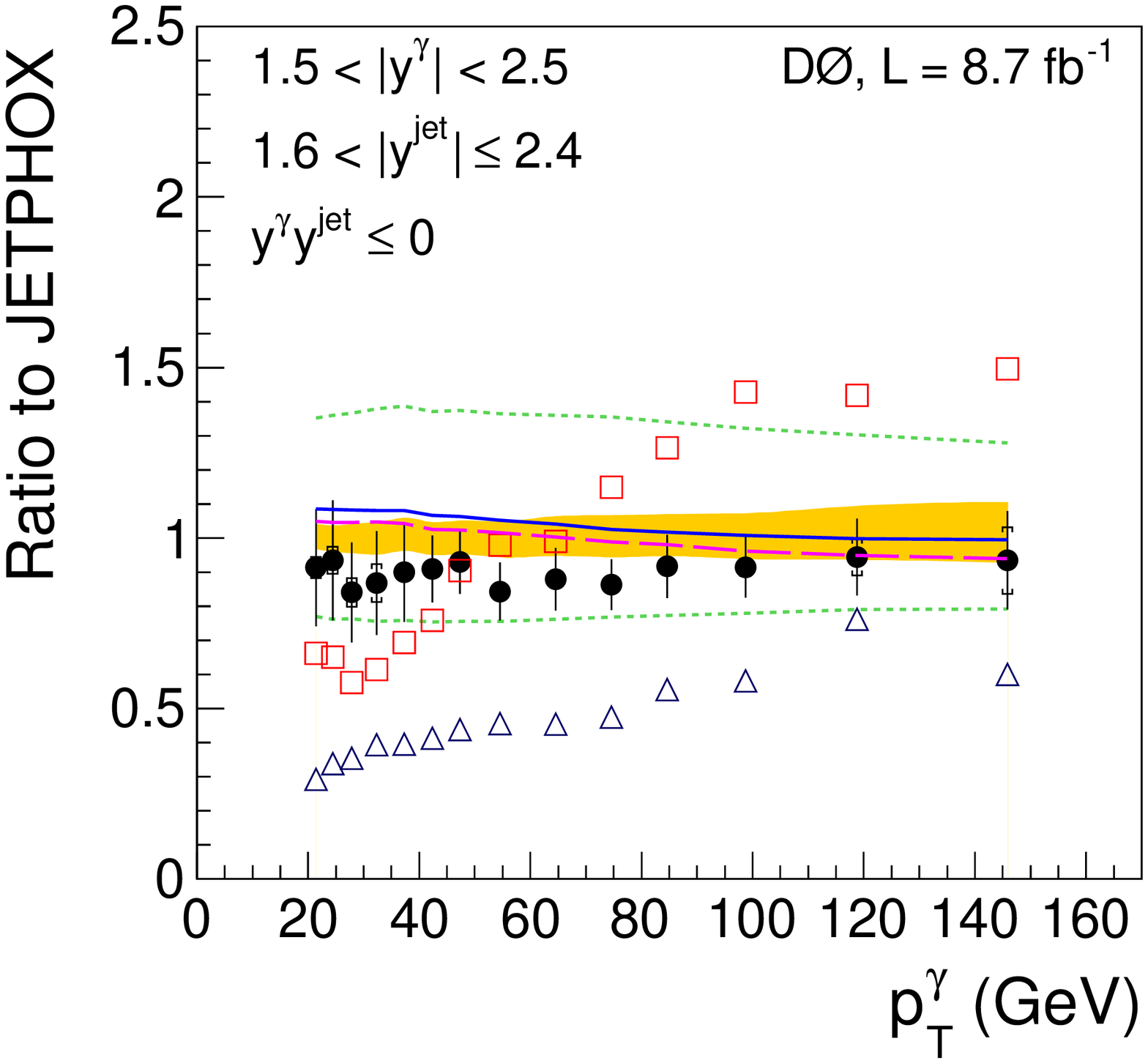}
\vskip-7mm
\includegraphics[scale=0.33]{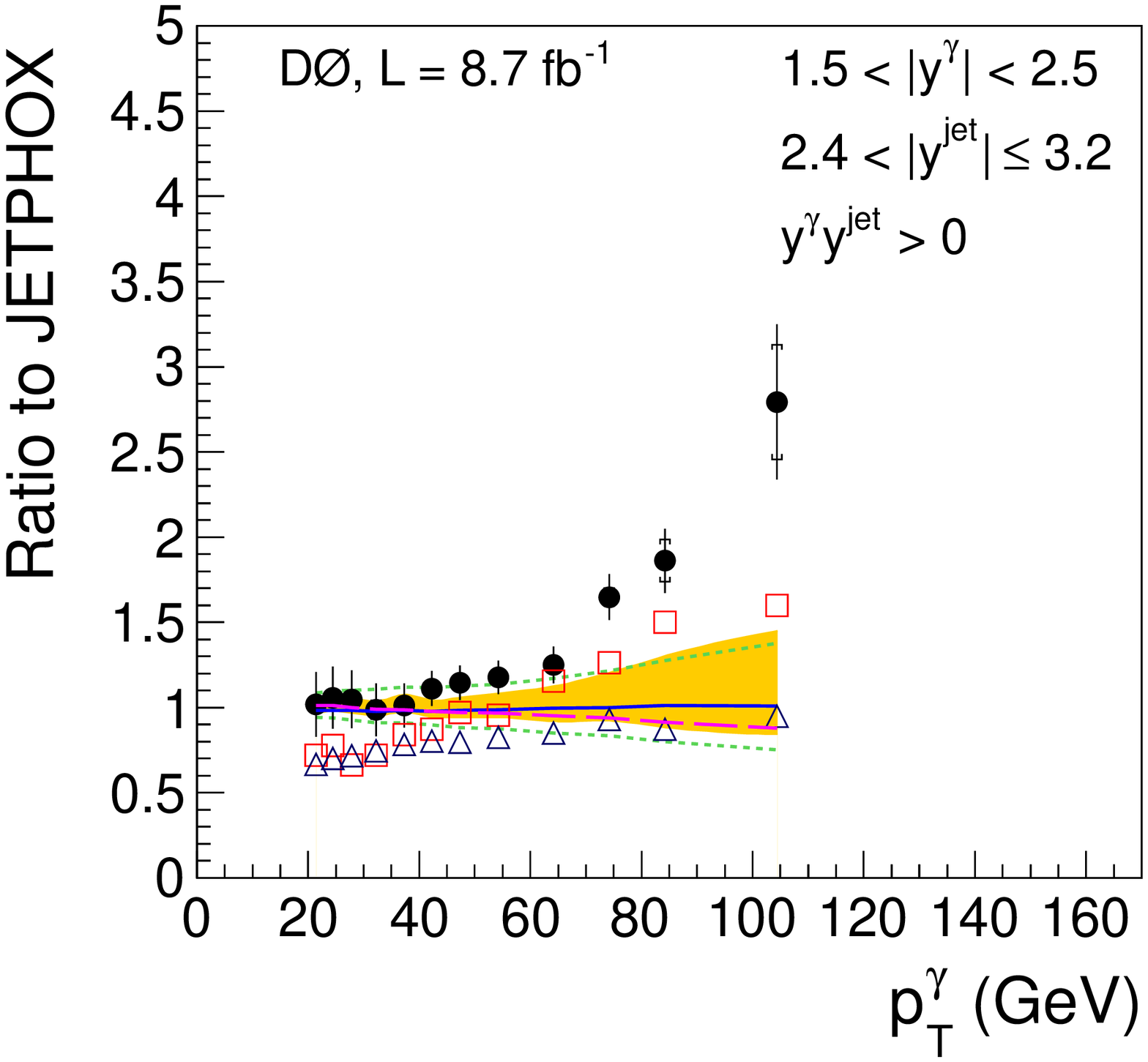}
\includegraphics[scale=0.33]{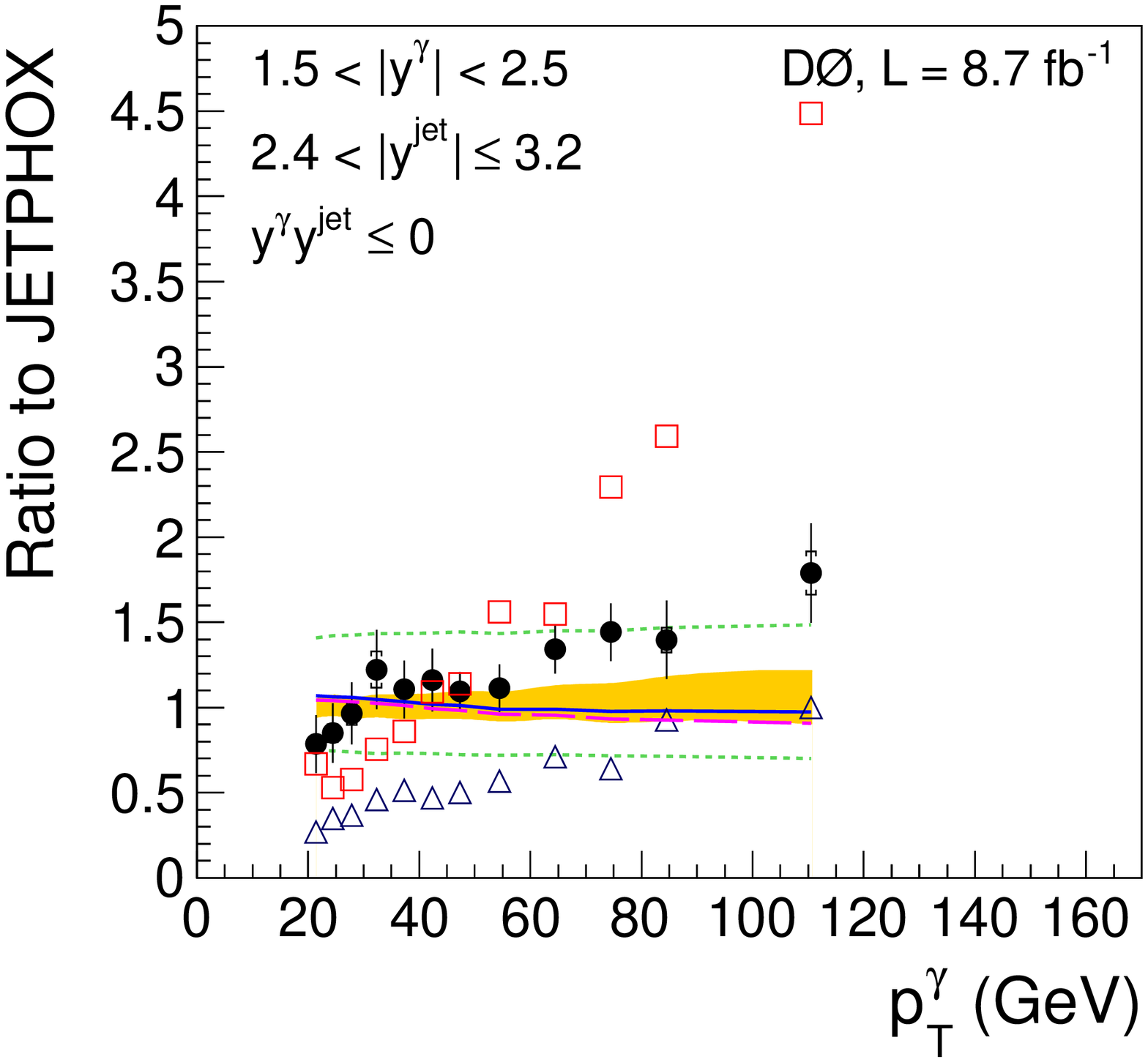}

\vskip-5mm
\caption{(color online)
Same as Fig.~\ref{fig:DT_reg1} but for events with forward photons.
A common $11.2\%$ normalization uncertainty on the data points is not shown.
}
\label{fig:DT_reg2}
\end{figure*}

To make a more detailed comparison, the ratio of the measured cross section 
to the pQCD NLO prediction is calculated in each interval. The results are shown 
in Figs.~\ref{fig:DT_reg1} and \ref{fig:DT_reg2}. 
The normalization uncertainty (6.8\% for events with central photons and 11.2\% for forward photons)
are not included in the figures. Ratios of the {\sc jetphox} predictions with MSTW2008NLO \cite{MSTW} 
and NNPDFv2.1 \cite{NNPDF} PDF sets to those with CT10 PDF set are also shown. The results are also compared
to the predictions from {\sc sherpa} and {\sc pythia}.
The {\sc jetphox} scale uncertainties are $10\%-15\%$ for events with central photons and jets,
but increase to $35\%-40\%$ for events with forward photons and more forward jets.
The CT10 PDF uncertainties usually increase with $\Ptg$ and may reach $40\%-50\%$ 
in some regions of the phase space,
e.g., at high $\Ptg$ with forward photons and either $y^\gamma y^{\rm jet}\leq0$ and $|y^{\rm jet}|\leq0.8$,
or $y^\gamma y^{\rm jet}\gt0$ and $2.4<|y^{\rm jet}|\leq3.2$.

For central photons, the pQCD NLO theory agrees with data except for small $\Ptg$ in almost 
all jet rapidity regions, and except for high $\Ptg$ with very forward jets 
($2.4<|y^{\mathrm{jet}}|\leq3.2$) and opposite-sign photon-jet rapidities. 
Qualitatively, these results are very similar to those obtained by ATLAS Collaboration \cite{Atlas_gj}.
Due to small size of the fragmentation photon contribution ($<\!10\%$) and a weak dependence
of theoretical scale uncertainties on $\Ptg$, 
a possible explanation is the mismodeling of the gluon PDF.
The shapes of cross sections predicted by {\sc sherpa} agree with the
data but are typically slightly low with
a significant exception  for events with very forward jets where the {\sc sherpa} predictions
agree well with data at 
$20 \leq \Ptg\lesssim 50$ GeV, and are much larger at higher $\Ptg$.
Predictions from {\sc pythia} are about a factor of $1.3-2$ below the measured data points.
For events with forward photons, the NLO theory  agrees with data within theoretical and experimental uncertainties,
except for the region $\Ptg>70$ GeV in the same-sign events with very forward jets.


\section{Summary}
\label{sec:summary}

The triple differential cross section 
$\mathrm{d^3}\sigma / \mathrm{d}\Ptg\mathrm{d}y^{\gamma}\mathrm{d}y^{\mathrm{jet}} $ for 
the associated inclusive photon and jet production process $p\bar{p}\rightarrow \gamma +\mathrm{jet} +X$ is measured for events 
with central ($|y^\gamma|\lt 1.0$) and forward ($1.5\lt |y^\gamma|\lt 2.5$) photons  
in four jet rapidity intervals ($|y^\text{jet}|\leq 0.8$, $0.8<|y^\text{jet}|\leq 1.6$, 
$1.6<|y^\text{jet}|\leq 2.4$, and $2.4<|y^\text{jet}|\leq 3.2$), for 
configurations with same and for opposite sign of photon and jet rapidities. 

The pQCD NLO predictions describe data with central photons
in almost all jet rapidity regions except low $\Ptg$ ($<40$~GeV) and the opposite-sign rapidity events at high $\Ptg$
with very forward jets ($2.4<|y^{\mathrm{jet}}|<3.2$).
They also describe data with forward photons 
except for the same-sign rapidity events with $\Ptg>70$ GeV and $2.4<|y^{\mathrm{jet}}|\leq3.2$.
The measured cross sections typically have similar or smaller uncertainties than 
the NLO PDF and scale uncertainties.
These measurements provide valuable information for tuning QCD theory predictions 
and particularly can be used as valuable input to global fits to gluon and other PDFs.

%
We thank the staffs at Fermilab and collaborating institutions,
and acknowledge support from the
DOE and NSF (USA);
CEA and CNRS/IN2P3 (France);
MON, NRC KI and RFBR (Russia);
CNPq, FAPERJ, FAPESP and FUNDUNESP (Brazil);
DAE and DST (India);
Colciencias (Colombia);
CONACyT (Mexico);
NRF (Korea);
FOM (The Netherlands);
STFC and the Royal Society (United Kingdom);
MSMT and GACR (Czech Republic);
BMBF and DFG (Germany);
SFI (Ireland);
The Swedish Research Council (Sweden);
and
CAS and CNSF (China).
%

\begin{table*}
\appendix
\section*{Appendix: Measured cross sections}
\centering
\caption{The \gpj cross section \tcs in bins of $\Ptg$ for $|y^\gamma|<1.0$ and $|y^{\rm jet}|\leq0.8$, $y^{\gamma}y^{\rm jet} \gt 0$
together with statistical ($\delta_{\text{stat}}$) and systematic ($\delta_{\text{syst}}$) uncertainties, and the NLO prediction 
together with scale ($\delta_{\rm scale}$) and PDF ($\delta_{\rm pdf}$) uncertainties. 
A common normalization uncertainty of $6.8\%$ is included in $\delta_{\text{syst}}$ for all points.}
\label{tab:cross1}
\begin{tabular}{ccccccccc} \hline\hline \\ [-2.5ex]
 ~$\Ptg$ bin~ & ~$\langle \Ptg \rangle$~ & \multicolumn{7}{c}{\tcs (pb/GeV)} \\[0.1ex] \cline{3-9} \\ [-2.5ex]
  ~(GeV)~     & ~(GeV)~    & ~Data~  & ~$\delta_{\rm stat} (\%)$~ & ~$\delta_{\rm syst}(\%)$~ & ~$\delta_{\rm tot}(\%)$ ~ & ~NLO~ & ~$\delta_{\rm scale}(\%)$~ & ~$\delta_{\rm pdf}(\%)$~ \\\hline
   20 --  23  &   21.4  & $  5.52\times10^{1} $ &   2.4  &  15.3  &  15.5 & $  7.61\times10^{1} $ &  ~$+8.7/-\!\!6.7$~  & ~$+4.6/-\!\!5.4$~ \\
   23 --  26  &   24.4  & $  3.69\times10^{1} $ &   2.7  &  14.4  &  14.7 & $  4.61\times10^{1} $ &  ~$+9.7/-\!\!7.5$~  & ~$+4.8/-\!\!4.6$~ \\
   26 --  30  &   27.9  & $  2.30\times10^{1} $ &   2.9  &  14.3  &  14.6 & $  2.68\times10^{1} $ &  ~$+10.9/-\!\!8.3$~  & ~$+5.7/-\!\!3.6$~ \\
   30 --  35  &   32.3  & $  1.31\times10^{1} $ &   3.3  &  12.4  &  12.8 & $  1.43\times10^{1} $ &  ~$+11.6/-\!\!8.9$~  & ~$+4.0/-\!\!4.3$~ \\
   35 --  40  &   37.3  & $  6.87\times10^{0} $ &   1.3  &  10.0  &  10.1 & $  7.60\times10^{0} $ &  ~$+11.2/-\!\!10.3$~  & ~$+3.6/-\!\!4.3$~ \\
   40 --  45  &   42.4  & $  3.96\times10^{0} $ &   1.3  &   9.3  &   9.4 & $  4.34\times10^{0} $ &  ~$+11.8/-\!\!10.4$~  & ~$+4.4/-\!\!2.7$~ \\
   45 --  50  &   47.4  & $  2.44\times10^{0} $ &   1.3  &   9.0  &   9.1 & $  2.64\times10^{0} $ &  ~$+11.0/-\!\!11.0$~  & ~$+1.8/-\!\!5.2$~ \\
   50 --  60  &   54.6  & $  1.28\times10^{0} $ &   1.3  &   8.2  &   8.3 & $  1.39\times10^{0} $ &  ~$+12.1/-\!\!10.9$~  & ~$+2.9/-\!\!4.1$~ \\
   60 --  70  &   64.7  & $  6.03\times10^{-1} $ &   1.3  &   8.3  &   8.4 & $  6.40\times10^{-1} $ &  ~$+11.3/-\!\!11.4$~  & ~$+2.4/-\!\!4.5$~ \\
   70 --  80  &   74.7  & $  3.05\times10^{-1} $ &   1.3  &   8.3  &   8.4 & $  3.25\times10^{-1} $ &  ~$+12.1/-\!\!10.6$~  & ~$+5.0/-\!\!2.2$~ \\
   80 --  90  &   84.7  & $  1.73\times10^{-1} $ &   1.4  &   8.4  &   8.5 & $  1.80\times10^{-1} $ &  ~$+11.4/-\!\!10.4$~  & ~$+2.9/-\!\!3.8$~ \\
   90 -- 110  &   99.1  & $  8.04\times10^{-2} $ &   1.4  &   8.4  &   8.5 & $  8.46\times10^{-2} $ &  ~$+10.8/-\!\!10.3$~  & ~$+3.4/-\!\!4.2$~ \\
  110 -- 130  &  119.2  & $  3.27\times10^{-2} $ &   1.6  &   8.5  &   8.6 & $  3.38\times10^{-2} $ &  ~$+10.9/-\!\!10.4$~  & ~$+4.1/-\!\!3.4$~ \\
  130 -- 150  &  139.3  & $  1.44\times10^{-2} $ &   1.9  &   8.6  &   8.8 & $  1.53\times10^{-2} $ &  ~$+10.2/-\!\!10.2$~  & ~$+3.9/-\!\!4.8$~ \\
  150 -- 170  &  159.4  & $  6.95\times10^{-3} $ &   2.4  &   8.6  &   8.9 & $  7.47\times10^{-3} $ &  ~$+10.1/-\!\!10.3$~  & ~$+4.1/-\!\!4.5$~ \\
  170 -- 200  &  183.7  & $  3.16\times10^{-3} $ &   2.7  &   8.7  &   9.2 & $  3.38\times10^{-3} $ &  ~$+9.1/-\!\!10.7$~  & ~$+3.5/-\!\!6.1$~ \\
  200 -- 230  &  213.8  & $  1.28\times10^{-3} $ &   4.0  &   8.9  &   9.7 & $  1.37\times10^{-3} $ &  ~$+9.0/-\!\!10.5$~  & ~$+4.3/-\!\!6.3$~ \\
  230 -- 300  &  259.6  & $  3.88\times10^{-4} $ &   4.7  &   9.1  &  10.2 & $  3.83\times10^{-4} $ &  ~$+8.8/-\!\!10.3$~  & ~$+6.7/-\!\!4.4$~ \\
  300 -- 400  &  340.5  & $  3.95\times10^{-5} $ &  11.9  &   9.4  &  15.2 & $  4.62\times10^{-5} $ &  ~$+10.0/-\!\!11.5$~  & ~$+8.5/-\!\!7.1$~ \\
\hline\hline
\end{tabular}
\end{table*}

\begin{table*}
\centering
\caption{Same as in Table~\ref{tab:cross1} but for $|y^\gamma|<1.0$ and $0.8\lt|y^{\rm jet}|\leq1.6$, $y^{\gamma}y^{\rm jet} \gt 0$.}
\label{tab:cross2}
\begin{tabular}{ccccccccc} \hline\hline \\ [-2.5ex]
 ~$\Ptg$ bin~ & ~$\langle \Ptg \rangle$~ & \multicolumn{7}{c}{\tcs (pb/GeV)} \\[0.1ex] \cline{3-9} \\ [-2.5ex]
  ~(GeV)~     & ~(GeV)~    & ~Data~  & ~$\delta_{\rm stat} (\%)$~ & ~$\delta_{\rm syst}(\%)$~ & ~$\delta_{\rm tot}(\%)$ ~ & ~NLO~ & ~$\delta_{\rm scale}(\%)$~ & ~$\delta_{\rm pdf}(\%)$~ \\\hline
   20 --  23  &   21.4  & $  3.70\times10^{1} $ &   2.4  &  15.7  &  15.9 & $  6.22\times10^{1} $ &  ~$+9.5/-\!\!6.6$~  & ~$+4.2/-\!\!4.5$~ \\
   23 --  26  &   24.4  & $  2.32\times10^{1} $ &   2.9  &  15.0  &  15.2 & $  3.72\times10^{1} $ &  ~$+10.9/-\!\!7.9$~  & ~$+5.1/-\!\!2.5$~ \\
   26 --  30  &   27.9  & $  1.45\times10^{1} $ &   3.1  &  15.2  &  15.5 & $  2.16\times10^{1} $ &  ~$+10.9/-\!\!9.1$~  & ~$+2.5/-\!\!5.1$~ \\
   30 --  35  &   32.3  & $  8.44\times10^{0} $ &   3.7  &  12.6  &  13.1 & $  1.13\times10^{1} $ &  ~$+12.1/-\!\!9.2$~  & ~$+3.7/-\!\!3.0$~ \\
   35 --  40  &   37.3  & $  4.79\times10^{0} $ &   1.3  &  10.5  &  10.6 & $  5.92\times10^{0} $ &  ~$+11.7/-\!\!10.3$~  & ~$+3.2/-\!\!3.0$~ \\
   40 --  45  &   42.4  & $  2.84\times10^{0} $ &   1.3  &   9.7  &   9.8 & $  3.36\times10^{0} $ &  ~$+11.4/-\!\!10.7$~  & ~$+2.1/-\!\!3.9$~ \\
   45 --  50  &   47.4  & $  1.71\times10^{0} $ &   1.3  &   9.3  &   9.4 & $  2.01\times10^{0} $ &  ~$+11.4/-\!\!10.8$~  & ~$+2.4/-\!\!2.4$~ \\
   50 --  60  &   54.6  & $  8.87\times10^{-1} $ &   1.3  &   8.4  &   8.5 & $  1.04\times10^{0} $ &  ~$+11.9/-\!\!10.8$~  & ~$+2.4/-\!\!3.0$~ \\
   60 --  70  &   64.6  & $  4.04\times10^{-1} $ &   1.3  &   8.6  &   8.7 & $  4.67\times10^{-1} $ &  ~$+11.6/-\!\!10.8$~  & ~$+3.5/-\!\!2.2$~ \\
   70 --  80  &   74.7  & $  2.06\times10^{-1} $ &   1.4  &   8.5  &   8.6 & $  2.33\times10^{-1} $ &  ~$+11.4/-\!\!10.3$~  & ~$+2.7/-\!\!3.2$~ \\
   80 --  90  &   84.7  & $  1.09\times10^{-1} $ &   1.4  &   8.6  &   8.7 & $  1.24\times10^{-1} $ &  ~$+10.3/-\!\!10.0$~  & ~$+2.6/-\!\!3.0$~ \\
   90 -- 110  &   99.0  & $  5.00\times10^{-2} $ &   1.4  &   8.6  &   8.7 & $  5.57\times10^{-2} $ &  ~$+11.2/-\!\!9.7$~  & ~$+4.4/-\!\!3.1$~ \\
  110 -- 130  &  119.1  & $  1.85\times10^{-2} $ &   1.8  &   8.8  &   8.9 & $  2.04\times10^{-2} $ &  ~$+11.3/-\!\!9.8$~  & ~$+5.4/-\!\!1.8$~ \\
  130 -- 150  &  139.2  & $  7.75\times10^{-3} $ &   2.3  &   9.0  &   9.3 & $  8.31\times10^{-3} $ &  ~$+9.9/-\!\!10.8$~  & ~$+3.7/-\!\!4.2$~ \\
  150 -- 170  &  159.3  & $  3.24\times10^{-3} $ &   3.2  &   9.3  &   9.8 & $  3.57\times10^{-3} $ &  ~$+10.6/-\!\!10.5$~  & ~$+4.6/-\!\!4.5$~ \\
  170 -- 200  &  183.6  & $  1.22\times10^{-3} $ &   4.1  &   9.2  &  10.1 & $  1.35\times10^{-3} $ &  ~$+10.3/-\!\!10.3$~  & ~$+7.4/-\!\!3.5$~ \\
  200 -- 230  &  213.8  & $  4.51\times10^{-4} $ &   6.5  &   9.4  &  11.5 & $  4.40\times10^{-4} $ &  ~$+12.2/-\!\!11.3$~  & ~$+9.6/-\!\!5.6$~ \\
  230 -- 400  &  285.9  & $  3.80\times10^{-5} $ &   9.7  &  10.4  &  14.2 & $  3.67\times10^{-5} $ &  ~$+10.2/-\!\!11.4$~  & ~$+11.4/-\!\!7.1$~ \\
\hline\hline
\end{tabular}
\end{table*}

\begin{table*}
\centering
\caption{Same as in Table~\ref{tab:cross1} but for $|y^\gamma|<1.0$ and $1.6\lt|y^{\rm jet}|\leq2.4$, $y^{\gamma}y^{\rm jet} \gt 0$.}
\label{tab:cross3}
\begin{tabular}{ccccccccc} \hline\hline \\ [-2.5ex]
 ~$\Ptg$ bin~ & ~$\langle \Ptg \rangle$~ & \multicolumn{7}{c}{\tcs (pb/GeV)} \\[0.1ex] \cline{3-9} \\ [-2.5ex]
  ~(GeV)~     & ~(GeV)~    & ~Data~  & ~$\delta_{\rm stat} (\%)$~ & ~$\delta_{\rm syst}(\%)$~ & ~$\delta_{\rm tot}(\%)$ ~ & ~NLO~ & ~$\delta_{\rm scale}(\%)$~ & ~$\delta_{\rm pdf}(\%)$~ \\\hline
   20 --  23  &   21.4  & $  2.26\times10^{1} $ &   2.9  &  17.0  &  17.3 & $  3.43\times10^{1} $ &  ~$+11.6/-\!\!9.3$~  & ~$+1.8/-\!\!4.7$~ \\
   23 --  26  &   24.4  & $  1.58\times10^{1} $ &   3.4  &  15.3  &  15.7 & $  2.01\times10^{1} $ &  ~$+12.3/-\!\!9.7$~  & ~$+1.7/-\!\!3.8$~ \\
   26 --  30  &   27.9  & $  9.45\times10^{0} $ &   3.8  &  15.7  &  16.1 & $  1.13\times10^{1} $ &  ~$+13.4/-\!\!10.3$~  & ~$+2.4/-\!\!3.0$~ \\
   30 --  35  &   32.3  & $  5.52\times10^{0} $ &   4.6  &  13.1  &  13.9 & $  5.73\times10^{0} $ &  ~$+13.8/-\!\!11.2$~  & ~$+2.1/-\!\!3.4$~ \\
   35 --  40  &   37.3  & $  2.63\times10^{0} $ &   1.3  &  11.6  &  11.7 & $  2.88\times10^{0} $ &  ~$+14.2/-\!\!11.4$~  & ~$+3.9/-\!\!1.3$~ \\
   40 --  45  &   42.4  & $  1.48\times10^{0} $ &   1.3  &  10.1  &  10.1 & $  1.57\times10^{0} $ &  ~$+13.7/-\!\!11.7$~  & ~$+2.7/-\!\!3.4$~ \\
   45 --  50  &   47.4  & $  8.61\times10^{-1} $ &   1.3  &   9.8  &   9.9 & $  9.05\times10^{-1} $ &  ~$+13.5/-\!\!12.1$~  & ~$+3.4/-\!\!2.1$~ \\
   50 --  60  &   54.5  & $  4.23\times10^{-1} $ &   1.3  &   9.0  &   9.1 & $  4.45\times10^{-1} $ &  ~$+11.4/-\!\!11.9$~  & ~$+1.4/-\!\!4.3$~ \\
   60 --  70  &   64.6  & $  1.76\times10^{-1} $ &   1.4  &   9.1  &   9.2 & $  1.82\times10^{-1} $ &  ~$+13.0/-\!\!11.7$~  & ~$+3.1/-\!\!4.0$~ \\
   70 --  80  &   74.6  & $  7.89\times10^{-2} $ &   1.5  &   9.0  &   9.1 & $  8.07\times10^{-2} $ &  ~$+12.7/-\!\!10.9$~  & ~$+6.0/-\!\!2.1$~ \\
   80 --  90  &   84.7  & $  3.87\times10^{-2} $ &   1.8  &   9.2  &   9.4 & $  3.86\times10^{-2} $ &  ~$+12.5/-\!\!11.5$~  & ~$+4.0/-\!\!5.3$~ \\
   90 -- 110  &   98.8  & $  1.48\times10^{-2} $ &   1.9  &   9.5  &   9.7 & $  1.43\times10^{-2} $ &  ~$+12.1/-\!\!10.3$~  & ~$+5.9/-\!\!3.7$~ \\
  110 -- 130  &  118.9  & $  4.28\times10^{-3} $ &   3.0  &  10.1  &  10.6 & $  3.91\times10^{-3} $ &  ~$+12.3/-\!\!13.2$~  & ~$+7.5/-\!\!5.7$~ \\
  130 -- 150  &  139.0  & $  1.28\times10^{-3} $ &   5.3  &  10.3  &  11.5 & $  1.10\times10^{-3} $ &  ~$+13.5/-\!\!12.7$~  & ~$+10.1/-\!\!5.5$~ \\
  150 -- 170  &  159.1  & $  4.45\times10^{-4} $ &   8.7  &  10.9  &  14.0 & $  3.20\times10^{-4} $ &  ~$+15.5/-\!\!13.2$~  & ~$+14.7/-\!\!6.5$~ \\
  170 -- 300  &  206.9  & $  2.82\times10^{-5} $ &  13.7  &  14.3  &  19.8 & $  1.98\times10^{-5} $ &  ~$+18.7/-\!\!16.1$~  & ~$+21.6/-\!\!9.0$~ \\
\hline\hline
\end{tabular}
\end{table*}

\begin{table*}
\centering
\caption{Same as in Table~\ref{tab:cross1} but for $|y^\gamma|<1.0$ and $2.4\lt|y^{\rm jet}|\leq3.2$, $y^{\gamma}y^{\rm jet} \gt 0$.}
\label{tab:cross4}
\begin{tabular}{ccccccccc} \hline\hline \\ [-2.5ex]
 ~$\Ptg$ bin~ & ~$\langle \Ptg \rangle$~ & \multicolumn{7}{c}{\tcs (pb/GeV)} \\[0.1ex] \cline{3-9} \\ [-2.5ex]
  ~(GeV)~     & ~(GeV)~    & ~Data~  & ~$\delta_{\rm stat} (\%)$~ & ~$\delta_{\rm syst}(\%)$~ & ~$\delta_{\rm tot}(\%)$ ~ & ~NLO~ & ~$\delta_{\rm scale}(\%)$~ & ~$\delta_{\rm pdf}(\%)$~ \\\hline
   20 --  23  &   21.4  & $  8.09\times10^{0} $ &   3.9  &  19.0  &  19.4 & $  1.32\times10^{1} $ &  ~$+16.1/-\!\!11.8$~  & ~$+3.4/-\!\!3.8$~ \\
   23 --  26  &   24.4  & $  5.44\times10^{0} $ &   4.9  &  16.6  &  17.3 & $  7.40\times10^{0} $ &  ~$+17.4/-\!\!12.6$~  & ~$+2.8/-\!\!4.4$~ \\
   26 --  30  &   27.9  & $  2.95\times10^{0} $ &   6.0  &  16.8  &  17.9 & $  3.91\times10^{0} $ &  ~$+18.3/-\!\!13.7$~  & ~$+3.7/-\!\!3.7$~ \\
   30 --  35  &   32.3  & $  1.61\times10^{0} $ &   7.5  &  13.7  &  15.6 & $  1.81\times10^{0} $ &  ~$+18.1/-\!\!13.8$~  & ~$+3.4/-\!\!4.7$~ \\
   35 --  40  &   37.3  & $  8.15\times10^{-1} $ &   1.4  &  12.2  &  12.3 & $  8.13\times10^{-1} $ &  ~$+18.7/-\!\!15.3$~  & ~$+6.4/-\!\!5.4$~ \\
   40 --  45  &   42.3  & $  4.22\times10^{-1} $ &   1.4  &  11.2  &  11.2 & $  3.89\times10^{-1} $ &  ~$+18.1/-\!\!15.1$~  & ~$+4.5/-\!\!4.9$~ \\
   45 --  50  &   47.3  & $  2.16\times10^{-1} $ &   1.4  &  10.4  &  10.5 & $  1.95\times10^{-1} $ &  ~$+18.5/-\!\!14.9$~  & ~$+6.8/-\!\!4.4$~ \\
   50 --  60  &   54.5  & $  8.67\times10^{-2} $ &   1.5  &   9.7  &   9.9 & $  7.86\times10^{-2} $ &  ~$+18.3/-\!\!15.4$~  & ~$+7.5/-\!\!6.0$~ \\
   60 --  70  &   64.5  & $  2.78\times10^{-2} $ &   1.9  &  10.5  &  10.7 & $  2.34\times10^{-2} $ &  ~$+19.2/-\!\!16.4$~  & ~$+10.8/-\!\!5.4$~ \\
   70 --  80  &   74.6  & $  8.96\times10^{-3} $ &   2.7  &  11.0  &  11.3 & $  7.39\times10^{-3} $ &  ~$+21.4/-\!\!17.6$~  & ~$+12.7/-\!\!8.8$~ \\
   80 --  90  &   84.6  & $  3.17\times10^{-3} $ &   4.3  &  12.6  &  13.3 & $  2.39\times10^{-3} $ &  ~$+24.4/-\!\!18.6$~  & ~$+18.2/-\!\!7.1$~ \\
   90 -- 110  &   98.5  & $  6.47\times10^{-4} $ &   6.6  &  15.7  &  17.1 & $  5.20\times10^{-4} $ &  ~$+28.5/-\!\!20.7$~  & ~$+24.7/-\!\!8.2$~ \\
  110 -- 200  &  134.9  & $  1.93\times10^{-5} $ &  17.1  &  14.7  &  22.5 & $  1.38\times10^{-5} $ &  ~$+40.6/-\!\!26.6$~  & ~$+38.6/-\!\!11.0$~ \\
\hline\hline
\end{tabular}
\end{table*}

\begin{table*}
\centering
\caption{Same as in Table~\ref{tab:cross1} but for $|y^\gamma|<1.0$ and $|y^{\rm jet}|\leq0.8$, $y^{\gamma}y^{\rm jet} \le 0$.}
\label{tab:cross5}
\begin{tabular}{ccccccccc} \hline\hline \\ [-2.5ex]
 ~$\Ptg$ bin~ & ~$\langle \Ptg \rangle$~ & \multicolumn{7}{c}{\tcs (pb/GeV)} \\[0.1ex] \cline{3-9} \\ [-2.5ex]
  ~(GeV)~     & ~(GeV)~    & ~Data~  & ~$\delta_{\rm stat} (\%)$~ & ~$\delta_{\rm syst}(\%)$~ & ~$\delta_{\rm tot}(\%)$ ~ & ~NLO~ & ~$\delta_{\rm scale}(\%)$~ & ~$\delta_{\rm pdf}(\%)$~ \\\hline
   20 --  23  &   21.4  & $  4.66\times10^{1} $ &   2.4  &  15.2  &  15.4 & $  6.11\times10^{1} $ &  ~$+16.2/-\!\!3.6$~  & ~$+29.1/-\!\!0.0$~ \\
   23 --  26  &   24.4  & $  3.04\times10^{1} $ &   2.8  &  14.4  &  14.7 & $  3.69\times10^{1} $ &  ~$+17.6/-\!\!4.7$~  & ~$+29.8/-\!\!0.0$~ \\
   26 --  30  &   27.9  & $  1.89\times10^{1} $ &   3.0  &  14.5  &  14.8 & $  2.14\times10^{1} $ &  ~$+18.0/-\!\!5.5$~  & ~$+28.2/-\!\!0.0$~ \\
   30 --  35  &   32.3  & $  1.02\times10^{1} $ &   3.6  &  12.3  &  12.8 & $  1.13\times10^{1} $ &  ~$+18.9/-\!\!6.0$~  & ~$+27.8/-\!\!0.0$~ \\
   35 --  40  &   37.3  & $  5.67\times10^{0} $ &   1.3  &   9.9  &  10.0 & $  6.03\times10^{0} $ &  ~$+18.6/-\!\!8.4$~  & ~$+24.0/-\!\!0.0$~ \\
   40 --  45  &   42.4  & $  3.31\times10^{0} $ &   1.3  &   9.3  &   9.4 & $  3.46\times10^{0} $ &  ~$+17.6/-\!\!8.9$~  & ~$+22.0/-\!\!0.0$~ \\
   45 --  50  &   47.4  & $  2.04\times10^{0} $ &   1.3  &   9.1  &   9.2 & $  2.08\times10^{0} $ &  ~$+17.8/-\!\!9.0$~  & ~$+22.6/-\!\!0.0$~ \\
   50 --  60  &   54.6  & $  1.06\times10^{0} $ &   1.3  &   8.2  &   8.3 & $  1.11\times10^{0} $ &  ~$+16.9/-\!\!9.5$~  & ~$+16.7/-\!\!0.0$~ \\
   60 --  70  &   64.7  & $  5.03\times10^{-1} $ &   1.3  &   8.3  &   8.4 & $  5.08\times10^{-1} $ &  ~$+16.9/-\!\!9.8$~  & ~$+17.4/-\!\!0.0$~ \\
   70 --  80  &   74.7  & $  2.55\times10^{-1} $ &   1.4  &   8.3  &   8.4 & $  2.60\times10^{-1} $ &  ~$+16.0/-\!\!9.5$~  & ~$+14.7/-\!\!0.0$~ \\
   80 --  90  &   84.7  & $  1.43\times10^{-1} $ &   1.4  &   8.3  &   8.4 & $  1.45\times10^{-1} $ &  ~$+15.6/-\!\!9.5$~  & ~$+13.6/-\!\!0.2$~ \\
   90 -- 110  &   99.1  & $  6.84\times10^{-2} $ &   1.4  &   8.3  &   8.4 & $  6.89\times10^{-2} $ &  ~$+14.2/-\!\!9.4$~  & ~$+12.8/-\!\!0.4$~ \\
  110 -- 130  &  119.2  & $  2.79\times10^{-2} $ &   1.6  &   8.4  &   8.6 & $  2.80\times10^{-2} $ &  ~$+13.8/-\!\!9.3$~  & ~$+14.2/-\!\!0.0$~ \\
  130 -- 150  &  139.3  & $  1.28\times10^{-2} $ &   2.0  &   8.5  &   8.7 & $  1.31\times10^{-2} $ &  ~$+12.3/-\!\!9.7$~  & ~$+9.0/-\!\!1.8$~ \\
  150 -- 170  &  159.4  & $  6.40\times10^{-3} $ &   2.4  &   8.6  &   8.9 & $  6.55\times10^{-3} $ &  ~$+11.9/-\!\!9.7$~  & ~$+9.2/-\!\!1.8$~ \\
  170 -- 200  &  183.8  & $  2.95\times10^{-3} $ &   2.8  &   8.7  &   9.1 & $  3.08\times10^{-3} $ &  ~$+10.7/-\!\!10.0$~  & ~$+7.2/-\!\!3.1$~ \\
  200 -- 230  &  213.9  & $  1.34\times10^{-3} $ &   3.9  &   8.8  &   9.6 & $  1.31\times10^{-3} $ &  ~$+9.9/-\!\!9.5$~  & ~$+6.1/-\!\!3.7$~ \\
  230 -- 300  &  259.8  & $  4.18\times10^{-4} $ &   4.6  &   9.0  &  10.1 & $  3.95\times10^{-4} $ &  ~$+8.4/-\!\!9.3$~  & ~$+6.9/-\!\!3.8$~ \\
  300 -- 400  &  341.1  & $  5.04\times10^{-5} $ &  10.5  &   9.6  &  14.2 & $  5.38\times10^{-5} $ &  ~$+8.9/-\!\!11.0$~  & ~$+7.5/-\!\!6.2$~ \\
\hline\hline
\end{tabular}
\end{table*}

\begin{table*}
\centering
\caption{Same as in Table~\ref{tab:cross1} but for $|y^\gamma|<1.0$ and $0.8\lt|y^{\rm jet}|\leq1.6$, $y^{\gamma}y^{\rm jet} \le 0$.}
\label{tab:cross6}
\begin{tabular}{ccccccccc} \hline\hline \\ [-2.5ex]
 ~$\Ptg$ bin~ & ~$\langle \Ptg \rangle$~ & \multicolumn{7}{c}{\tcs (pb/GeV)} \\[0.1ex] \cline{3-9} \\ [-2.5ex]
  ~(GeV)~     & ~(GeV)~    & ~Data~  & ~$\delta_{\rm stat} (\%)$~ & ~$\delta_{\rm syst}(\%)$~ & ~$\delta_{\rm tot}(\%)$ ~ & ~NLO~ & ~$\delta_{\rm scale}(\%)$~ & ~$\delta_{\rm pdf}(\%)$~ \\\hline
   20 --  23  &   21.4  & $  2.34\times10^{1} $ &   2.5  &  15.2  &  15.4 & $  4.17\times10^{1} $ &  ~$+14.0/-\!\!12.4$~  & ~$+5.0/-\!\!5.0$~ \\
   23 --  26  &   24.4  & $  1.57\times10^{1} $ &   3.0  &  15.2  &  15.5 & $  2.47\times10^{1} $ &  ~$+15.0/-\!\!13.7$~  & ~$+3.6/-\!\!4.6$~ \\
   26 --  30  &   27.9  & $  9.71\times10^{0} $ &   3.3  &  15.4  &  15.7 & $  1.39\times10^{1} $ &  ~$+17.0/-\!\!13.6$~  & ~$+4.4/-\!\!5.1$~ \\
   30 --  35  &   32.3  & $  5.81\times10^{0} $ &   4.0  &  12.7  &  13.3 & $  7.12\times10^{0} $ &  ~$+20.0/-\!\!13.2$~  & ~$+3.6/-\!\!3.5$~ \\
   35 --  40  &   37.3  & $  3.08\times10^{0} $ &   1.3  &  10.2  &  10.3 & $  3.67\times10^{0} $ &  ~$+20.4/-\!\!13.8$~  & ~$+3.7/-\!\!3.4$~ \\
   40 --  45  &   42.4  & $  1.81\times10^{0} $ &   1.3  &   9.6  &   9.7 & $  2.05\times10^{0} $ &  ~$+17.7/-\!\!14.0$~  & ~$+4.0/-\!\!3.5$~ \\
   45 --  50  &   47.4  & $  1.10\times10^{0} $ &   1.3  &   9.2  &   9.3 & $  1.22\times10^{0} $ &  ~$+17.4/-\!\!13.8$~  & ~$+4.1/-\!\!3.1$~ \\
   50 --  60  &   54.6  & $  5.73\times10^{-1} $ &   1.3  &   8.4  &   8.5 & $  6.29\times10^{-1} $ &  ~$+17.3/-\!\!13.8$~  & ~$+4.4/-\!\!3.4$~ \\
   60 --  70  &   64.6  & $  2.62\times10^{-1} $ &   1.3  &   8.4  &   8.5 & $  2.81\times10^{-1} $ &  ~$+16.5/-\!\!13.5$~  & ~$+5.2/-\!\!2.7$~ \\
   70 --  80  &   74.7  & $  1.35\times10^{-1} $ &   1.4  &   8.4  &   8.5 & $  1.41\times10^{-1} $ &  ~$+16.0/-\!\!13.1$~  & ~$+5.0/-\!\!3.7$~ \\
   80 --  90  &   84.7  & $  7.33\times10^{-2} $ &   1.5  &   8.4  &   8.5 & $  7.62\times10^{-2} $ &  ~$+15.5/-\!\!12.0$~  & ~$+6.8/-\!\!2.8$~ \\
   90 -- 110  &   99.0  & $  3.46\times10^{-2} $ &   1.5  &   8.5  &   8.6 & $  3.54\times10^{-2} $ &  ~$+14.9/-\!\!12.4$~  & ~$+5.7/-\!\!4.0$~ \\
  110 -- 130  &  119.2  & $  1.32\times10^{-2} $ &   1.9  &   8.6  &   8.8 & $  1.40\times10^{-2} $ &  ~$+13.0/-\!\!12.2$~  & ~$+3.3/-\!\!6.1$~ \\
  130 -- 150  &  139.3  & $  5.76\times10^{-3} $ &   2.5  &   8.7  &   9.0 & $  6.09\times10^{-3} $ &  ~$+12.4/-\!\!12.8$~  & ~$+4.9/-\!\!5.1$~ \\
  150 -- 170  &  159.4  & $  2.86\times10^{-3} $ &   3.3  &   8.9  &   9.5 & $  2.85\times10^{-3} $ &  ~$+11.7/-\!\!11.7$~  & ~$+5.3/-\!\!4.6$~ \\
  170 -- 200  &  183.7  & $  1.20\times10^{-3} $ &   4.1  &   9.0  &   9.9 & $  1.20\times10^{-3} $ &  ~$+12.2/-\!\!11.8$~  & ~$+8.0/-\!\!5.7$~ \\
  200 -- 230  &  213.9  & $  4.69\times10^{-4} $ &   6.4  &   9.5  &  11.5 & $  4.41\times10^{-4} $ &  ~$+11.4/-\!\!11.3$~  & ~$+8.3/-\!\!3.5$~ \\
  230 -- 400  &  289.5  & $  5.02\times10^{-5} $ &   8.6  &   9.6  &  12.9 & $  4.80\times10^{-5} $ &  ~$+9.6/-\!\!12.7$~  & ~$+7.2/-\!\!9.8$~ \\
 \hline\hline
\end{tabular}
\end{table*}

\begin{table*}
\centering
\caption{Same as in Table~\ref{tab:cross1} but for $|y^\gamma|<1.0$ and $1.6\lt|y^{\rm jet}|\leq2.4$, $y^{\gamma}y^{\rm jet} \le 0$.}
\label{tab:cross7}
\begin{tabular}{ccccccccc} \hline\hline \\ [-2.5ex]
 ~$\Ptg$ bin~ & ~$\langle \Ptg \rangle$~ & \multicolumn{7}{c}{\tcs (pb/GeV)} \\[0.1ex] \cline{3-9} \\ [-2.5ex]
  ~(GeV)~     & ~(GeV)~    & ~Data~  & ~$\delta_{\rm stat} (\%)$~ & ~$\delta_{\rm syst}(\%)$~ & ~$\delta_{\rm tot}(\%)$ ~ & ~NLO~ & ~$\delta_{\rm scale}(\%)$~ & ~$\delta_{\rm pdf}(\%)$~ \\\hline
   20 --  23  &   21.4  & $  1.44\times10^{1} $ &   3.1  &  16.6  &  16.9 & $  2.07\times10^{1} $ &  ~$+20.1/-\!\!15.0$~  & ~$+3.1/-\!\!4.4$~ \\
   23 --  26  &   24.4  & $  9.11\times10^{0} $ &   3.9  &  15.8  &  16.2 & $  1.19\times10^{1} $ &  ~$+22.1/-\!\!14.6$~  & ~$+3.7/-\!\!3.2$~ \\
   26 --  30  &   27.9  & $  5.85\times10^{0} $ &   4.3  &  16.4  &  16.9 & $  6.61\times10^{0} $ &  ~$+23.3/-\!\!16.8$~  & ~$+5.4/-\!\!3.1$~ \\
   30 --  35  &   32.3  & $  3.19\times10^{0} $ &   5.4  &  13.2  &  14.3 & $  3.30\times10^{0} $ &  ~$+22.7/-\!\!17.3$~  & ~$+1.4/-\!\!7.1$~ \\
   35 --  40  &   37.3  & $  1.65\times10^{0} $ &   1.4  &  10.9  &  11.0 & $  1.62\times10^{0} $ &  ~$+22.3/-\!\!16.5$~  & ~$+8.7/-\!\!3.4$~ \\
   40 --  45  &   42.4  & $  8.87\times10^{-1} $ &   1.3  &   9.9  &  10.0 & $  8.82\times10^{-1} $ &  ~$+22.4/-\!\!18.2$~  & ~$+3.3/-\!\!9.2$~ \\
   45 --  50  &   47.4  & $  5.15\times10^{-1} $ &   1.3  &   9.5  &   9.6 & $  4.99\times10^{-1} $ &  ~$+22.4/-\!\!16.9$~  & ~$+3.5/-\!\!4.8$~ \\
   50 --  60  &   54.5  & $  2.60\times10^{-1} $ &   1.3  &   9.0  &   9.1 & $  2.44\times10^{-1} $ &  ~$+22.2/-\!\!16.6$~  & ~$+4.3/-\!\!4.6$~ \\
   60 --  70  &   64.6  & $  1.07\times10^{-1} $ &   1.4  &   9.1  &   9.2 & $  1.01\times10^{-1} $ &  ~$+21.5/-\!\!16.4$~  & ~$+4.9/-\!\!5.1$~ \\
   70 --  80  &   74.7  & $  4.98\times10^{-2} $ &   1.6  &   9.3  &   9.4 & $  4.68\times10^{-2} $ &  ~$+20.4/-\!\!16.4$~  & ~$+4.3/-\!\!6.8$~ \\
   80 --  90  &   84.7  & $  2.46\times10^{-2} $ &   1.9  &   9.4  &   9.6 & $  2.29\times10^{-2} $ &  ~$+18.6/-\!\!15.7$~  & ~$+5.9/-\!\!5.4$~ \\
   90 -- 110  &   98.9  & $  1.01\times10^{-2} $ &   2.1  &   9.5  &   9.7 & $  9.15\times10^{-3} $ &  ~$+18.8/-\!\!15.4$~  & ~$+5.7/-\!\!6.0$~ \\
  110 -- 130  &  119.0  & $  2.95\times10^{-3} $ &   3.4  &   9.7  &  10.3 & $  2.77\times10^{-3} $ &  ~$+19.0/-\!\!15.2$~  & ~$+8.9/-\!\!3.8$~ \\
  130 -- 150  &  139.1  & $  9.77\times10^{-4} $ &   5.6  &   9.7  &  11.2 & $  9.10\times10^{-4} $ &  ~$+19.0/-\!\!15.5$~  & ~$+9.3/-\!\!6.8$~ \\
  150 -- 170  &  159.2  & $  3.97\times10^{-4} $ &   8.7  &  10.5  &  13.6 & $  3.09\times10^{-4} $ &  ~$+19.0/-\!\!15.4$~  & ~$+11.8/-\!\!5.7$~ \\
  170 -- 300  &  209.4  & $  3.14\times10^{-5} $ &  12.4  &  13.0  &  18.0 & $  2.50\times10^{-5} $ &  ~$+20.8/-\!\!17.0$~  & ~$+15.4/-\!\!7.3$~ \\
\hline\hline
\end{tabular}
\end{table*}

\begin{table*}
\centering
\caption{Same as in Table~\ref{tab:cross1} but for $|y^\gamma|<1.0$ and $2.4\lt|y^{\rm jet}|\leq3.2$, $y^{\gamma}y^{\rm jet} \le 0$.}
\label{tab:cross8}
\begin{tabular}{ccccccccc} \hline\hline \\ [-2.5ex]
 ~$\Ptg$ bin~ & ~$\langle \Ptg \rangle$~ & \multicolumn{7}{c}{\tcs (pb/GeV)} \\[0.1ex] \cline{3-9} \\ [-2.5ex]
  ~(GeV)~     & ~(GeV)~    & ~Data~  & ~$\delta_{\rm stat} (\%)$~ & ~$\delta_{\rm syst}(\%)$~ & ~$\delta_{\rm tot}(\%)$ ~ & ~NLO~ & ~$\delta_{\rm scale}(\%)$~ & ~$\delta_{\rm pdf}(\%)$~ \\\hline
   20 --  23  &   21.4  & $  4.38\times10^{0} $ &   4.4  &  18.8  &  19.3 & $  8.14\times10^{0} $ &  ~$+27.6/-\!\!18.4$~  & ~$+4.3/-\!\!3.8$~ \\
   23 --  26  &   24.4  & $  3.25\times10^{0} $ &   5.5  &  17.1  &  18.0 & $  4.52\times10^{0} $ &  ~$+28.4/-\!\!19.0$~  & ~$+4.7/-\!\!4.0$~ \\
   26 --  30  &   27.8  & $  1.93\times10^{0} $ &   6.6  &  16.9  &  18.2 & $  2.38\times10^{0} $ &  ~$+29.5/-\!\!19.7$~  & ~$+4.9/-\!\!4.2$~ \\
   30 --  35  &   32.3  & $  1.04\times10^{0} $ &   8.8  &  13.6  &  16.2 & $  1.09\times10^{0} $ &  ~$+30.1/-\!\!20.0$~  & ~$+5.5/-\!\!4.8$~ \\
   35 --  40  &   37.3  & $  5.38\times10^{-1} $ &   1.4  &  11.6  &  11.7 & $  4.85\times10^{-1} $ &  ~$+29.6/-\!\!21.1$~  & ~$+6.1/-\!\!5.7$~ \\
   40 --  45  &   42.3  & $  2.75\times10^{-1} $ &   1.4  &  12.1  &  12.2 & $  2.33\times10^{-1} $ &  ~$+29.2/-\!\!20.6$~  & ~$+6.1/-\!\!6.0$~ \\
   45 --  50  &   47.3  & $  1.47\times10^{-1} $ &   1.5  &  10.3  &  10.4 & $  1.18\times10^{-1} $ &  ~$+29.5/-\!\!20.8$~  & ~$+8.2/-\!\!5.3$~ \\
   50 --  60  &   54.5  & $  5.89\times10^{-2} $ &   1.5  &   9.2  &   9.3 & $  4.84\times10^{-2} $ &  ~$+29.4/-\!\!21.2$~  & ~$+9.2/-\!\!6.0$~ \\
   60 --  70  &   64.5  & $  1.97\times10^{-2} $ &   2.0  &   9.8  &  10.0 & $  1.50\times10^{-2} $ &  ~$+30.9/-\!\!21.6$~  & ~$+12.5/-\!\!6.7$~ \\
   70 --  80  &   74.6  & $  7.00\times10^{-3} $ &   3.0  &  10.7  &  11.1 & $  5.09\times10^{-3} $ &  ~$+31.2/-\!\!21.8$~  & ~$+14.0/-\!\!7.4$~ \\
   80 --  90  &   84.6  & $  2.61\times10^{-3} $ &   4.5  &  11.5  &  12.3 & $  1.81\times10^{-3} $ &  ~$+32.4/-\!\!22.4$~  & ~$+16.5/-\!\!7.8$~ \\
   90 -- 110  &   98.6  & $  7.66\times10^{-4} $ &   6.0  &  11.5  &  12.9 & $  4.55\times10^{-4} $ &  ~$+34.9/-\!\!23.9$~  & ~$+20.2/-\!\!9.0$~ \\
  110 -- 200  &  136.8  & $  3.87\times10^{-5} $ &  11.5  &  13.2  &  17.5 & $  1.68\times10^{-5} $ &  ~$+43.0/-\!\!27.6$~  & ~$+31.0/-\!\!9.6$~ \\
\hline\hline
\end{tabular}
\end{table*}

\begin{table*}
\centering
\caption{The \gpj cross section \tcs in bins of $\Ptg$ for $1.5<|y^\gamma|<2.5$ and $|y^{\rm jet}|\leq0.8$, $y^{\gamma}y^{\rm jet} \gt 0$ together with statistical ($\delta_{\text{stat}}$) and systematic ($\delta_{\text{syst}}$) uncertainties, and the NLO prediction together with scale ($\delta_{\rm scale}$) and PDF ($\delta_{\rm pdf}$) uncertainties. 
A common normalization uncertainty of $11.2\%$ is included in $\delta_{\text{syst}}$ for all points.}
\label{tab:cross9}
\begin{tabular}{ccccccccc} \hline\hline \\ [-2.5ex]
 ~$\Ptg$ bin~ & ~$\langle \Ptg \rangle$~ & \multicolumn{7}{c}{\tcs (pb/GeV)} \\[0.1ex] \cline{3-9} \\ [-2.5ex]
  ~(GeV)~     & ~(GeV)~    & ~Data~  & ~$\delta_{\rm stat} (\%)$~ & ~$\delta_{\rm syst}(\%)$~ & ~$\delta_{\rm tot}(\%)$ ~ & ~NLO~ & ~$\delta_{\rm scale}(\%)$~ & ~$\delta_{\rm pdf}(\%)$~ \\\hline
   20 --  23  &   21.4  & $  5.67\times10^{1} $ &   2.2  &  18.2  &  18.4 & $  5.69\times10^{1} $ &  ~$+14.5/-\!\!11.0$~  & ~$+4.3/-\!\!3.9$~ \\
   23 --  26  &   24.4  & $  3.46\times10^{1} $ &   2.5  &  17.0  &  17.1 & $  3.41\times10^{1} $ &  ~$+15.4/-\!\!11.2$~  & ~$+3.8/-\!\!3.4$~ \\
   26 --  30  &   27.9  & $  2.00\times10^{1} $ &   2.6  &  16.9  &  17.1 & $  1.96\times10^{1} $ &  ~$+16.5/-\!\!11.8$~  & ~$+2.6/-\!\!3.4$~ \\
   30 --  35  &   32.3  & $  1.02\times10^{1} $ &   3.0  &  16.6  &  16.9 & $  1.01\times10^{1} $ &  ~$+17.0/-\!\!13.0$~  & ~$+3.5/-\!\!2.7$~ \\
   35 --  40  &   37.3  & $  4.64\times10^{0} $ &   1.3  &  15.0  &  15.0 & $  5.23\times10^{0} $ &  ~$+17.2/-\!\!13.2$~  & ~$+2.5/-\!\!3.1$~ \\
   40 --  45  &   42.4  & $  2.56\times10^{0} $ &   1.2  &  12.3  &  12.3 & $  2.87\times10^{0} $ &  ~$+17.0/-\!\!13.4$~  & ~$+2.3/-\!\!3.4$~ \\
   45 --  50  &   47.4  & $  1.47\times10^{0} $ &   1.3  &  11.8  &  11.9 & $  1.65\times10^{0} $ &  ~$+17.4/-\!\!13.9$~  & ~$+3.6/-\!\!2.5$~ \\
   50 --  60  &   54.6  & $  7.00\times10^{-1} $ &   1.2  &  11.7  &  11.8 & $  8.05\times10^{-1} $ &  ~$+16.8/-\!\!14.0$~  & ~$+1.8/-\!\!4.4$~ \\
   60 --  70  &   64.6  & $  2.86\times10^{-1} $ &   1.3  &  11.8  &  11.8 & $  3.24\times10^{-1} $ &  ~$+16.8/-\!\!14.1$~  & ~$+2.0/-\!\!3.8$~ \\
   70 --  80  &   74.6  & $  1.26\times10^{-1} $ &   1.4  &  10.9  &  11.0 & $  1.42\times10^{-1} $ &  ~$+16.5/-\!\!13.3$~  & ~$+3.0/-\!\!2.4$~ \\
   80 --  90  &   84.7  & $  5.94\times10^{-2} $ &   1.4  &  10.6  &  10.7 & $  6.55\times10^{-2} $ &  ~$+16.6/-\!\!13.2$~  & ~$+5.0/-\!\!1.3$~ \\
   90 -- 110  &   98.8  & $  2.14\times10^{-2} $ &   1.5  &  11.2  &  11.3 & $  2.41\times10^{-2} $ &  ~$+14.9/-\!\!13.1$~  & ~$+4.0/-\!\!2.3$~ \\
  110 -- 130  &  118.8  & $  5.64\times10^{-3} $ &   2.2  &  11.0  &  11.2 & $  6.39\times10^{-3} $ &  ~$+14.0/-\!\!12.8$~  & ~$+4.3/-\!\!3.7$~ \\
  130 -- 150  &  138.9  & $  1.57\times10^{-3} $ &   4.1  &  11.3  &  12.1 & $  1.82\times10^{-3} $ &  ~$+13.3/-\!\!13.1$~  & ~$+4.6/-\!\!5.9$~ \\
  150 -- 170  &  158.9  & $  5.00\times10^{-4} $ &   6.4  &  11.8  &  13.4 & $  5.41\times10^{-4} $ &  ~$+13.3/-\!\!12.7$~  & ~$+7.1/-\!\!5.6$~ \\
  170 -- 230  &  191.6  & $  8.15\times10^{-5} $ &  10.3  &  12.4  &  16.1 & $  7.51\times10^{-5} $ &  ~$+12.6/-\!\!13.2$~  & ~$+9.2/-\!\!10.8$~ \\
\hline\hline
\end{tabular}
\end{table*}

\begin{table*}
\centering
\caption{Same as in Table~\ref{tab:cross9} but of $\Ptg$ for $1.5<|y^\gamma|<2.5$ and $0.8\lt|y^{\rm jet}|\leq1.6$, $y^{\gamma}y^{\rm jet} \gt 0$.}
\label{tab:cross10}
\begin{tabular}{ccccccccc} \hline\hline \\ [-2.5ex]
 ~$\Ptg$ bin~ & ~$\langle \Ptg \rangle$~ & \multicolumn{7}{c}{\tcs (pb/GeV)} \\[0.1ex] \cline{3-9} \\ [-2.5ex]
  ~(GeV)~     & ~(GeV)~    & ~Data~  & ~$\delta_{\rm stat} (\%)$~ & ~$\delta_{\rm syst}(\%)$~ & ~$\delta_{\rm tot}(\%)$ ~ & ~NLO~ & ~$\delta_{\rm scale}(\%)$~ & ~$\delta_{\rm pdf}(\%)$~ \\\hline
   20 --  23  &   21.4  & $  6.20\times10^{1} $ &   2.1  &  18.2  &  18.4 & $  7.12\times10^{1} $ &  ~$+11.8/-\!\!8.4$~  & ~$+3.5/-\!\!4.6$~ \\
   23 --  26  &   24.4  & $  3.89\times10^{1} $ &   2.3  &  16.9  &  17.1 & $  4.29\times10^{1} $ &  ~$+12.2/-\!\!8.7$~  & ~$+3.6/-\!\!3.9$~ \\
   26 --  30  &   27.9  & $  2.13\times10^{1} $ &   2.5  &  16.8  &  17.0 & $  2.46\times10^{1} $ &  ~$+13.0/-\!\!9.5$~  & ~$+2.6/-\!\!4.8$~ \\
   30 --  35  &   32.3  & $  1.16\times10^{1} $ &   2.9  &  16.4  &  16.6 & $  1.28\times10^{1} $ &  ~$+12.8/-\!\!10.7$~  & ~$+2.2/-\!\!4.5$~ \\
   35 --  40  &   37.3  & $  5.61\times10^{0} $ &   1.3  &  14.8  &  14.9 & $  6.53\times10^{0} $ &  ~$+13.5/-\!\!11.0$~  & ~$+2.5/-\!\!3.9$~ \\
   40 --  45  &   42.3  & $  3.11\times10^{0} $ &   1.2  &  12.3  &  12.4 & $  3.56\times10^{0} $ &  ~$+13.8/-\!\!11.1$~  & ~$+2.7/-\!\!4.0$~ \\
   45 --  50  &   47.4  & $  1.78\times10^{0} $ &   1.3  &  11.9  &  11.9 & $  2.02\times10^{0} $ &  ~$+13.9/-\!\!11.7$~  & ~$+3.9/-\!\!2.0$~ \\
   50 --  60  &   54.5  & $  8.27\times10^{-1} $ &   1.2  &  11.8  &  11.8 & $  9.65\times10^{-1} $ &  ~$+12.8/-\!\!11.6$~  & ~$+2.1/-\!\!3.4$~ \\
   60 --  70  &   64.6  & $  3.22\times10^{-1} $ &   1.3  &  11.9  &  11.9 & $  3.70\times10^{-1} $ &  ~$+13.6/-\!\!11.0$~  & ~$+3.4/-\!\!3.3$~ \\
   70 --  80  &   74.6  & $  1.34\times10^{-1} $ &   1.4  &  11.1  &  11.2 & $  1.51\times10^{-1} $ &  ~$+12.0/-\!\!11.7$~  & ~$+3.4/-\!\!2.1$~ \\
   80 --  90  &   84.6  & $  5.91\times10^{-2} $ &   1.5  &  10.8  &  10.9 & $  6.55\times10^{-2} $ &  ~$+11.7/-\!\!11.6$~  & ~$+3.5/-\!\!2.8$~ \\
   90 -- 110  &   98.6  & $  1.92\times10^{-2} $ &   1.5  &  11.5  &  11.6 & $  2.11\times10^{-2} $ &  ~$+12.1/-\!\!11.6$~  & ~$+4.2/-\!\!3.5$~ \\
  110 -- 130  &  118.8  & $  4.09\times10^{-3} $ &   2.6  &  11.3  &  11.5 & $  4.49\times10^{-3} $ &  ~$+11.9/-\!\!12.0$~  & ~$+5.9/-\!\!4.9$~ \\
  130 -- 150  &  138.8  & $  8.66\times10^{-4} $ &   5.4  &  12.0  &  13.2 & $  9.80\times10^{-4} $ &  ~$+12.9/-\!\!12.4$~  & ~$+10.8/-\!\!6.2$~ \\
  150 -- 230  &  175.1  & $  6.97\times10^{-5} $ &   9.2  &  12.3  &  15.4 & $  6.59\times10^{-5} $ &  ~$+12.5/-\!\!13.4$~  & ~$+14.0/-\!\!10.2$~ \\
\hline\hline
\end{tabular}
\end{table*}

\begin{table*}
\centering
\caption{Same as in Table~\ref{tab:cross9} but for $1.5<|y^\gamma|<2.5$ and $1.6\lt|y^{\rm jet}|\leq2.4$, $y^{\gamma}y^{\rm jet} \gt 0$.}
\label{tab:cross11}
\begin{tabular}{ccccccccc} \hline\hline \\ [-2.5ex]
 ~$\Ptg$ bin~ & ~$\langle \Ptg \rangle$~ & \multicolumn{7}{c}{\tcs (pb/GeV)} \\[0.1ex] \cline{3-9} \\ [-2.5ex]
  ~(GeV)~     & ~(GeV)~    & ~Data~  & ~$\delta_{\rm stat} (\%)$~ & ~$\delta_{\rm syst}(\%)$~ & ~$\delta_{\rm tot}(\%)$ ~ & ~NLO~ & ~$\delta_{\rm scale}(\%)$~ & ~$\delta_{\rm pdf}(\%)$~ \\\hline
   20 --  23  &   21.4  & $  6.17\times10^{1} $ &   2.3  &  18.2  &  18.3 & $  6.40\times10^{1} $ &  ~$+9.2/-\!\!8.2$~  & ~$+2.5/-\!\!5.4$~ \\
   23 --  26  &   24.4  & $  3.92\times10^{1} $ &   2.6  &  16.7  &  16.9 & $  3.76\times10^{1} $ &  ~$+9.7/-\!\!7.3$~  & ~$+2.2/-\!\!5.6$~ \\
   26 --  30  &   27.9  & $  1.98\times10^{1} $ &   2.9  &  16.7  &  17.0 & $  2.12\times10^{1} $ &  ~$+10.6/-\!\!7.8$~  & ~$+3.4/-\!\!3.8$~ \\
   30 --  35  &   32.3  & $  1.12\times10^{1} $ &   3.3  &  16.3  &  16.6 & $  1.06\times10^{1} $ &  ~$+10.6/-\!\!9.0$~  & ~$+2.6/-\!\!3.9$~ \\
   35 --  40  &   37.3  & $  4.96\times10^{0} $ &   1.3  &  14.8  &  14.9 & $  5.14\times10^{0} $ &  ~$+11.8/-\!\!9.1$~  & ~$+3.5/-\!\!2.9$~ \\
   40 --  45  &   42.3  & $  2.60\times10^{0} $ &   1.2  &  12.3  &  12.4 & $  2.67\times10^{0} $ &  ~$+11.0/-\!\!10.1$~  & ~$+3.2/-\!\!4.0$~ \\
   45 --  50  &   47.4  & $  1.42\times10^{0} $ &   1.3  &  11.9  &  12.0 & $  1.41\times10^{0} $ &  ~$+11.4/-\!\!10.5$~  & ~$+2.9/-\!\!3.9$~ \\
   50 --  60  &   54.5  & $  6.00\times10^{-1} $ &   1.2  &  12.0  &  12.1 & $  6.01\times10^{-1} $ &  ~$+10.8/-\!\!10.4$~  & ~$+4.0/-\!\!3.3$~ \\
   60 --  70  &   64.5  & $  2.00\times10^{-1} $ &   1.3  &  12.2  &  12.3 & $  1.93\times10^{-1} $ &  ~$+11.5/-\!\!11.1$~  & ~$+2.8/-\!\!5.2$~ \\
   70 --  80  &   74.5  & $  7.03\times10^{-2} $ &   1.5  &  11.3  &  11.4 & $  6.37\times10^{-2} $ &  ~$+11.7/-\!\!11.0$~  & ~$+4.9/-\!\!4.6$~ \\
   80 --  90  &   84.6  & $  2.49\times10^{-2} $ &   1.9  &  11.1  &  11.3 & $  2.19\times10^{-2} $ &  ~$+11.4/-\!\!11.4$~  & ~$+6.0/-\!\!5.8$~ \\
   90 -- 110  &   98.4  & $  5.72\times10^{-3} $ &   2.4  &  11.6  &  11.9 & $  5.06\times10^{-3} $ &  ~$+14.8/-\!\!12.4$~  & ~$+10.4/-\!\!6.0$~ \\
  110 -- 130  &  118.4  & $  7.10\times10^{-4} $ &   6.2  &  12.2  &  13.7 & $  5.99\times10^{-4} $ &  ~$+15.2/-\!\!14.0$~  & ~$+16.8/-\!\!9.6$~ \\
  130 -- 170  &  144.3  & $  4.08\times10^{-5} $ &  18.2  &  13.2  &  22.5 & $  3.52\times10^{-5} $ &  ~$+19.9/-\!\!16.7$~  & ~$+32.7/-\!\!15.4$~ \\
\hline\hline
\end{tabular}

\vspace*{10mm}

\centering
\caption{Same as in Table~\ref{tab:cross9} but for $1.5<|y^\gamma|<2.5$ and $2.4\lt|y^{\rm jet}|\leq3.2$, $y^{\gamma}y^{\rm jet} \gt 0$.}
\label{tab:cross12}
\begin{tabular}{ccccccccc} \hline\hline \\ [-2.5ex]
 ~$\Ptg$ bin~ & ~$\langle \Ptg \rangle$~ & \multicolumn{7}{c}{\tcs (pb/GeV)} \\[0.1ex] \cline{3-9} \\ [-2.5ex]
  ~(GeV)~     & ~(GeV)~    & ~Data~  & ~$\delta_{\rm stat} (\%)$~ & ~$\delta_{\rm syst}(\%)$~ & ~$\delta_{\rm tot}(\%)$ ~ & ~NLO~ & ~$\delta_{\rm scale}(\%)$~ & ~$\delta_{\rm pdf}(\%)$~ \\\hline
   20 --  23  &   21.4  & $  3.29\times10^{1} $ &   2.9  &  19.8  &  20.1 & $  3.23\times10^{1} $ &  ~$+8.6/-\!\!5.9$~  & ~$+3.7/-\!\!3.8$~ \\
   23 --  26  &   24.4  & $  1.90\times10^{1} $ &   3.4  &  18.0  &  18.4 & $  1.80\times10^{1} $ &  ~$+9.8/-\!\!6.1$~  & ~$+7.1/-\!\!2.0$~ \\
   26 --  30  &   27.9  & $  9.87\times10^{0} $ &   3.9  &  17.7  &  18.1 & $  9.42\times10^{0} $ &  ~$+10.2/-\!\!7.3$~  & ~$+5.1/-\!\!3.4$~ \\
   30 --  35  &   32.3  & $  4.09\times10^{0} $ &   5.3  &  17.1  &  17.9 & $  4.14\times10^{0} $ &  ~$+10.5/-\!\!9.1$~  & ~$+3.7/-\!\!4.5$~ \\
   35 --  40  &   37.3  & $  1.71\times10^{0} $ &   1.3  &  15.6  &  15.7 & $  1.69\times10^{0} $ &  ~$+12.1/-\!\!8.9$~  & ~$+7.5/-\!\!2.7$~ \\
   40 --  45  &   42.3  & $  8.11\times10^{-1} $ &   1.3  &  12.8  &  12.9 & $  7.30\times10^{-1} $ &  ~$+11.6/-\!\!10.3$~  & ~$+4.0/-\!\!5.7$~ \\
   45 --  50  &   47.3  & $  3.64\times10^{-1} $ &   1.3  &  12.7  &  12.8 & $  3.18\times10^{-1} $ &  ~$+12.7/-\!\!11.8$~  & ~$+5.9/-\!\!5.8$~ \\
   50 --  60  &   54.2  & $  1.19\times10^{-1} $ &   1.3  &  12.5  &  12.6 & $  1.01\times10^{-1} $ &  ~$+13.5/-\!\!12.6$~  & ~$+8.3/-\!\!5.9$~ \\
   60 --  70  &   64.2  & $  2.47\times10^{-2} $ &   1.8  &  12.8  &  12.9 & $  1.97\times10^{-2} $ &  ~$+17.2/-\!\!15.0$~  & ~$+12.9/-\!\!8.2$~ \\
   70 --  80  &   74.2  & $  6.23\times10^{-3} $ &   3.3  &  12.8  &  13.2 & $  3.78\times10^{-3} $ &  ~$+21.6/-\!\!16.8$~  & ~$+21.1/-\!\!7.9$~ \\
   80 --  90  &   84.2  & $  1.30\times10^{-3} $ &   6.7  &  13.7  &  15.2 & $  6.98\times10^{-4} $ &  ~$+27.5/-\!\!20.2$~  & ~$+30.9/-\!\!11.9$~ \\
   90 -- 170  &  104.4  & $  5.15\times10^{-5} $ &  12.0  &  17.2  &  21.0 & $  1.84\times10^{-5} $ &  ~$+37.8/-\!\!25.0$~  & ~$+45.0/-\!\!15.7$~ \\
\hline\hline
\end{tabular}

\vspace*{10mm}

\centering
\caption{Same as in Table~\ref{tab:cross9} but for $1.5<|y^\gamma|<2.5$ and $|y^{\rm jet}|\leq0.8$, $y^{\gamma}y^{\rm jet} \le 0$.}
\label{tab:cross13}
\begin{tabular}{ccccccccc} \hline\hline \\ [-2.5ex]
 ~$\Ptg$ bin~ & ~$\langle \Ptg \rangle$~ & \multicolumn{7}{c}{\tcs (pb/GeV)} \\[0.1ex] \cline{3-9} \\ [-2.5ex]
  ~(GeV)~     & ~(GeV)~    & ~Data~  & ~$\delta_{\rm stat} (\%)$~ & ~$\delta_{\rm syst}(\%)$~ & ~$\delta_{\rm tot}(\%)$ ~ & ~NLO~ & ~$\delta_{\rm scale}(\%)$~ & ~$\delta_{\rm pdf}(\%)$~ \\\hline
   20 --  23  &   21.4  & $  3.57\times10^{1} $ &   2.3  &  18.8  &  18.9 & $  3.56\times10^{1} $ &  ~$+26.3/-\!\!12.0$~  & ~$+23.2/-\!\!0.0$~ \\
   23 --  26  &   24.4  & $  2.11\times10^{1} $ &   2.7  &  17.4  &  17.6 & $  2.11\times10^{1} $ &  ~$+27.5/-\!\!12.5$~  & ~$+24.3/-\!\!0.0$~ \\
   26 --  30  &   27.9  & $  1.22\times10^{1} $ &   2.9  &  17.4  &  17.6 & $  1.20\times10^{1} $ &  ~$+28.5/-\!\!12.9$~  & ~$+23.5/-\!\!0.0$~ \\
   30 --  35  &   32.3  & $  6.29\times10^{0} $ &   3.4  &  17.3  &  17.7 & $  6.12\times10^{0} $ &  ~$+28.2/-\!\!14.4$~  & ~$+21.4/-\!\!0.0$~ \\
   35 --  40  &   37.3  & $  2.84\times10^{0} $ &   1.3  &  15.6  &  15.7 & $  3.09\times10^{0} $ &  ~$+28.7/-\!\!14.6$~  & ~$+20.2/-\!\!0.0$~ \\
   40 --  45  &   42.4  & $  1.51\times10^{0} $ &   1.2  &  12.6  &  12.7 & $  1.68\times10^{0} $ &  ~$+28.4/-\!\!15.0$~  & ~$+17.8/-\!\!0.0$~ \\
   45 --  50  &   47.4  & $  8.72\times10^{-1} $ &   1.3  &  12.0  &  12.0 & $  9.68\times10^{-1} $ &  ~$+27.2/-\!\!16.0$~  & ~$+14.7/-\!\!0.0$~ \\
   50 --  60  &   54.5  & $  4.14\times10^{-1} $ &   1.2  &  12.0  &  12.1 & $  4.69\times10^{-1} $ &  ~$+26.6/-\!\!15.9$~  & ~$+12.3/-\!\!0.0$~ \\
   60 --  70  &   64.6  & $  1.72\times10^{-1} $ &   1.3  &  12.1  &  12.2 & $  1.90\times10^{-1} $ &  ~$+24.4/-\!\!16.3$~  & ~$+7.0/-\!\!0.1$~ \\
   70 --  80  &   74.6  & $  7.57\times10^{-2} $ &   1.4  &  11.2  &  11.3 & $  8.45\times10^{-2} $ &  ~$+22.1/-\!\!16.9$~  & ~$+2.0/-\!\!3.6$~ \\
   80 --  90  &   84.7  & $  3.62\times10^{-2} $ &   1.5  &  10.6  &  10.7 & $  4.01\times10^{-2} $ &  ~$+20.4/-\!\!17.3$~  & ~$+0.7/-\!\!7.3$~ \\
   90 -- 110  &   98.8  & $  1.34\times10^{-2} $ &   1.5  &  11.0  &  11.1 & $  1.53\times10^{-2} $ &  ~$+17.1/-\!\!17.8$~  & ~$+0.0/-\!\!15.5$~ \\
  110 -- 130  &  118.9  & $  3.80\times10^{-3} $ &   2.3  &  11.0  &  11.2 & $  4.42\times10^{-3} $ &  ~$+13.0/-\!\!18.3$~  & ~$+0.0/-\!\!24.5$~ \\
  130 -- 150  &  139.0  & $  1.20\times10^{-3} $ &   4.2  &  11.4  &  12.2 & $  1.39\times10^{-3} $ &  ~$+10.3/-\!\!18.2$~  & ~$+0.0/-\!\!28.8$~ \\
  150 -- 170  &  159.0  & $  3.40\times10^{-4} $ &   7.0  &  11.8  &  13.7 & $  4.65\times10^{-4} $ &  ~$+8.2/-\!\!18.4$~  & ~$+0.0/-\!\!33.1$~ \\
  170 -- 230  &  192.4  & $  6.69\times10^{-5} $ &  10.2  &  12.2  &  16.0 & $  8.17\times10^{-5} $ &  ~$+2.4/-\!\!22.3$~  & ~$+0.0/-\!\!59.8$~ \\
\hline\hline
\end{tabular}
\end{table*}

\begin{table*}
\centering
\caption{Same as in Table~\ref{tab:cross9} but for $1.5<|y^\gamma|<2.5$ and $0.8\lt|y^{\rm jet}|\leq1.6$, $y^{\gamma}y^{\rm jet} \le 0$.}
\label{tab:cross14}
\begin{tabular}{ccccccccc} \hline\hline \\ [-2.5ex]
 ~$\Ptg$ bin~ & ~$\langle \Ptg \rangle$~ & \multicolumn{7}{c}{\tcs (pb/GeV)} \\[0.1ex] \cline{3-9} \\ [-2.5ex]
  ~(GeV)~     & ~(GeV)~    & ~Data~  & ~$\delta_{\rm stat} (\%)$~ & ~$\delta_{\rm syst}(\%)$~ & ~$\delta_{\rm tot}(\%)$ ~ & ~NLO~ & ~$\delta_{\rm scale}(\%)$~ & ~$\delta_{\rm pdf}(\%)$~ \\\hline
   20 --  23  &   21.4  & $  1.34\times10^{1} $ &   2.6  &  19.2  &  19.4 & $  2.15\times10^{1} $ &  ~$+28.7/-\!\!19.5$~  & ~$+2.9/-\!\!3.3$~ \\
   23 --  26  &   24.4  & $  9.24\times10^{0} $ &   3.1  &  17.9  &  18.1 & $  1.26\times10^{1} $ &  ~$+29.7/-\!\!19.9$~  & ~$+2.7/-\!\!3.0$~ \\
   26 --  30  &   27.9  & $  4.94\times10^{0} $ &   3.5  &  17.8  &  18.1 & $  6.99\times10^{0} $ &  ~$+30.5/-\!\!20.4$~  & ~$+2.9/-\!\!2.9$~ \\
   30 --  35  &   32.3  & $  2.83\times10^{0} $ &   4.1  &  17.8  &  18.2 & $  3.46\times10^{0} $ &  ~$+30.9/-\!\!21.3$~  & ~$+2.4/-\!\!3.0$~ \\
   35 --  40  &   37.3  & $  1.35\times10^{0} $ &   1.3  &  16.1  &  16.2 & $  1.70\times10^{0} $ &  ~$+31.0/-\!\!21.4$~  & ~$+2.5/-\!\!3.5$~ \\
   40 --  45  &   42.4  & $  7.11\times10^{-1} $ &   1.3  &  13.0  &  13.0 & $  8.98\times10^{-1} $ &  ~$+30.8/-\!\!21.6$~  & ~$+2.4/-\!\!4.2$~ \\
   45 --  50  &   47.4  & $  4.00\times10^{-1} $ &   1.3  &  12.6  &  12.6 & $  4.99\times10^{-1} $ &  ~$+31.9/-\!\!21.7$~  & ~$+3.5/-\!\!2.8$~ \\
   50 --  60  &   54.5  & $  1.87\times10^{-1} $ &   1.2  &  12.4  &  12.5 & $  2.34\times10^{-1} $ &  ~$+30.5/-\!\!21.2$~  & ~$+3.5/-\!\!3.6$~ \\
   60 --  70  &   64.6  & $  7.40\times10^{-2} $ &   1.3  &  12.4  &  12.5 & $  9.09\times10^{-2} $ &  ~$+29.6/-\!\!20.8$~  & ~$+3.5/-\!\!3.9$~ \\
   70 --  80  &   74.6  & $  3.18\times10^{-2} $ &   1.5  &  11.7  &  11.8 & $  3.90\times10^{-2} $ &  ~$+28.2/-\!\!20.3$~  & ~$+3.9/-\!\!4.4$~ \\
   80 --  90  &   84.7  & $  1.51\times10^{-2} $ &   1.7  &  11.2  &  11.3 & $  1.79\times10^{-2} $ &  ~$+27.4/-\!\!19.5$~  & ~$+5.1/-\!\!3.3$~ \\
   90 -- 110  &   98.8  & $  5.45\times10^{-3} $ &   1.8  &  12.5  &  12.6 & $  6.62\times10^{-3} $ &  ~$+25.5/-\!\!18.3$~  & ~$+4.8/-\!\!4.0$~ \\
  110 -- 130  &  118.9  & $  1.60\times10^{-3} $ &   2.9  &  11.3  &  11.7 & $  1.83\times10^{-3} $ &  ~$+22.2/-\!\!17.7$~  & ~$+4.4/-\!\!5.2$~ \\
  130 -- 150  &  138.9  & $  4.39\times10^{-4} $ &   5.6  &  11.7  &  13.0 & $  5.47\times10^{-4} $ &  ~$+21.5/-\!\!16.4$~  & ~$+5.6/-\!\!5.8$~ \\
  150 -- 230  &  176.6  & $  5.65\times10^{-5} $ &   7.6  &  13.1  &  15.1 & $  6.55\times10^{-5} $ &  ~$+18.0/-\!\!16.3$~  & ~$+8.1/-\!\!6.5$~ \\
\hline\hline
\end{tabular}

\vspace*{10mm}

\centering
\caption{Same as in Table~\ref{tab:cross9} but for $1.5<|y^\gamma|<2.5$ and $1.6\lt|y^{\rm jet}|\leq2.4$, $y^{\gamma}y^{\rm jet} \le 0$.}
\label{tab:cross15}
\begin{tabular}{ccccccccc} \hline\hline \\ [-2.5ex]
 ~$\Ptg$ bin~ & ~$\langle \Ptg \rangle$~ & \multicolumn{7}{c}{\tcs (pb/GeV)} \\[0.1ex] \cline{3-9} \\ [-2.5ex]
  ~(GeV)~     & ~(GeV)~    & ~Data~  & ~$\delta_{\rm stat} (\%)$~ & ~$\delta_{\rm syst}(\%)$~ & ~$\delta_{\rm tot}(\%)$ ~ & ~NLO~ & ~$\delta_{\rm scale}(\%)$~ & ~$\delta_{\rm pdf}(\%)$~ \\\hline
   20 --  23  &   21.4  & $  9.79\times10^{0} $ &   3.4  &  20.4  &  20.7 & $  1.07\times10^{1} $ &  ~$+35.2/-\!\!23.1$~  & ~$+3.9/-\!\!3.1$~ \\
   23 --  26  &   24.4  & $  5.79\times10^{0} $ &   4.2  &  19.1  &  19.6 & $  6.19\times10^{0} $ &  ~$+36.0/-\!\!23.8$~  & ~$+3.4/-\!\!3.8$~ \\
   26 --  30  &   27.9  & $  2.84\times10^{0} $ &   4.9  &  18.6  &  19.3 & $  3.38\times10^{0} $ &  ~$+36.5/-\!\!23.7$~  & ~$+3.8/-\!\!4.2$~ \\
   30 --  35  &   32.3  & $  1.40\times10^{0} $ &   6.4  &  18.2  &  19.3 & $  1.61\times10^{0} $ &  ~$+37.9/-\!\!24.5$~  & ~$+4.4/-\!\!4.6$~ \\
   35 --  40  &   37.3  & $  6.81\times10^{-1} $ &   1.3  &  17.4  &  17.5 & $  7.56\times10^{-1} $ &  ~$+38.7/-\!\!24.1$~  & ~$+5.6/-\!\!3.6$~ \\
   40 --  45  &   42.3  & $  3.50\times10^{-1} $ &   1.3  &  13.7  &  13.7 & $  3.85\times10^{-1} $ &  ~$+37.0/-\!\!24.7$~  & ~$+4.4/-\!\!4.8$~ \\
   45 --  50  &   47.4  & $  1.91\times10^{-1} $ &   1.3  &  12.9  &  13.0 & $  2.06\times10^{-1} $ &  ~$+37.4/-\!\!24.5$~  & ~$+4.9/-\!\!4.6$~ \\
   50 --  60  &   54.5  & $  7.73\times10^{-2} $ &   1.3  &  13.3  &  13.4 & $  9.17\times10^{-2} $ &  ~$+36.4/-\!\!24.5$~  & ~$+4.6/-\!\!5.5$~ \\
   60 --  70  &   64.6  & $  2.88\times10^{-2} $ &   1.5  &  13.5  &  13.6 & $  3.27\times10^{-2} $ &  ~$+36.0/-\!\!23.9$~  & ~$+6.2/-\!\!5.0$~ \\
   70 --  80  &   74.6  & $  1.11\times10^{-2} $ &   1.8  &  12.1  &  12.3 & $  1.29\times10^{-2} $ &  ~$+35.5/-\!\!23.3$~  & ~$+6.5/-\!\!5.4$~ \\
   80 --  90  &   84.6  & $  4.96\times10^{-3} $ &   2.4  &  13.3  &  13.5 & $  5.41\times10^{-3} $ &  ~$+34.0/-\!\!22.7$~  & ~$+6.7/-\!\!5.1$~ \\
   90 -- 110  &   98.7  & $  1.59\times10^{-3} $ &   2.8  &  13.0  &  13.3 & $  1.74\times10^{-3} $ &  ~$+32.2/-\!\!22.2$~  & ~$+7.1/-\!\!6.0$~ \\
  110 -- 130  &  118.8  & $  3.54\times10^{-4} $ &   5.7  &  14.0  &  15.1 & $  3.76\times10^{-4} $ &  ~$+30.3/-\!\!21.0$~  & ~$+9.3/-\!\!6.2$~ \\
  130 -- 170  &  145.8  & $  5.09\times10^{-5} $ &  10.5  &  16.1  &  19.2 & $  5.46\times10^{-5} $ &  ~$+27.9/-\!\!20.9$~  & ~$+10.3/-\!\!7.0$~ \\
\hline\hline
\end{tabular}

\vspace*{10mm}

\centering
\caption{Same as in Table~\ref{tab:cross9} but for $1.5<|y^\gamma|<2.5$ and $2.4\lt|y^{\rm jet}|\leq3.2$, $y^{\gamma}y^{\rm jet} \le 0$.}
\label{tab:cross16}
\begin{tabular}{ccccccccc} \hline\hline \\ [-2.5ex]
 ~$\Ptg$ bin~ & ~$\langle \Ptg \rangle$~ & \multicolumn{7}{c}{\tcs (pb/GeV)} \\ [0.1ex] \cline{3-9} \\ [-2.5ex]
  ~(GeV)~     & ~(GeV)~    & ~Data~  & ~$\delta_{\rm stat} (\%)$~ & ~$\delta_{\rm syst}(\%)$~ & ~$\delta_{\rm tot}(\%)$ ~ & ~NLO~ & ~$\delta_{\rm scale}(\%)$~ & ~$\delta_{\rm pdf}(\%)$~ \\\hline
   20 --  23  &   21.4  & $  3.45\times10^{0} $ &   4.8  &  23.3  &  23.8 & $  4.40\times10^{0} $ &  ~$+41.0/-\!\!25.3$~  & ~$+4.7/-\!\!5.5$~ \\
   23 --  26  &   24.4  & $  2.06\times10^{0} $ &   6.1  &  22.4  &  23.3 & $  2.42\times10^{0} $ &  ~$+42.0/-\!\!25.6$~  & ~$+7.0/-\!\!4.6$~ \\
   26 --  30  &   27.9  & $  1.22\times10^{0} $ &   6.9  &  21.7  &  22.7 & $  1.27\times10^{0} $ &  ~$+42.6/-\!\!26.2$~  & ~$+5.5/-\!\!5.8$~ \\
   30 --  35  &   32.3  & $  6.82\times10^{-1} $ &   8.8  &  19.8  &  21.7 & $  5.58\times10^{-1} $ &  ~$+43.4/-\!\!26.9$~  & ~$+7.1/-\!\!5.3$~ \\
   35 --  40  &   37.3  & $  2.64\times10^{-1} $ &   1.3  &  18.4  &  18.5 & $  2.39\times10^{-1} $ &  ~$+43.3/-\!\!26.8$~  & ~$+7.3/-\!\!6.5$~ \\
   40 --  45  &   42.3  & $  1.26\times10^{-1} $ &   1.4  &  17.9  &  18.0 & $  1.09\times10^{-1} $ &  ~$+43.8/-\!\!27.2$~  & ~$+7.9/-\!\!6.2$~ \\
   45 --  50  &   47.3  & $  5.73\times10^{-2} $ &   1.5  &  14.8  &  14.9 & $  5.23\times10^{-2} $ &  ~$+44.4/-\!\!27.8$~  & ~$+9.1/-\!\!6.3$~ \\
   50 --  60  &   54.4  & $  2.23\times10^{-2} $ &   1.5  &  16.3  &  16.3 & $  2.00\times10^{-2} $ &  ~$+43.3/-\!\!28.2$~  & ~$+8.9/-\!\!7.7$~ \\
   60 --  70  &   64.5  & $  7.36\times10^{-3} $ &   2.1  &  13.9  &  14.1 & $  5.49\times10^{-3} $ &  ~$+45.1/-\!\!27.9$~  & ~$+12.6/-\!\!6.3$~ \\
   70 --  80  &   74.5  & $  2.36\times10^{-3} $ &   3.2  &  15.0  &  15.3 & $  1.64\times10^{-3} $ &  ~$+44.9/-\!\!28.3$~  & ~$+13.9/-\!\!8.6$~ \\
   80 --  90  &   84.5  & $  7.09\times10^{-4} $ &   5.2  &  17.6  &  18.4 & $  5.08\times10^{-4} $ &  ~$+46.9/-\!\!28.6$~  & ~$+17.9/-\!\!7.4$~ \\
   90 -- 170  &  110.5  & $  5.45\times10^{-5} $ &   7.0  &  17.9  &  19.2 & $  3.05\times10^{-5} $ &  ~$+48.6/-\!\!29.9$~  & ~$+21.5/-\!\!9.4$~ \\
\hline\hline
\end{tabular}
\end{table*}

\clearpage
\bibliography{prd_tgj}

\begin{thebibliography}{99}
\bibitem{d0_coordinate}
The polar angle $\theta$ is defined
with respect to the positive $z$ axis, which is along the proton beam direction,
while the azimuthal angle $\phi$ is defined with respect to the $x$ axis.
Rapidity is related to the polar scattering angle $\theta$ by
$y=0.5\ln[(1+\beta\cos\theta)/(1-\beta\cos\theta)]$, with $\beta = |\vec{p}| / E$.
Also, $|\eta_{\rm det}| = -\ln[\tan(\theta_{\rm det}/2)]$ and $|\phi_{\rm det}|$ 
are the pseudorapidity and the azimuthal angle defined with respect to the geometric center of the D0 detector.
\end{thebibliography}
\bibliographystyle{apsrev}

\end{document}